\documentclass{aa}  

\usepackage{graphicx}
\usepackage{txfonts}
\begin{document}

   \title{Impact of stellar population synthesis choices on forward modelling-based redshift distribution estimates}

   \author{Luca Tortorelli
          \inst{1}
          \and
          Jamie McCullough\inst{1,2,3}
          \and
          Daniel Gruen\inst{1,4}
          }

   \institute{Universit\"ats-Sternwarte, Fakult\"at f\"ur Physik, Ludwig-Maximilians-Universit\"at M\"unchen, Scheinerstr. 1, 81679 M\"unchen, Germany\\
              \email{luca.tortorelli@physik.lmu.de}
         \and
             Kavli Institute for Particle Astrophysics and Cosmology, Department of Physics, Stanford University, Stanford, CA, USA
         \and
             SLAC National Accelerator Laboratory, Menlo Park, CA, USA
         \and
             Excellence Cluster ORIGINS, Boltzmannstr. 2, 85748 Garching, Germany
             }

    \date{Received XXXX; accepted XXXX}

% \abstract{}{}{}{}{} 
% 5 {} token are mandatory
 
  \abstract
   {The forward modelling of galaxy surveys has recently gathered interest as one of the primary methods to achieve the required precision on the estimate of the redshift distributions for stage IV surveys, allowing them to perform cosmological tests with unprecedented accuracy. One of the key aspects of forward modelling a galaxy survey is the connection between the physical properties drawn from a galaxy population model and the intrinsic galaxy spectral energy distributions (SEDs), achieved through stellar population synthesis (SPS) codes (e.g. \textsc{FSPS}).  However, SPS requires a large number of detailed assumptions on the constituents of galaxies, for which the model choice or parameter values are currently uncertain.} 
   {In this work, we perform a sensitivity study of the impact that the variations of the SED modelling choices have on the mean and scatter of the tomographic galaxy redshift distributions.} 
   {We assumed the \textsc{Prospector-}$\beta$ model as the fiducial input galaxy population model and used its SPS parameters to build 9-bands $ugriZYJHK_s$ observed-frame magnitudes of a fiducial sample of galaxies. We then built samples of galaxy magnitudes by varying one SED modelling choice at a time. We modelled the colour-redshift relation of these galaxy samples using the self-organising map (SOM) approach that optimally groups similar redshifts galaxies by their multidimensional colours.  We placed galaxies in the SOM cells according to their simulated observed-frame colours and used their cell assignment to build colour-selected tomographic bins. Finally, we compared each variant's binned redshift distributions against the estimates obtained for the original \textsc{Prospector-}$\beta$ model.}
   {We find that the SED components related to the initial mass function,  as well as the active galactic nuclei,  the gas physics, and the attenuation law substantially bias the mean and the scatter of the tomographic redshift distributions with respect to those estimated with the fiducial model.}
  {For the uncertainty of these choices currently present in the literature and regardless of the applied stellar mass function based re-weighting strategy, the bias in the mean and the scatter of the tomographic redshift distributions are greater than the precision requirements set by next-generation Stage IV galaxy surveys,  such as the \textit{Vera C. Rubin} Observatory’s Legacy Survey of Space and Time (LSST) and Euclid.}


   \titlerunning{Impact of SPS choices on forward modelling-based $N(z)$ estimates}

   \maketitle
%
%-------------------------------------------------------------------

\section{Introduction}
\label{sect:introduction}

In recent years, the results from galaxy surveys have begun to enter the realm of precision cosmology, a prerogative that had previously been unique to studies based on cosmic microwave background data. Next-generation cosmological surveys, such as the \textit{Vera C. Rubin} Observatory’s Legacy Survey of Space and Time (hereafter, Rubin-LSST, \citealt{Ivezic2019}), Euclid \citep{Racca2016}, and the Roman space telescope \citep{Dore2019}, are aimed at providing the best tests yet of models for the dark energy,  including constraints on $\sigma_8$ and $\Omega_M$ that either reconcile or further strengthen the existing tension between the cosmological parameters estimated from the early Universe probes and those from the late Universe (e.g. \citealt{Planck2020,Heymans2021,Abbott2022}). 

To perform such cosmological measurements with the large-scale structure distribution, we need to measure the statistics of the density field evolving over time. Two of the main probes used for this analysis are the weak gravitational lensing shear field and the galaxy clustering. Weak lensing (see \citealt{Mandelbaum2018} for a comprehensive review) is widely considered as one of the most promising probes to study the growth of dark matter structure and the dark energy equation of state because light from galaxies gets continuously lensed by the intervening matter; as a result, this gets imprinted on the measured shape of galaxies. By measuring the latter for millions of objects, we can reconstruct the intervening mass distribution and the statistics of the density field. Galaxy clustering relies on using galaxies as a biased tracer \citep{Desjacques2018} of the underlying matter field at all physical scales. Galaxy clustering provides information about the large scale structure distribution of matter in the Universe, which, in turn, allows us to obtain information about the structure growth and the expansion history of the Universe. For both probes, the cosmological information is usually extracted using the angular two-point correlation function that offers a glimpse into the spatial distribution of galaxies in tomographic photometric redshift bins (e.g. \citealt{Rodriguez-Monroy2022}).

Both weak lensing and galaxy clustering measurements are a challenging endeavour because they require a strong control of the observational systematic effects \citep{Massey2013,Shapiro2013,Mandelbaum2015}. This is especially true for the systematics related to the accurate measurement of galaxy shapes \citep{Kacprzak2012,Kacprzak2014}, the spatially varying survey properties \citep{Rodriguez-Monroy2022}, and those related to the robust estimates of galaxy redshifts \citep{Crocce2016,Myles2021}.  Indeed,  accurate knowledge of the redshift bins of the galaxy samples observed by cosmological galaxy surveys is required to obtain precise cosmological parameter estimates. 

This is the reason why stage IV experiments \citep{Albrecht2006} have stringent requirements on the tomographic redshift distribution uncertainties. The latter is generally modelled in terms of uncertainty on the mean redshift of each tomographic bin, since cosmological parameters are most sensitive to shifts in the redshift mean $\bar{z}$ \citep{Amon2022,vandenbusch2022,Li2023,Dalal2023} and uncertainty in redshift scatter, $\sigma_z$, within a bin, which has a second-order effect on the amplitude of lensing signals; however, it has a first-order effect on the strength of angular galaxy clustering. In the case of Rubin-LSST, the requirements set down by the LSST Dark Energy Science Collaboration (LSST DESC, \citealt{LSSTDESC2018}) do foresee that for weak lensing measurements, the systematic uncertainty in the mean redshift of each source tomographic bin should not exceed $0.002 (1 + z)$ in the year 1 (Y1) analysis. We note that this becomes $0.001(1+z)$ in the year 10 (Y10) analysis. For galaxy clustering, the systematic uncertainty on the mean should not exceed $0.005 (1 + z)$ in the Y1 and $0.003(1+z)$ in the Y10 analysis. In addition, the systematic uncertainty in the source redshift scatter, $\sigma_z$, should not exceed $0.006(1 + z)$ in the Y1 and $0.003(1+z)$ in the Y10 weak lensing analysis, while for galaxy clustering these numbers lower to $0.1(1 + z)$ for Y1 and $0.03(1+z)$ for Y10.

Rubin-LSST is poised to measure broad-band fluxes for billions of objects up to the $i$-band depth of 26.3 for the Y10 catalogue \citep{Ivezic2019}. Obtaining a redshift for each object via spectroscopy is therefore not a viable solution and we need to rely on photometric redshift (photo-z) estimates. Unfortunately, the photo-z calibration accuracies of existing methods are still not able to achieve the Rubin-LSST requirements for the weak lensing and galaxy clustering analysis. Although the most recent methods \citep{Wright2020,Myles2021} when applied to Stage III dark energy experiments (under idealised conditions) yield uncertainties on the mean redshift that are of the order of $0.001-000.6$, there are a number of effects that significantly degrade the redshift characterisation. Indeed, effects such as spectroscopic incompleteness, outliers in the calibration redshift, sample variance, blending, and astrophysical systematics may cause systematic errors in the mean redshift and scatter that are $\sim 7-10 \times$ larger than the Stage IV requirements \citep{Newman2022}.

Two main factors concur with the lack of the required accuracy on the redshift distribution estimates (see \citealt{Newman2022} for a comprehensive review). The first is the challenge of estimating the redshift of an individual galaxy precisely since estimates are generally based on measurements in only a few broad noisy photometric bands where very few spectral features can be used to constrain the galaxy redshift, leading to degeneracies between multiple fits to a galaxy spectral energy distribution (SED, \citealt{Buchs2019,Wright2019,Wang2023}). The second factor is having a good knowledge of the galaxy population, which (given the intrinsic uncertainty described above) sets an informative prior on the redshift of a photometrically observed galaxy. This is a truly fundamental problem because often the spectroscopic redshift samples on which photo-z methods are trained, or physical models of the galaxy population are built, systematically miss populations of galaxies, giving rise to biased colour-redshift relations.

The forward modelling of galaxy surveys offers an alternative approach to the problem of accurate galaxy redshift distribution estimates. The forward process involves the detailed modelling of all the observational and instrumental effects of a galaxy survey. This includes the modelling of the galaxy population (i.e.  the intrinsic distribution of redshift-evolving physical properties of galaxies),   modelling of the galaxy stellar populations (i.e.  the connection between the physical properties and the galaxy SEDs),  mapping of intrinsic properties and fluxes to the observed ones by means of image (see \citealt{Plazas2020} for a review) and spectra simulators \citep{Fagioli2018,Fagioli2020}, and the characterisation of the selection function of galaxies observed in a survey, conditioned on these physical properties and observational and instrumental effects. This method has already shown promising results in characterising survey redshift distributions and galaxy population properties when applied to photometric and spectroscopic galaxy surveys \citep{Fagioli2018,Tortorelli2018,Fagioli2020,Tortorelli2021,Kacprzak2020,Herbel2017,Bruderer2016,Alsing2023,Fortuni2023,Leistedt2023,Alsing2024,Moser2024}. 

The ongoing forward modelling efforts model the galaxy population physical properties, such as redshifts, stellar masses, and star formation histories and metallicities,  either via parametric relations (e.g. \citealt{Tortorelli2020,Wang2023,Alsing2023,Moser2024}) or machine learning models \citep{Alsing2024}. The galaxy stellar population models, which in this work are referred to the set of prescriptions that are used to generate a galaxy spectrum  based on its physical properties (e.g. the stellar templates, the prescription for the dust attenuation, active galactic nuclei emission, initial mass function, velocity dispersion), are instead created using three main methods: empirically determined templates (e.g. \citealt{Blanton2007,Brown2014} templates), stellar population synthesis (SPS) codes (e.g. Flexible Stellar Population Synthesis \textsc{FSPS} \citealt{Conroy2009,Conroy2010}) and, equivalently but computationally much more feasible, emulators of SPS-generated spectra, such as \textsc{SPECULATOR} \citep{Alsing2020}, \textsc{SPENDER} \citep{Melchior2023} and \textsc{DSPS} \citep{Hearin2023}.  The empirically determined templates are generally controlled by a limited number of parameters,  for instance the coefficients of the linear combination of templates \citep{Tortorelli2021},  while the SPS-based SEDs are more flexible, as they allow us to simulate the flux coming from a galaxy via the modelling of all its emission components (e.g.  stars, gas, dust, and active galactic nuclei).  In the literature, the population of SPS-based SEDs, especially those that employ \textsc{FSPS}, are often constructed either by drawing from the stellar population model parameter distributions employed in the \textsc{Prospector-}$\alpha$ \citep{Leja2017,Leja2018,Leja2019} and \textsc{Prospector-}$\beta$ models \citep{Wang2023,Wang2023b}, or by using the default values implemented in a code like \textsc{FSPS}.  However, it has been shown in \cite{Conroy2009} that the choices of the parameters of the stellar population components affect the galaxy colours up to the magnitude level. Since the galaxy colours are used to assign a galaxy to a redshift bin for weak lensing and galaxy clustering studies, as well as to estimate the redshift distribution of a bin via the colour-redshift relation, studying the impact of the choices of the stellar population parameters and components on the SPS-based forward modelling estimates of the galaxy redshift distributions is of crucial importance.

In this work, we evaluated the impact of stellar population parameter choices on the mean and scatter of the redshift distributions of tomographic redshift bins to evaluate whether the variations induced by those choices are within stage IV cosmological survey requirements. The primary purpose of this is to identify choices that are significantly affecting redshift distribution estimates; hence, they need to be better informed by data in order not to limit cosmological studies from these surveys.

To do this, we constructed galaxy observed-frame SEDs using \textsc{FSPS} and generated mock apparent AB magnitudes integrating the SEDs in the $ugriZYJHK_s$ bands. %CFHTLS u-band, Subaru griz bands and Ultra-VISTA YJHKs bands. 
Those bands are the ones used to train the \cite{Masters2015,Masters2017} self-organising map (SOM, \citealt{Kohonen2001}) that robustly maps the empirical distribution of galaxies in the COSMOS field to the multi-dimensional colour space spanned by the filters mentioned above. We use the remapped version of the SOM presented in \cite{McCullough2023}, therefore the $ugriZYJHK_s$ bands refer to those of the KiDS-Viking-450 survey \citep{Wright2019}.

Overall, SOMs have been used to calibrate the colour-redshift relation with spectroscopic quality redshift estimates \citep{Masters2015,Masters2017,Masters2019,Stanford2021,McCullough2023} and for the redshift calibration of the weak lensing source galaxies \citep{Hildebrandt2020,Myles2021}. In this work, we used the mock apparent AB magnitudes to place galaxies in the remapped SOM and used the redshift bin definition in \cite{McCullough2023} to construct the redshift distributions. We then computed the mean and scatter of redshift in each of those tomographic bins for different stellar population parameter choices to evaluate whether their impact is within stage IV cosmological surveys requirements.

The \textsc{FSPS} code requires physical properties of galaxies as input,  such as star formation histories (SFHs), metallicities,  and redshifts. Therefore, we need a model that serves both as a galaxy population model, from which these physical properties should be drawn, and as a stellar population model, from which the stellar population parameters used to build the SED can be drawn instead. We use the \textsc{Prospector-}$\beta$ model \citep{Wang2023} as our input galaxy and stellar population model, which has been proven to provide a realistic representation of the galaxy population as shown by its use to robustly measure the physical properties of galaxies observed with both the Hubble Space Telescope (HST) and the James Webb Telescope (JWST, \citealt{Wang2024}).  We used the mock colours,  mean, and the scatter of the tomographic redshift distributions obtained from this model as our fiducial set of properties. We then built new samples of galaxies varying one stellar population component at a time. Additionally, we also considered the case where we vary all of them at the same time for each galaxy within the observationally meaningful ranges set for every component. We varied all the relevant stellar population components implemented in \textsc{FSPS}, from those connected to stellar physics to those related to the active galactic nucleus (AGN) model and the gas emission. We then evaluated the differences in colours, means and scatters of the tomographic bins against the fiducial ones. We also applied re-weighting strategies based on fitting the stellar mass function to test whether any SED mis-modelling might be compensated for by matching the observed abundance of galaxies in a survey.

In Sect. \ref{sect:gal_pop_model}, we describe the \textsc{Prospector-}$\beta$ galaxy population model from which we sampled the physical properties of galaxies and the stellar population parameters that are used by \textsc{FSPS} to generate the galaxy SEDs. Section \ref{sect:stellar_pop_components_description}
describes the stellar population components and their parameters that we vary in this study. The impact of each component on the rest-frame and the observed-frame galaxy colours is presented in Sect. \ref{sec:impact_on_colours}, while in Sect. \ref{sec:impact_on_colour_redshift} we evaluated the impact that the same components have on the SOM-based colour-redshift relation, and in particular on the SOM cell assignments. Section \ref{sect:impact_on_tomo_bins} describes the SED modelling impact on the tomographic redshift distribution means and scatters, while Sect. \ref{sect:discussion} contains the discussion on the implications of the bias induced by the stellar population components. Section \ref{sect:conclusions} summarises the main findings of this study and provides future directions on SPS-based forward modelling studies.

\begin{figure*}
   \centering
   {\includegraphics[width=15cm]{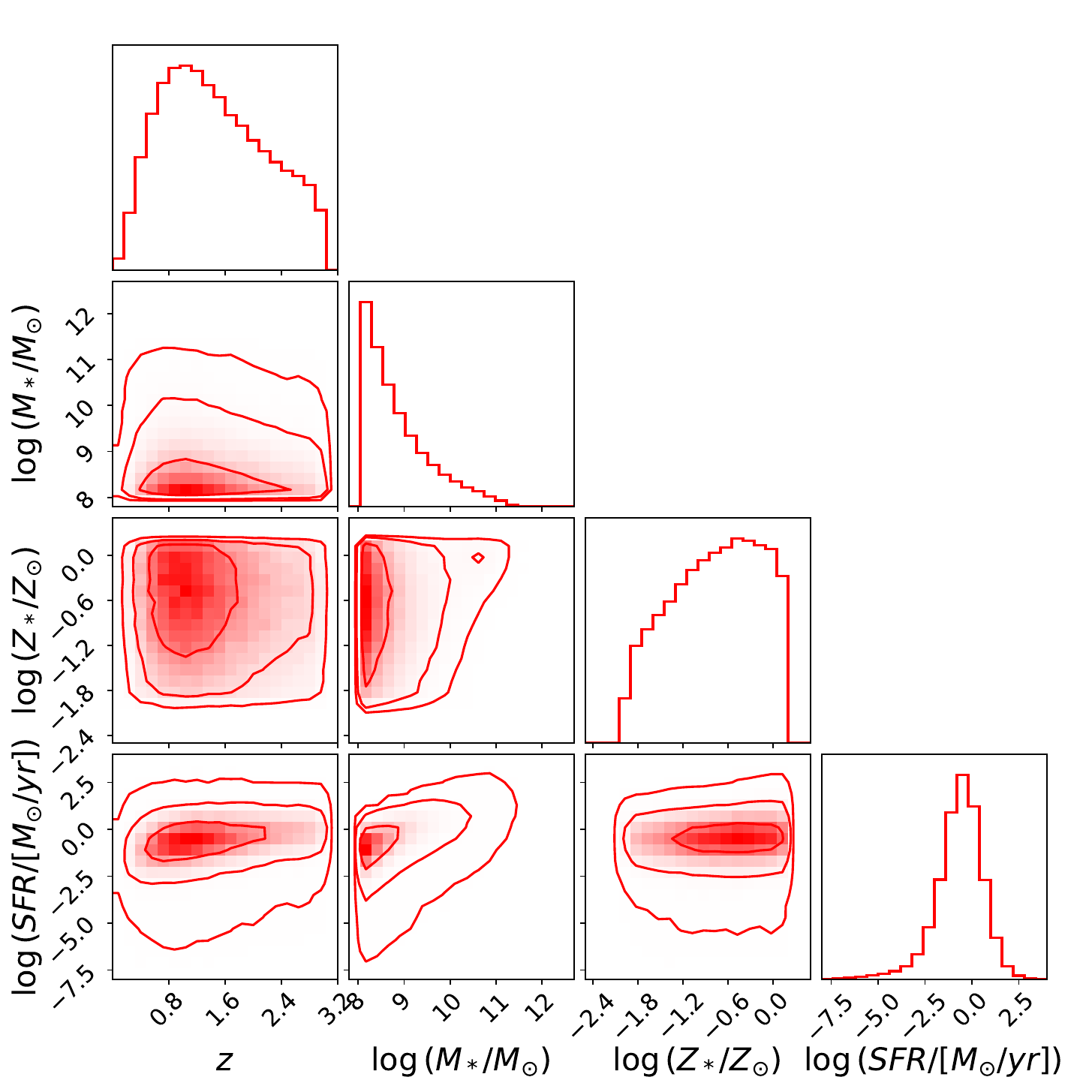}}
      \caption{Distribution of physical properties of galaxies drawn from the \textsc{Prospector-}$\beta$ model. The redshifts and stellar masses are jointly drawn from the sum of two Schechter functions \citep{Leja2020}, the star formation rates (averaged over the last $100$ Myr) are computed from the non-parametric star formation histories, while the stellar metallicities are drawn from a clipped Gaussian approximating the stellar mass-stellar metallicity relationship of \cite{Gallazzi2005}.}
         \label{fig:tortorelli_2024_figure1}
\end{figure*}

\section{Galaxy population model}
\label{sect:gal_pop_model}

Generating galaxy SEDs with SPS codes requires as input a number of physical properties of galaxies, most notably the galaxy star formation and metallicity histories. These histories are used by codes like \textsc{FSPS}, together with a set of stellar population prescriptions, to build up the composite galaxy input spectra.

To generate a realistic representation of the galaxy population properties, we used the \textsc{Prospector-}$\beta$ galaxy population model \citep{Wang2023,Wang2023b}. When used as a prior, this model has been proven to be successful in recovering stellar ages, star formation rates (SFR) and rest-frame colours in the redshift range $0 .2 \lesssim z \lesssim 15$ of galaxies from the UNCOVER survey \citep{Bezanson2022}, which uses HST and JWST data in the Abell 2744 cluster field \citep{Wang2024}. The \textsc{Prospector-}$\beta$ model is tightly linked to \textsc{FSPS} to generate galaxy SEDs as both galaxy and stellar population model, with most of the stellar population components and parameter choices related to those implemented in \textsc{FSPS} itself.

The \textsc{Prospector-}$\beta$ model is a refinement of the \textsc{Prospector-}$\alpha$ model, presented in \cite{Leja2017,Leja2018,Leja2019} and used to measure the stellar mass function \citep{Leja2020} and the star forming sequence \citep{Leja2022} of $0.2 < z < 3.0$ galaxies in the 3D-HST \citep{Skelton2014} and COSMOS-2015 survey \citep{Laigle2016}. The \textsc{Prospector-}$\alpha$ model has also been used in \cite{Alsing2020} in conjunction with \textsc{FSPS} to generate the rest-frame SEDs on which the \textsc{SPECULATOR} emulator has been trained. 

The \textsc{Prospector-}$\beta$ model aims at providing a realistic representation of the galaxy population in the Universe through the use of an observationally informative prior on the galaxy stellar mass function and a non-parametric SFH matched to the cosmic star formation rate density (SFRD).  The stellar mass function adopts a Schechter-like behaviour of the distribution of galaxy stellar masses \citep{Marchesini2009,Baldry2012,Muzzin2013,Moustakas2013,Tomczak2014,Grazian2015,Song2016,Weigel2016,Davidzon2017, Wright2018,Weaver2023}.  In the \textsc{Prospector-}$\beta$ model, the evolution of the mass function $\Phi(\mathcal{M},z)$ is modelled as the sum of two Schechter functions \citep{Leja2020}, where the logarithmic form of a single Schechter function $\Phi(\mathcal{M})$ at fixed redshift $z$ is 
\begin{equation}
    \Phi(\mathcal{M}) = \ln{(10)} \ \phi_* 10^{(\mathcal{M} - \mathcal{M^{*}})(\alpha+1)} \exp{(-10^{\mathcal{M} - \mathcal{M^{*}}})} \ ,
\end{equation}
with $\mathcal{M} = \log{M_*}$ and $\phi_*$, $\mathcal{M^*}$ and $\alpha$ evolving with redshift. In our work, we set the minimum and maximum stellar masses for the mass function to $M_{*,\rm min}=10^8 M_{\odot}$ and $M_{*,\rm max}=10^{12.5} M_{\odot}$. The stellar mass function in \cite{Leja2020} is continuous in redshift and defined only between $0.2 \le z \le 3.0$, while for $z <0.2$ and $z>3.0$, the model adopts a stellar mass function fixed at $z=0.2$ and $z=3.0$ parameter values, respectively. The public release of the \textsc{Prospector-}$\beta$ model allows for the use of higher redshift mass functions (e.g.  \citealt{Tacchella2018}). However, in our work we generated samples of galaxies in the redshift range $0.2 \le z \le 3.0$ that is similarly used by other galaxy population models,  for instance CosmoDC2 \citep{Korytov2019}.

The stellar masses drawn from $\Phi(\mathcal{M},z)$ are then used to put constraints on the SFH via a joint prior.  In \textsc{Prospector-}$\beta$,  the SFH is non-parametric and is described by the mass formed in seven logarithmically spaced time bins,  where the ratio between SFRs in adjacent bins $\log{(\mathrm{SFR}_n)/\mathrm{SFR}_{n+1}}$ is drawn from a Student-t distribution \citep{Leja2019b}.  \textsc{Prospector-}$\beta$ matches the expectation value in each bin in lookback time to the cosmic SFRD in \cite{Behroozi2019},  in order to be consistent with a global peak in SFR at cosmic noon and a systematic trend with mass.  It also allows for a mass dependence on the start of the age bins \citep{Wang2023}
\begin{equation}
    \log{(t_{\mathrm{start}}/\mathrm{Gyr})} = \log{(t_{\mathrm{univ}}(z)/\mathrm{Gyr})} + \delta_m \ ,
\end{equation}
where the $\delta_m$ value depends on the galaxy stellar mass. This new prior tries to match the expectation that high-mass (low-mass) galaxies form earlier (later). However, it does not account for the quiescent and star forming bimodality, approximating the double-peaked distribution of SFRs at a given mass and redshift as a wide single-peaked distribution. The justification for this approximation provided in \cite{Wang2023} is that the quiescent fraction, computed using the specific star formation rate (sSFR), roughly matches with the observed trend at $z < 3$ \citep{Leja2022}, which also provides a further reason for generating mock galaxies in this redshift range.

The galaxy stellar metallicity prescription $P(Z_{*}|M_{*})$ allows to sample galaxy stellar metallicities $\log(Z_*/Z_{\odot})$ conditioned on their stellar masses $\log(M_*/M_{\odot})$. Stellar metallicity affects the optical-to-near-infrared flux ratios and helps to set the normalisation and shape of the SED blueward of the Lyman limit, therefore properly modelling it is of great importance. The stellar metallicity is modelled as a clipped Gaussian distribution where the mean and the standard deviation at fixed stellar mass are taken from the $z=0$ SDSS stellar mass-stellar metallicity relationship of \cite{Gallazzi2005}. The clipped range of possible galaxy stellar metallicities is set to match that of the MIST isochrones \citep{Dotter2016,Choi2016,Paxton2011,Paxton2013,Paxton2015} used in \textsc{FSPS}, namely $-1.98 < \log{(Z_*/Z_{\odot})} < 0.19$. The standard deviation as a function of stellar mass is defined as the $84\mathrm{th}-16\mathrm{th}$ percentile range from the \cite{Gallazzi2005} relationship. This definition of the standard deviation is roughly twice the observed one from the $z=0$ relationship, and is adopted in order to account for potential unknown systematics or redshift evolution. The distribution of physical properties of galaxies drawn from the \textsc{Prospector-}$\beta$ model is shown in Fig. \ref{fig:tortorelli_2024_figure1}.

The \textsc{Prospector-}$\beta$ model also provides the prescriptions for the stellar population components and their parameters that are required to generate SEDs with \textsc{FSPS}. For some prescriptions, \textsc{Prospector-}$\beta$ uses the default implementation in \textsc{FSPS}, while for others, namely the emission from the gas component, the dust attenuation and emission, and the AGN component, it uses specific prescriptions.

The galaxy gas component is modelled as a two-component emission, the nebular continuum and the nebular emission lines. The stellar ionising continuum is self-consistently used to power nebular lines and continuum emission, following the \textsc{CLOUDY} code implementation within \textsc{FSPS} \citep{Byler2017}. The two parameters that govern the gas emission are the gas-phase metallicity $\log(Z_{\mathrm{gas}}/Z_{\odot})$ and the gas ionisation parameter $\log U$, which is the ratio of ionising photons to the total hydrogen density.  In \textsc{Prospector-}$\beta$,  the gas-phase metallicity is decoupled from the stellar one \citep{Shapley2015,Steidel2016}, and it is drawn from a flat prior in the range $-2.0 < \log(Z_{\mathrm{gas}}/Z_{\odot}) < 0.5$, following the observations that the gas in the star forming galaxies at higher redshifts has metallicity abundances that may differ significantly from their stellar abundances \citep{Shapley2015,Steidel2016}.  The ionisation parameter is instead kept fixed at $\log U=-2$, although, as mentioned in \cite{Leja2019}, gas in star forming galaxies at higher redshift might experience a stronger ionising radiation field that requires raising the ionisation parameter to $\log U=-1$ \citep{Cohn2018}. 

The dust attenuation is modelled with two components \citep{Charlot2000}, a birth-cloud and a diffuse dust screen. The birth cloud component adds extra attenuation towards young stars, simulating their embedding in molecular clouds and HII regions. It affects nebular emission as well. The birth cloud attenuation component scales as
\begin{equation}
\hat{\tau}_{\lambda,1} = \hat{\tau}_1 (\lambda / 5500 \ \AA)^{-1.0}
\end{equation}
and only attenuates  stars that have formed in the last $10\ \rm Myr$, which is the typical timescale for the disruption of a molecular cloud \citep{Blitz1980}.  In \cite{Charlot2000} the birth cloud attenuation scales as $\lambda^{-0.7}$,  but the wavelength dependence varies across studies (e.g.  \citealt{Nelson2019}).  In \textsc{Prospector-}$\beta$ the authors adopt a shallower dependence with wavelength, $\lambda^{-1.0}$, which was introduced in \cite{Leja2017} to be an average value for local galaxies of different stellar masses and SFRs.  The diffuse component, instead, attenuates all the light from the stars and the nebular emission of a galaxy. Its wavelength dependence is set by the \cite{Noll2009} prescription,
\begin{equation}
    \hat{\tau}_{\lambda,2} = \frac{\hat{\tau}_{2}}{4.05} [k^{'}(\lambda) + D(\lambda)] \left ( \frac{\lambda}{\lambda_V} \right)^n \ ,
\end{equation}
which combines the fixed \cite{Calzetti2000} dust attenuation curve $k^{'}(\lambda)$ with a Lorentzian-like Drude profile $D(\lambda)$ describing the UV dust bump, whose strength is linked to the diffuse dust attenuation index $n$ \citep{Noll2009,Kriek2013}. The prescription in \cite{Noll2009} is meant to provide a robust fit to the individual galaxy SEDs and is motivated by the observations that show that a single attenuation curve, such as the \cite{Calzetti2000} curve, is insufficient to describe the diversity of attenuations found in star forming galaxies, particularly at high redshift. Therefore, a more complex law is required to describe the dust attenuation from a generic galaxy, with the \cite{Calzetti2000} law providing a reasonable basis as an average for large samples.  The diffuse dust attenuation index $n$ expresses a power-law modifier to the slope of the \cite{Calzetti2000} dust attenuation curve, allowing for dust attenuations with a flatter slope than this widely used attenuation law.

The optical depth of the diffuse dust $\hat{\tau}_2$ is drawn from a truncated ($\hat{\tau}_{2,\mathrm{min}}=0$, $\hat{\tau}_{2,\mathrm{max}}=4$) normal distribution with mean $\mu_{\hat{\tau}_2} = 0.3$ and standard deviation $\sigma_{\hat{\tau}_2}=1$, while the optical depth of the birth-cloud component $\hat{\tau}_1$ is drawn from a truncated normal distribution of its ratio with $\hat{\tau}_2$, having mean $\mu_{\hat{\tau}_1/\hat{\tau}_2}=1$, standard deviation $\sigma_{\hat{\tau}_1/\hat{\tau}_2}=0.3$, and range $0 < \hat{\tau}_1/\hat{\tau}_2 < 2$. The diffuse dust attenuation index $n$ is uniformly distributed in the range $-1.0 < n < 0.4$.

\begin{figure*}
\includegraphics[width=\hsize]{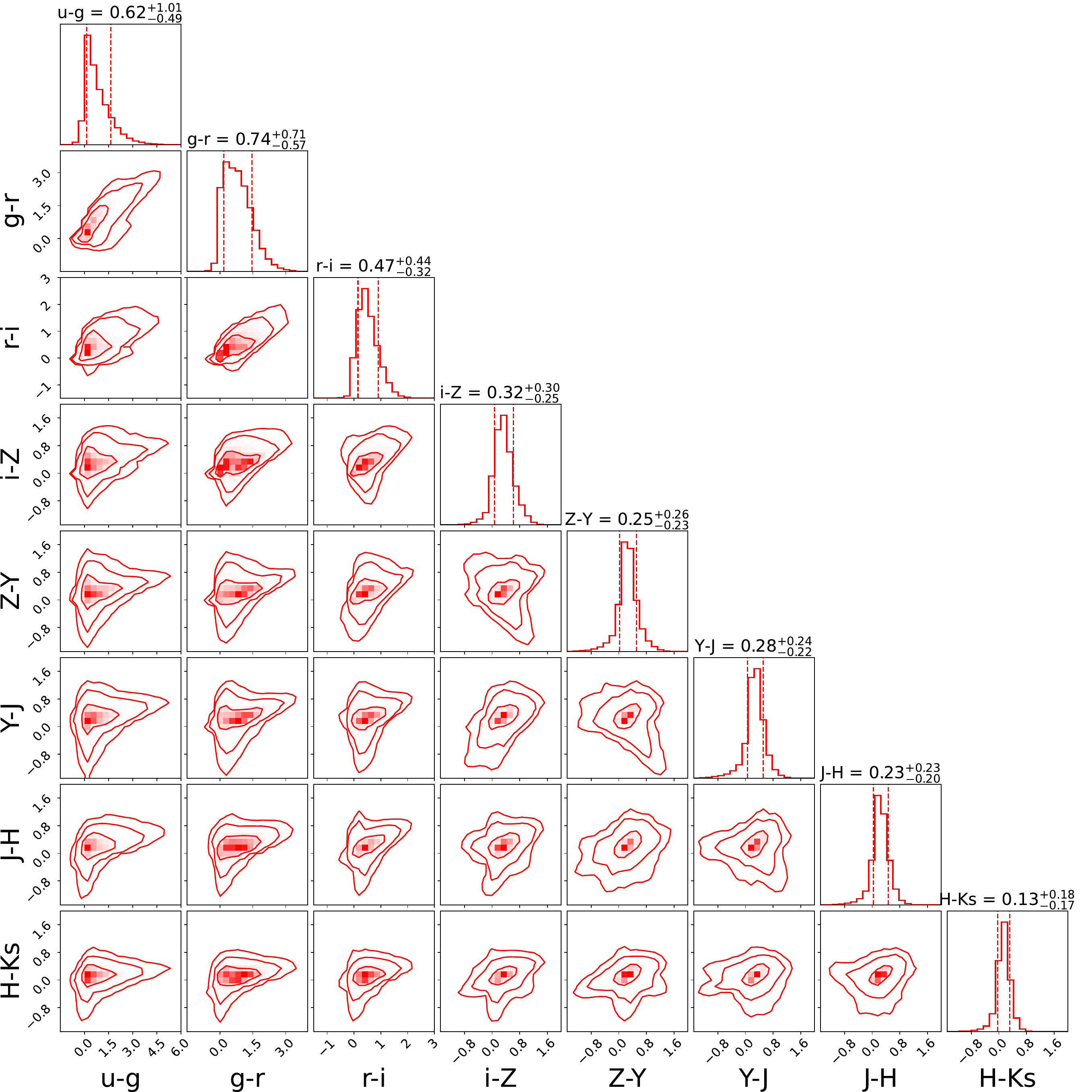}
   %\resizebox{\hsize}{!}{\includegraphics{colours_fiducial.pdf}}
      \caption{Distribution of model colours obtained from the sample of $10^6$ galaxies drawn from the \textsc{Prospector-}$\beta$ model. The colours are obtained using mock apparent magnitudes in the KiDS-VIKING $ugriZYJHK_s$ bands. We display the 16th, 50th, and 84th quantile values for each colour at the top of the corresponding 1D histogram. The dotted lines in each 1D histogram are the 16th and 84th quantile values. The 2D contours enclose the $68\%$, $95\%,$ and $99\%$ of the values.}
         \label{fig:tortorelli_2024_figure2}
\end{figure*}

\textsc{Prospector-}$\beta$ includes also a prescription to model the emission of infrared radiation coming from the dust that has been heated by the starlight. This makes use of the \cite{Draine2007} dust emission templates that are based on the silicate-graphite-polycyclic aromatic hydrocarbon (PAH) model of interstellar dust \citep{Mathis1977,Draine1984}. This model has three free parameters: $U_{\rm min}$, representing the minimum starlight intensity to which the dust mass is exposed, $\gamma_e$, representing the fraction of dust mass which is exposed to this minimum starlight intensity and $Q_{\mathrm{PAH}}$, describing the the fraction of total dust mass in PAHs. $U_{\rm min}$ and $\gamma_e$ have uniform priors in the ranges $0.1 < U_{min} < 15$ and $0 < \gamma_e < 0.15$, respectively, while $Q_{\mathrm{PAH}}$ is drawn from a truncated normal distribution having mean $\mu_{Q_{\mathrm{PAH}}}=2$ and $\sigma_{Q_{\mathrm{PAH}}}=2$.

Accurate forward modelling of a galaxy SED requires also the addition of an AGN component. In the \textsc{Prospector-}$\beta$ framework, this is achieved using AGN templates from the \cite{Nenkova2008,Nenkova2008b} CLUMPY models. These templates do not reproduce the broad and narrow-line region emission lines commonly found in quasar spectra, but they are created using radiative approximations by shining an AGN broken power-law spectrum through a clumpy dust torus medium. This model is characterised by two variables: $f_{\mathrm{AGN}}$, describing total luminosity of the AGN expressed as a fraction of the bolometric stellar luminosity, and $\tau_{\mathrm{AGN}}$, which is the optical depth of an individual dust clump at $5500 \ \AA$. In the \textsc{Prospector-}$\beta$ model, a log-uniform, redshift independent prior is adopted for both parameters, specifically in the ranges $-5 < \log{(f_{\mathrm{AGN}})} < \log{3}$ and $\log{5} < \log{(\tau_{\mathrm{AGN}})} < \log{150}$.

For our work, we sampled $10^6$ sets of galaxy physical and stellar population properties from the \textsc{Prospector-}$\beta$ model prior described above and we used those to generate galaxy observed-frame SEDs with \textsc{FSPS}. We then integrated the SEDs in the KiDS-VIKING $ugriZYJHK_s$ bands to obtain a sample of mock apparent AB magnitudes and colours. The colours and the redshift distributions of samples selected by colour-magnitude from the \textsc{Prospector-}$\beta$ model constitute our fiducial estimates against which the impact of the modifications of the stellar population components is evaluated. Figure \ref{fig:tortorelli_2024_figure2} shows the distribution of colours resulting from the \textsc{Prospector-}$\beta$ model.

\section{Description of the stellar population components}
\label{sect:stellar_pop_components_description}

The evaluation of the impact of the stellar population modelling components on the galaxy tomographic redshift distributions is motivated by the fact that there are still numerous phases of stellar evolution that are subject of intense research, being not well understood on both theoretical and observational grounds (see \citealt{Conroy2009} for an in-depth discussion). This evaluation is carried out by modifying the stellar population components and their parameters to generate new sets of colours, and therefore redshift distributions through the colour-redshift relation, which are compared against the fiducial estimates.

We generated galaxy magnitude samples in the $ugriZYJHK_s$ bands, separately varying all the stellar population modelling components implemented in \textsc{FSPS} at fixed physical properties of galaxies, for a total of 21 sets of apparent magnitudes per galaxy. We also considered the case where all the modelling components are varied at the same time within the ranges set for each individual component. We labelled this augmented version as `P-$\beta$+'. The components we varied are: the gas emission, dust attenuation and emission, AGN component, initial mass function (IMF),  velocity dispersion,  specific frequency of blue straggler stars,  fraction of blue horizontal branch stars,  post-asymptotic giant branch phase,  circumstellar asymptotic giant branch dust emission,  thermally pulsating asymptotic giant branch normalisation scheme,  presence of Wolf-Rayet stars, and the inter-galactic medium absorption. The new sets of magnitudes were then used to generate the galaxy colours needed to model the colour-redshift relation with the SOM (see Sect. \ref{sec:impact_on_colour_redshift}).

The following sections provide a description of the physical details and the prescriptions we used to model the stellar population components we varied in our work. We also report the observationally motivated ranges over which the stellar population component parameters were varied.

\subsection{Gas emission}
\label{sect:gas_emission_description}

The radiation emitted by young stellar populations ionises the surrounding gas in galaxies, producing nebular emission. The latter has two components: a nebular continuum, generated by Bremsstrahlung, recombination and two-photon emission processes, and a nebular line emission, generated by recombination and electronic transitions from forbidden and fine-structure lines. The amount of emission from the two components depends mostly on the strength of the ionising field and on the metallicity of the gas. This is why many codes rely solely on these two parameters to predict the continuum and the line emission fluxes.  The \textsc{FSPS} code includes the nebular contribution on top of the stellar population one by means of the \textsc{CLOUDY} \citep{Ferland2017} spectral synthesis code for astrophysical plasmas.  The \textsc{CLOUDY} code measures the number of ionising photons from single stellar populations (SSPs) younger than $10 \ \mathrm{Myr}$, removes them from the final SED, and uses their energy to compute line luminosities and nebular continua. These are slow computations due to the complex set of equations they have to solve, therefore \textsc{FSPS} implements a grid of pre-computed line luminosities and nebular continua from \cite{Byler2017} that depends on SSP age, SSP and gas-phase metallicity, and ionisation parameter $U$.

The nebular line emission can contribute between $20\%$ and $60\%$ to the UV-optical rest-frame broadband fluxes and is responsible for nearly all the optical emission lines present in the spectra of star forming galaxies \citep{Anders2003,Reines2010}. The higher the sSFR of the galaxy, the higher is the contribution of the emission lines to the broad-band photometry. The gas metallicity impacts the strength of the emission lines, while the ionisation parameter has an effect on the relative intensity of emission lines \citep{Leja2017}, contributing less to Balmer lines (e.g. $\mathrm{H\alpha}$, $\mathrm{H\beta}$), and more to forbidden lines (e.g. $\mathrm{[OII]}$, $\mathrm{[OIII]}$). The nebular continuum emission contributes instead more significantly to the near-infrared fluxes \citep{Byler2017}. 

The observations that the gas-phase metallicity depends on the galaxy stellar mass dates back to the early results of SDSS \citep{Tremonti2004}. Recent, higher redshift studies, have shown that the gas-phase metallicity evolves with lookback time \citep{Bellstedt2021} and that there is a more complex dependence between gas-phase metallicity, stellar mass, and SFR \citep{Magrini2012,Curti2020,Bellstedt2021}. In \textsc{Prospector-}$\alpha$, \textsc{Prospector-}$\beta$ and in general SPS-based forward modelling, various prescriptions are employed to assign physically motivated values to the gas-phase metallicity and to the ionisation parameter. In \cite{Alsing2023}, the metallicity of a galaxy is built up with the stellar mass production, such that the present-day value of the stellar metallicity is equal to the gas-phase metallicity. In \cite{Leja2017}, the stellar and the gas-phase metallicity are also set to be equal, without allowing for the chemical evolution of the galaxy,  under the assumption that the stars have the same metallicity as the gas cloud from which they had formed.  Additionally, there are other studies (e.g. \citealt{Leja2019,Wang2023}) that decouple the two metallicities, with the stellar one being drawn from the \cite{Gallazzi2005} mass-metallicity relation and the gas-phase metallicity drawn from a uniform distribution of sub-solar and super-solar abundances, with ranges taken from \cite{Byler2017}. The ionisation parameter is instead generally kept fixed at $\log{U}=-2,-1$ \citep{Leja2019,Wang2023} or related to the SFR as in \cite{Alsing2023}, where the authors adopt the \cite{Kaasinen2018} relation between the gas ionisation and the SFR. Decoupling the stellar and the gas-phase metallicity and setting the ionisation parameters to a higher value are choices that are motivated by observations suggesting that the gas in high-redshift star forming galaxies experiences stronger ionising fields \citep{Tripodi2024},  possibly due to the higher cosmic SFRD \citep{Madau2014}, and gas-phase metallicity abundances that differ significantly from the stellar ones \citep{Shapley2015,Steidel2016}.

To evaluate the impact of the different choices in the gas modelling on the tomographic redshift distributions, we generated three different samples of galaxies. In the first sample, which we labelled as `No gas emission', we evaluated the extreme case where we turned the nebular continuum and the nebular emission lines off, such that the emission from the galaxy contains only the stellar flux, the AGN dust torus and the dust emission. The second sample, labelled as `$Z_* = Z_{\mathrm{gas}}$', tests the prescription of equal values of the stellar and the gas-phase metallicity against decoupling the two, without allowing for a chemical evolution in the galaxy. The third sample tests the effect of varying the ionisation parameter with respect to the fixed $\log{U}=-2$ of the \textsc{Prospector-}$\beta$ model. We labelled this sample as `Variable $\log{U}$'. We drew the ionisation parameter for each galaxy from a uniform distribution in the range $-4 \le \log{U} \le -1$, a range that is used in \cite{Byler2017} and is consistent with ionisation parameters observed in local starburst galaxies \citep{Rigby2004}.

\subsection{Dust attenuation}
\label{sect:dust_attenuation_description}

Dust is a key component of the interstellar medium (ISM) of galaxies. It forms in the atmospheres of evolved stars or remnants of supernovae and gets released into the ISM \citep{Draine2003}, where it presents itself primarily in two forms, carbonaceous dust grains, like PAHs, and silicate dust grains. Observationally, dust has two main effects: it modifies the stellar continuum, particularly in the ultraviolet to near-infrared regime,  through the absorption and the scattering of starlight,  and it re-emits the absorbed energy in the infrared with an SED characterised by both a continuum and various emission and absorption features. 

Observations show that attenuation laws exhibit a wide range of behaviours and we might expect them to vary as function of galaxy types even among similar mass galaxies at different redshifts (see \citealt{Salim2020} for a recent review).  Therefore, the modelling of the internal dust attenuation may dramatically impact the physical properties we derive for individual galaxies (e.g. in SED fitting, \citealt{Kriek2013,Salim2016}) since it involves not only absorption and scattering out of the line of sight, but also the complex star-dust geometry in a galaxy and the scattering back into the line of sight. Additionally, we need to provide a prescriptions for the galaxy population modelling that encompass the wide range of behaviours seen in observations.

In Sect. \ref{sect:gal_pop_model}, we introduce the two-component dust attenuation model used by \textsc{Prospector-}$\beta$ in which the radiation from all stars is subject to the attenuation by a diffuse dust component, with stars below the dispersal time of the birth cloud ($10 \ \mathrm{Myr}$, \citealt{Blitz1980}) seeing an additional source of wavelength-dependent attenuation. To evaluate the impact of different dust attenuation prescription on the tomographic redshift distribution means and scatters, we generated samples of galaxies for two additional widely-used dust attenuation laws: the \cite{Calzetti2000} and the Milky Way (MW) attenuation laws. We labelled these samples as `Calzetti $k(\lambda)$' and `MW $k(\lambda)$', respectively. Both of these attenuation laws are applied to all starlight equally in \textsc{FSPS}.

The \cite{Calzetti2000} attenuation law is controlled by a single parameter $\hat{\tau_{2}}$ that is related to the ratio of total to selective extinction $R_V = A_V / E(B-V)$. This parameter sets the overall normalisation of the attenuation curve which has a $\lambda^{-1}$ dependence in the $6300 \ \AA \ \leq \lambda \leq 22000 \ \AA$ range and a third degree polynomial wavelength-dependence in the  $1200 \ \AA \ \leq \lambda < 6300 \ \AA$ range (stronger effect towards bluer wavelengths). We sampled $\hat{\tau_{2}}$ from the same distribution as that in the \textsc{Prospector-}$\beta$ model, meaning a truncated normal distribution with mean $\mu_{\hat{\tau_2}} = 0.3$ and standard deviation $\sigma_{\hat{\tau_2}}=1$.

The MW attenuation law follows the prescription in \cite{Cardelli1989}, which is parametrised in \textsc{FSPS} by means of two parameters: the first one controls the ratio of total to selective extinction $R_V = A_V / E(B-V)$, while the second parameter controls the strength of the $2175 \ \AA$ UV bump, a spectral feature on the interstellar extinction curve that is widely seen in the MW and nearby galaxies, but whose origin remains largely unidentified \citep{WangYang2023}. Despite being widely assumed fixed at $R_V=3.1$, there have been indications in the literature that the the ratio of total to selective extinction deviates from this value, both in the MW itself \citep{Cardelli1989} and in the Large Magellanic Cloud \citep{Apellaniz2017}, as well as in samples of star forming galaxies from SDSS \citep{Sextl2023}. In order to conservatively select a range of values that has been measured through the observation of individual stars line of sight, we modelled the distribution of $R_V$ values using the latest results \citep{Zhang2023} from the LAMOST survey \citep{Zhao2012}. The measurements in the MW show that the $R_V$ distribution can be modelled with a Gaussian having mean $\mu_{R_V}=3.25$ and standard deviation $\sigma_{R_V}=0.25$. The strength of the UV bump is instead kept fixed at the \cite{Cardelli1989} determination for the MW.

\subsection{Dust emission}
\label{sect:dust_emission}

More than $30\%$ of the starlight energy in the Universe is re-radiated by dust at infrared and far-infrared wavelengths \citep{Bernstein2002}. When modelling the dust emission, its continuum spectrum is assumed to be close to that produced by a grey body with a single temperature (see e.g. \citealt{Hensley2021}). However, the specific grain composition (typically carbonaceous PAHs and silicate grains) gives rise to different emission and absorption features due to vibrational modes,  among others, in this otherwise featureless continuum (see \citealt{Draine2007} for a detailed discussion). 

\textsc{Prospector-}$\beta$, and subsequently \textsc{FSPS}, implements only the \cite{Draine2007} dust emission model that is parametrised by the radiation field strength and the fraction of grain mass in PAHs. Given that the dust emission starts to make a significant contribution to the galaxy flux in the mid-infrared bands, we do not expect it to have large impact on the tomographic bins selected by optical and near-infrared colours. In order to evaluate whether the dust emission has any impact on the tomographic bins via the reddest wavelength bands, we created a sample of galaxies, labelled as `No dust emission', for which the extreme modelling choice is that the contribution of dust emission has been completely neglected.

\subsection{Active galactic nuclei}
\label{sect:agn_description}

It is known that AGNs can create a stronger emission than the regular nuclei of galaxies. This extra emission is nowadays universally accepted to be generated by the presence of a supermassive black-hole ($M_{\mathrm{BH}} \gtrsim 10^{6} M_{\odot}$) actively accreting matter. AGNs have very high luminosities ($L_{\mathrm{bol}} \lesssim 10^{48} \ \mathrm{erg / s^{-1}}$), strong evolution of their luminosity function with redshift \citep{Fotopoulou2016,Kulkarni2019}, and emission that covers basically the whole electromagnetic spectrum \citep{Padovani2017}. An AGN is detectable in the optical/UV due to emission from the accretion disc, mostly in the form of broad and narrow emission lines strongly ionised by the central engine, in the infrared due to thermal emission of the dust heated by the central engine, at X-ray energies due to the corona around the accretion disc, and in gamma-ray and radio due to non-thermal radiation related to the jet. For our study, the most interesting wavelength regions where the AGN emission is thought to make a contribution are the optical/UV and infrared. We expect AGN to modify galaxy colours with a degree that depends on the relative contribution of the AGN to the overall galaxy luminosity.

The AGN component implemented in \textsc{FSPS} comes from the CLUMPY AGN dust torus model \citep{Nenkova2008,Nenkova2008b}, which has proven successful in fitting the mid-infrared observations of AGN in the nearby universe \citep{Mor2009}. These templates model only the emission from the dust torus surrounding the accretion disc. However, AGN spectra in the optical/UV are also characterised by the power-law emission from the accretion disc and by the presence of broad and/or narrow emission lines coming from the strongly ionised material in the accretion disc. These lines are not modelled within the \cite{Nenkova2008,Nenkova2008b} templates and thus neither they are in \textsc{FSPS}.  Furthermore,  the prior values assigned to the $f_{\mathrm{AGN}}$ parameter (see Sect. \ref{sect:gal_pop_model}) in \textsc{Prospector-}$\beta$ lead to a fraction of galaxies hosting AGNs (according to the criterion $f_{\mathrm{AGN}}>0.1$, \citealt{Leja2018}) that is roughly $50\%$ smaller than what the observations predict \citep{Thorne2022b}. In \cite{Leja2018}, the authors point out that the $f_{\mathrm{AGN}}$ parameter prior do not aim to model the fraction of galaxies hosting an AGN, but rather to fit the SEDs of galaxies hosting AGNs. Therefore, the values are motivated by the range of observed values for the power-law distribution of black hole accretion rates \citep{Georgakakis2017}. Furthermore, as pointed out in \cite{Leja2018}, the range is quite uncertain due to the nonlinear correlation between the AGN bolometric luminosity and the accretion rate \citep{Shakura1973}, and the fact that the fraction of AGN bolometric luminosity re-processed into MIR emission by the surrounding environment is likely highly variable \citep{Urry1995}. A more informative prior range based on the comparison between the AGN and the galaxy luminosity function is therefore desirable, possibly based on what observations on large samples of galaxies predict \citep{Thorne2022b}.

To evaluate the impact of AGN models on the colour-redshift relation, we generated four different samples of galaxies. In one, we modified the \textsc{Prospector-}$\beta$ prior by setting the AGN luminosity fraction to zero $f_{\mathrm{AGN}}=0$, such that no AGN emission is present. We labelled this sample as `No AGN'. In the other three samples, we substituted the \cite{Nenkova2008,Nenkova2008b} templates with the parametric quasar SED model presented in \cite{Temple2021}. This model has been calibrated using the observed $ugrizYJHKW12$ colours of the SDSS DR16 quasar population \citep{Lyke2020} and is aimed at covering the full AGN contribution in the $912 \ \AA \ \le \lambda \le 30000 \ \AA$ rest-frame wavelength range. It models the blue continuum in the $900 \ \AA \ \le \lambda \le 10000 \ \AA$ range due to the low-frequency tail of the direct emission from the accretion disc, the emission from the hot dust torus in the $10000 \ \AA \ \le \lambda \le 30000 \ \AA$ range, the Balmer continuum at $\sim 3000 \ \AA$, the broad and narrow emission lines in the optical/UV region, and the Lyman-absorption suppression, which has the effect of setting the model flux to zero at all wavelengths below the Lyman limit at $\lambda_{\mathrm{LLS}} = 912 \ \AA$.

In swapping the \cite{Nenkova2008,Nenkova2008b} templates with the \cite{Temple2021} ones, the only parameter of the \textsc{Prospector-}$\beta$ model we needed is $f_{\mathrm{AGN}}$, the total luminosity of the AGN expressed as a fraction of the bolometric stellar luminosity. We computed the bolometric luminosity for the AGN component as $L_{\mathrm{bol,AGN}} = f_{\mathrm{AGN}} L_{\mathrm{bol,galaxy}}$ and we followed the prescription in the \cite{Temple2021} templates code repository\footnote{https://github.com/MJTemple/qsogen/} to relate the AGN bolometric luminosity to the monochromatic $3000 \ \AA$ continuum luminosity and to the absolute $i$-band magnitude at $z=2$, as defined by \cite{Richards2006}, two quantities that are necessary to the code to produce the AGN spectrum with the right units and normalisation. We generated three samples of galaxies with the \cite{Temple2021} templates. The first one, labelled as `Temple+21 QSO', uses the Baldwin effect \citep{Baldwin1977} to control the balance between strong, peaky, systemic emission and weak, highly skewed emission of the high-ionisation UV lines as function of redshift. In the second sample, labelled as `Temple+21 QSO RELB', the balance is randomly drawn to allow for more diversity in the population of galaxies\footnote{We follow the author suggestion in a private communication.}. In the third sample, we only used the narrow emission line templates to reproduce the population of galaxies hosting AGNs where the dust torus blocks the view to the broad-line region. We labelled this latest sample as `Temple+21 Nlr'. The motivation in using the \cite{Temple2021} templates resides in a better agreement of the latter with observed AGN colours with respect to the use of the \cite{Nenkova2008,Nenkova2008b} templates (see Appendix A), which is due to the latter containing only the contribution from the dust torus emission.

\begin{figure*}
   \resizebox{\hsize}{!}
            {\includegraphics{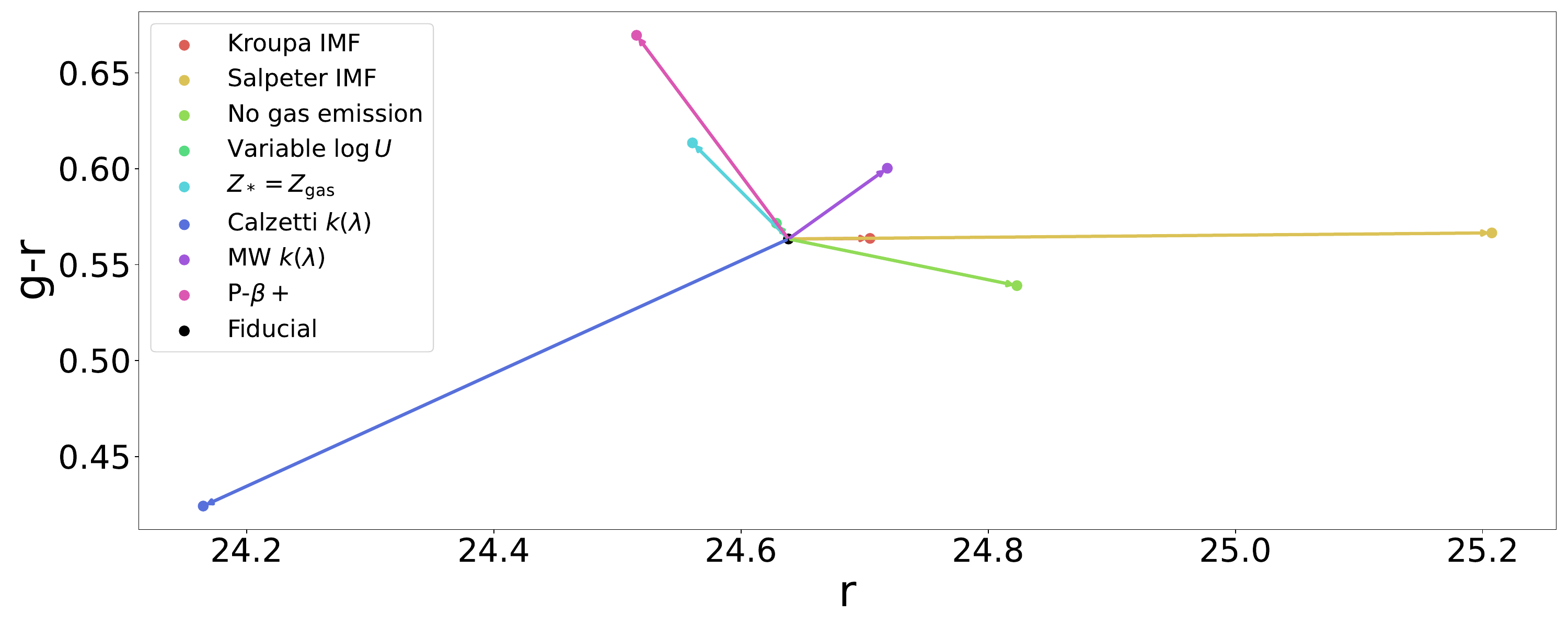}}
      \caption{Exemplary illustration of the directions in which a galaxy can move in colour-magnitude space when we vary a single stellar population component at a time with respect to the fiducial case (black point). We only show some of the variants that induce the most significant changes in the $g-r$ vs. $r$ colour-magnitude plane.}
         \label{fig:tortorelli_2024_figure3}
\end{figure*}

\subsection{Initial mass function}
\label{sect:imf_description}

The stellar IMF describes the distribution of birth masses of stars and it influences most observable properties of stellar populations and thus galaxies. The first observational characterisation of the IMF for stars in the Milky-Way is that of \cite{Salpeter1955}, who finds that the IMF can be described as a power-law of the stellar mass with slope $\Gamma = -2.35$ for $M_{*}\ge 0.5$. More modern IMFs have been described in the works of \cite{Kroupa2001} and \cite{Chabrier2003}, where the IMF power-law has the same slope than the \cite{Salpeter1955} one at high stellar masses, but fewer low-mass stars. These three IMFs are the most commonly used ones in extra-galactic studies and are generally referred to as MW-like IMFs. One of the key assumptions in modelling them has been their universality, such that galaxies in the close and far Universe were assumed to have the same IMF also found in the Milky Way, meaning that they do not depend on redshift, morphological type, local metallicity, star formation rate, or any other parameter. However, observations conducted in the last decade using spectroscopic, dynamical, and gravitational lensing measurements have instead shown the existence of a variety of IMFs. Massive ETGs, for instance, show deviations from the Milky Way in the form of more bottom-heavy IMFs (i.e.  excess of low-mass stars,  \citealt{Smith2020,Parikh2024}).  Other recent studies support instead the existence of more top-heavy (i.e.  more high mass stars produced) IMFs in nebular dominated galaxies in the early Universe \citep{Cameron2023}, as predicted by theoretical works \citep{Chon2021,Sneppen2022}.

In SPS, the IMF determines the relative weights that are assigned to the different parts of the isochrones. This implies that if a broad-band colour is dominated by a single phase of stellar evolution, then the IMF has little to not impact to it. However, if a broad-band colour is determined by a mixture of different evolutionary phases, then the IMF will down-weight or up-weight a specific phase, leading to IMF-sensitive changes in the broad-band colours. Typical colour changes induced by the IMF are of the order of 0.1 mag in the near-IR at intermediate ages (see Fig. 7 in \citealt{Conroy2009}). In addition to colours, the rate of luminosity evolution of a galaxy is also sensitive to IMF, being strongly related to the logarithmic slope of the IMF (see Fig. 8 in \citealt{Conroy2009}). Therefore, different IMFs can induce different predicted colours and mass-to-light ratios, which, in turn, lead to different predictions for physical properties of galaxies if a wrong IMF is assumed in the analysis.

\textsc{Prospector-}$\beta$ models the stellar populations with a Chabrier IMF \citep{Chabrier2003}. We therefore evaluated the impact of the IMF on the mean redshift of the tomographic bins by generating samples of galaxy properties and apparent magnitudes with two other kinds of IMF, a Salpeter \citep{Salpeter1955} and a Kroupa IMF \citep{Kroupa2001}, both implemented in \textsc{FSPS}. We labelled these two samples as `Salpeter IMF' and `Kroupa IMF', respectively.

\subsection{Velocity dispersion}
\label{sect:vel_disp}

The velocity dispersion of a galaxy, which quantifies the depth of its potential well, is not expected to make a significant contribution to broad-band colours. The observed velocity dispersion arises from the superposition of many individual stellar spectra that are Doppler shifted because of the star's motion within the galaxy. In an integrated spectrum, such as that of a galaxy, the observed velocity dispersion is going to be similar to that of the stars dominating the light from the galaxy. This implies that for integrated spectra dominated by multiple stellar populations and kinematics (e.g. bulge and disc component), the determination of the velocity dispersion is extremely complex. Hence, many studies determine the velocity dispersion only for spheroidal systems whose spectra are dominated by the light of red giant stars. 

The velocity dispersion effect is modelled in \textsc{FSPS} as a Gaussian broadening of the spectrum in velocity space in units of $\mathrm{km \ s^{-1}}$. The user provides as input the widths in terms of $\sigma$ of the smoothing Gaussian. In the fiducial case, the spectra are not smoothed by this Gaussian kernel and therefore they are at the dispersion set by the theoretical or empirical stars with which the stellar templates have been produced. To allow for a variability of the stellar velocity dispersion $\sigma_*$ values that is physically motivated, we used the redshift-dependent stellar mass-stellar velocity dispersion relation introduced in \cite{Zahid2016}:
\begin{equation}
\begin{split}
    \sigma_{\mathrm{disp}}(M_*) &= \sigma_b \left ( \frac{M_*}{M_b} \right)^{\alpha_1} \ \mathrm{for} \ M_* \le M_b \ , \\
    \sigma_{\mathrm{disp}}(M_*) &= \sigma_b \left ( \frac{M_*}{M_b} \right)^{\alpha_2} \ \mathrm{for} \ M_* > M_b \ ,
\end{split}
\end{equation}
where the redshift-dependent values of $M_b,\sigma_b,\alpha_1,\alpha_2$ are taken from Table 1 in \cite{Zahid2016}, obtained using Sloan Digital Sky Survey and Smithsonian Hectospec Lensing Survey data at $z<0.7$. The sample of galaxies generated with this prescription is labelled as `Variable $\sigma_{\mathrm{disp}}$'.

\subsection{Blue horizontal branch stars}
\label{sect:blue_hb}

The population of horizontal branch (HB) stars consists of objects with mass similar to that of the Sun that are burning helium in their cores via the triple-alpha process and hydrogen in a shell surrounding the core via the Carbon-Nitrogen-Oxygen cycle. These stars present a nearly constant bolometric luminosity \citep{Sweigart1987,Lee1990,Lee1994} which is largely driven by the constant mass of the Helium core when they enter this phase. A sub-population of the HB stars are the blue HB stars. They are a late stage in the evolution of low-mass stars ($0.8 M_{\odot} \lesssim M_* \lesssim 2.3 M_{\odot}$) found on the blue (hot) side of RR Lyrae, at the bottom of the instability strip \citep{Culpan2021}.

Observationally, blue HB stars are difficult to study for two main reasons: the paucity in the solar neighbourhood \citep{Jimenez1998} and the characteristics of their spectra. Most blue HB stars are expected to have spectra than resemble those of main-sequence A-type and late B-type stars \citep{Xue2008}. The main differences are their stronger Balmer jump, lack of metal spectral lines (indication of low-metallicity), and stronger and deeper Balmer lines \citep{Smith2010}, which can complicate the estimation of quantities such as stellar age and metallicity \citep{Lee2002}. These features make blue HB stars similar in broad-band colours to A- and B-type stars, only distinguishable with sufficient spectral resolution and signal-to-noise. 

The \textsc{FSPS} code includes the modelling of blue HB stars, however their contribution to the galaxy SED is turned off in \textsc{Prospector-}$\beta$. The blue HB stars contribution is added for stellar populations older than $5 \ \mathrm{Gyr}$ and is parametrised by a free parameter $f_{\mathrm{BHB}}$ that specifies the fraction of HB stars that are in this blue component. Following the prescription in \cite{Conroy2009}, we drew this parameter for each galaxy from a uniform distribution in the range $0 < f_{\mathrm{BHB}} < 0.5$, which we labelled as `Variable $f_{\mathrm{BHB}}$'. The addition of blue HB stars is expected to have a relatively stronger effect in the blue rest-frame colours rather than in the redder ones, not significantly altering the evolution of colours at late times, but rather giving rise to a constant blueward offset.

\subsection{Blue stragglers}
\label{sect:blue_stragglers}

Blue stragglers are old, population II main-sequence stars extending blueward of the main-sequence turnoff point at higher luminosity, firstly identified by \cite{Sandage1953}. These objects have been found in all stellar systems, from globular \citep{Piotto2004,Salinas2012} and open clusters \citep{Milone1994,Ahumada1995,Ahumada2007,Rain2021} to the field population of the Milky Way \citep{Preston2000,Santucci2015}. They are particularly abundant in open clusters \citep{Mathieu2015} and they seem to violate standard stellar evolution theory in which a co-eval population of stars should lie on a clearly defined curve in the Hertzsprung-Russell diagram determined solely by the initial mass. The fact that they lie off the main-sequence is an indication of an abnormal evolution. The general consensus is that blue stragglers start as normal main-sequence stars and then undergo some form of rejuvenation via acquisition of extra mass. The probable mechanisms for the latter are mass-transfer in a close binary \citep{McCrea1964} or dynamically induced stellar collisions and mergers \citep{Hills1976,Davies1994}. If the first process is dominant, then blue stragglers should have a non-negligible effect on the galaxy integrated light since they might be common throughout the galaxy. Conversely, if the second process is dominant, then their effect is negligible because collisions are only important in high density environments,  such as globular clusters, that contain only a small fraction of the galaxy stellar mass.

Blue stragglers cover a broader range in colours than normal main sequence stars and display A-type spectra. \cite{Li2007} found that blue stragglers makes broad-band colours bluer than $\sim 0.05 \ \mathrm{mag}$, while \cite{Xin2005} found that they contribute up to $\sim 0.2 \ \mathrm{mag}$ in the $B-V$ colours in old open clusters.  The \textsc{FSPS} code models the presence of blue stragglers in stellar populations older than $5 \ \mathrm{Gyr}$ via a parameter $S_{\mathrm{BS}} = N_{\mathrm{BS}} / N_{\mathrm{HB}}$, that defines the number of blue stragglers per unit HB stars. In \textsc{Prospector-}$\beta$, the blue stragglers contribution to the galaxy SED is turned off. In this work, we let this parameter vary uniformly in the range $0.1 < S_{\mathrm{SB}} < 5.0$ \citep{Piotto2004,Conroy2009} when generating a galaxy for our sample. This range covers typical values for both globular clusters \citep{Piotto2004} and field stars \citep{Preston2000}. The expectation is that they would contribute at most at the $\sim 0.1 \ \mathrm{mag}$ level \citep{Li2007,Xin2005,Conroy2009} in the blue bands given that they are one order of magnitude less luminous than HB stars. We labelled the sample of magnitudes generated varying the fraction of blue stragglers as `Variable $S_{\mathrm{BS}}$'.

\subsection{Asymptotic giant branch dust}
\label{sect:agb_dust}

Asymptotic giant branch (AGB) stars represent the last stage in the evolution of stars with masses in the range $0.1 M_{\odot} \lesssim M_{*} \lesssim 8 M_{\odot}$. They are very luminous due to helium shell flashes, contributing up to tens of percent to the integrated light of stellar populations \citep{Kelson2010,Melbourne2012,Conroy2013,Melbourne2013}, with photospheric spectra that reflect the mixing of inner core material with the outer layers occurring during the dredge-up process. During this phase, significant mass loss occurs as stars eject their envelopes and evolve towards the white dwarf sequence. The material that surrounds the star due to the mass loss is observed to be dust-rich \citep{Bedijn1987}. The degree of influence of this AGB dust emission on galaxy spectra is still a matter of debate \citep{Kelson2010,Chisari2012,Melbourne2013,Silva1998,Martini2013}, but what is known is that its effect manifests mostly in the mid-infrared. \cite{Villaume2015} found in particular that dust shells around AGB stars affects galaxy spectra at $\lambda \gtrsim 4 \ \mathrm{\mu m}$ by a factor of $\sim 10$ at intermediate ages $\sim 0.1 - 3 \ \mathrm{Gyr}$.

The \textsc{FSPS} code models the dusty circumstellar envelopes around AGB stars using the \cite{Villaume2015} model that makes use of the radiative transfer code \textsc{DUSTY} and of empirical prescription to assign dust shells to AGB stars in isochrones. This model is active per default in \textsc{FSPS}, and thus \textsc{Prospector-}$\beta$, and its amplitude is controlled by a free scaling parameter. We evaluated its effect on galaxy colours by generating a sample of objects for which the AGB dust model has been switched off. We labelled this sample as `No AGB dust model'.

\begin{figure*}
   \resizebox{\hsize}{!}
            {\includegraphics{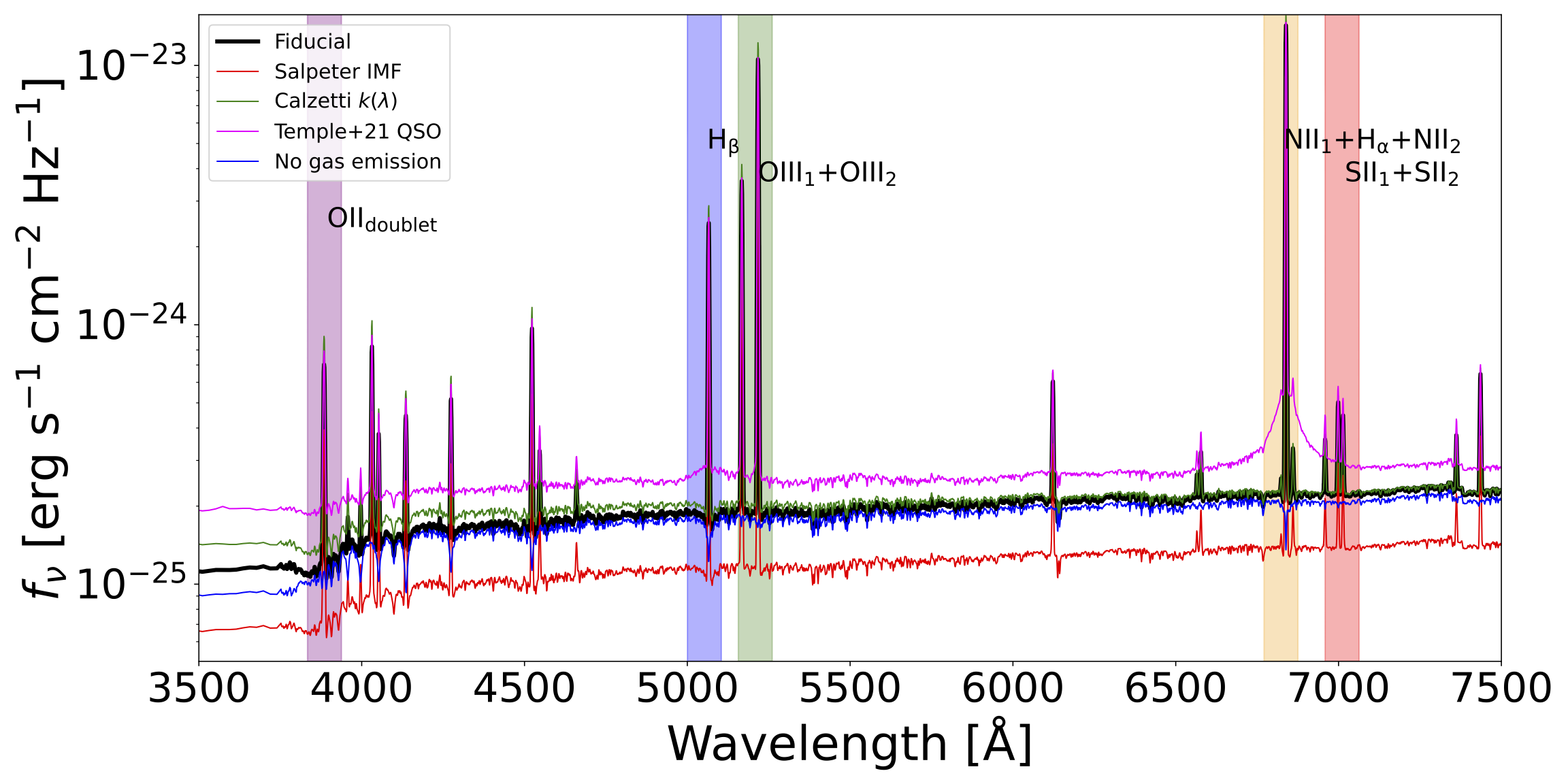}}
      \caption{Five different modelled observed-frame galaxy spectra belonging to intrinsically the same galaxy at $z=0.04$. The black line refers to the spectrum generated with the fiducial \textsc{Prospector-}$\beta$ model parameters, while the red, green, magenta and blue curves show the spectra generated varying the IMF (`Salpeter IMF'), the attenuation law (`Calzetti $k(\lambda)$'),  the AGN prescription (`Temple+21 QSO') and removing the gas component (`No gas emission'), keeping the other model parameters fixed. The vertical bands mark the regions where some of the most prominent emission lines appear. The emission line names are reported in black. The spectra units are \emph{not} arbitrarily rescaled to highlight the effect of varying the stellar population components at fixed physical properties of galaxies on the continuum flux.}
         \label{fig:tortorelli_2024_figure4}
\end{figure*}

\subsection{Post-asymptotic giant branch stars}
\label{sect:pagb}

Post-asymptotic giant branch (post-AGB) stars are a rapid phase in the evolutionary path of $0.1 M_{\odot} \lesssim M_{*} \lesssim 8 M_{\odot}$ stars that bridges the gap between the AGB phase and planetary nebula phase. It is one of the least understood phases because its duration ranges between $10^4-10^5 \ \mathrm{yr}$ for low-mass stars, down to only few decades or centuries to the most massive objects \citep{Pereira2007}. After the outer layers have been removed during the mass loss of the AGB phase, post-AGB stars increases their surface temperature at almost constant luminosity, moving horizontally along the Hertzsprung-Russel diagram. At some point, the temperature becomes high enough that the post-AGB star starts ionising its circumstellar material and become observable as a planetary nebula \citep{VanWinckel2003}.

Post-AGB stars have been detected using the IRAS $[12]-[25]$ versus $[25]-[60]$ colour–colour diagram and the $J-K$ versus $H-K$ diagram \citep{Garcia1997}, because they were expected to be located in a colour region between the AGB stars and the planetary nebulae. During the transition from the tip of the AGB to the planetary nebula phase, their spectra appear as those of M-, K-, G-, F-, A- and OB-type post-AGB supergiants for a short period \citep{VanWinckel2003}, surrounded by dust shells. When the temperature increases and the dust surrounding the post-AGB stars start to diffuse, the contribution of post-AGB stars becomes important to the UV flux from a galaxy. Post-AGB stars are also the ionising source of low-ionisation nuclear emission-line regions, particularly in early-type galaxies \citep{Belfiore2016,Byler2019}.

Post-AGB stars are implemented in \textsc{FSPS} using the \cite{Vassiliadis1994} tracks. The default behaviour of the code is to include these tracks as they are, but the user can freely change the weight given to the post-AGB phase. Therefore, we used this flexibility to generate a sample of galaxies where we turned off the contribution of post-AGB stars. We labelled this sample as `No Post-AGB stars'.

\subsection{Thermally pulsating asymptotic giant branch stars}
\label{sect:tpagb}

Thermally pulsating AGB stars (TP-AGB) represent short-living phases of the AGB which, together with AGB, HB and blue straggler stars, are important phases for SPS modelling due to their high bolometric luminosities compared to main sequence stars. TP-AGB arise from the instability that AGB stars undergo during the helium shell burning. When the latter receives new helium produced by the outer hydrogen shell fusion, the helium shell is not immediately able to adjust to the increased energy output, leading to a thermal runaway. This instability is typical of stellar populations older than $10^8 \ \mathrm{yr}$, hence having initial masses $\lesssim 5 \ M_{\odot}$. Therefore, TP-AGB stars tend to dominate the energy output redward of $\sim 1 \ \mathrm{\mu m}$ for stellar populations that are several Gyrs old \citep{Maraston2006}. Their contribution has been debated for decades, but recently \cite{Shiying2024} reported the detection of TP-AGB star signatures in the rest-frame near-infrared spectra of three young massive quiescent galaxies at $z=1-2$. 

TP-AGB stars are included in \textsc{FSPS} and their contribution can be altered using a weight factor. The default set of \textsc{FSPS} parameters uses the \cite{Villaume2015} TP-AGB normalisation scheme. However, \textsc{FSPS} allows for other two normalisation schemes to be used, the default \textsc{PADOVA} \citep{Girardi2000,Marigo2007,Marigo2008} isochrones and the \cite{Conroy2010} normalisation, which we used to generate two additional galaxy samples labelled as `TP-AGB Padova 2007' and `TP-AGB C\&G 2010', respectively. The \cite{Conroy2010} normalisation effect can be further controlled by shifting the bolometric luminosity $\log{L_{\mathrm{bol}}}$ and the effective temperature $\log{T_{\mathrm{eff}}}$ of the TP-AGB isochrones with respect to the calibrated values presented in \cite{Conroy2010}. The hotter the TP-AGB star, the bluer is the emission compared to its cooler sibling. Therefore, shifting $\log{L_{\mathrm{bol}}}$ has primarily an effect on the bands redwards of the V-band, while shifting $\log{T_{\mathrm{eff}}}$ impacts the V-K and bluer colours. We kept these values fixed to reflect the calibration in \cite{Conroy2010}. 

\subsection{Wolf-Rayet stars}
\label{sect:wolf_rayet}

Wolf-Rayet (WR) stars represent a late stage in the evolution of very massive O-type stars $\gtrsim 25 \ M_{\odot}$, with surface temperatures ranging from 20,000 K to around 210,000 K, hotter than almost all other type of star. WR occupy the same region of the Hertzsprung-Russel diagram as the hottest bright giants and supergiants. The strong ionising radiation from these stars, coupled with the expanding outer envelope of material expelled by their strong winds, leads to Doppler-broadened recombination emission lines of helium, carbon, nitrogen, oxygen, and/or hydrogen in the WR spectrum. This outflowing material and the explosion of WR stars in supernova Ib or Ic are important components of the enrichment of the interstellar medium of star forming galaxies.

WR spectra are characterised by broad emission lines and are classified according to the ratios of emission line strengths. Though short-lived ($\lesssim 10^{7} \ \mathrm{yr}$), the WR stars can dominate the UV emission from young starburst galaxies, while in the UVB photometry they cannot be distinguished from normal hot stars \citep{Crowther2007}. The \textsc{FSPS} code contains a spectral library to model the WR population, which is activated by default. To test its impact, we generated a sample of galaxies where the WR spectral library has been turned off and the main default library is used instead. We labelled this sample as `No WR'.

\subsection{Intergalactic medium absorption}
\label{sect:igm}

The bulk of baryonic cosmic matter resides in a dilute reservoir in the space between galaxies, known as intergalactic medium (IGM). The IGM has both cosmological and astrophysical implications: it can be used to test models of structure formation on the smaller comoving scales \citep{Viel2005}, its shape can bias cosmological parameter inferences from galaxy clustering \citep{Pritchard2007}, and it is the reservoir of gas from which galaxies feed. The higher the redshift of a galaxy, the more material its light has to travel through. This light is absorbed by the gaseous systems that lie along the line of sight which, at high redshift, are mostly in the form of neutral hydrogen. This effect is known as IGM absorption and at increasing redshift it can be so preponderant that all the light blueward of $\mathrm{Ly\alpha}$ at $1215.67 \ \AA$ becomes invisible to us.

The first modelling of the IGM absorption is detailed in \cite{Madau1995}. The author proposed a theoretical model which is able to reproduce the shape of the extinction curve as function of redshift, finding that the IGM transmission decreases with increasing redshift and that the scatter at a given redshift ranges between $20\%$ and $70\%$. Later models by \cite{Meiksin2006} and \cite{Inoue2014} updated \cite{Madau1995} prescription for the $\Lambda CDM $ model with weaker absorption at $3.5 \lesssim z \lesssim 5$ and stronger absorption at $z \gtrsim 6$.

\begin{figure*}
   \resizebox{\hsize}{!}{\includegraphics{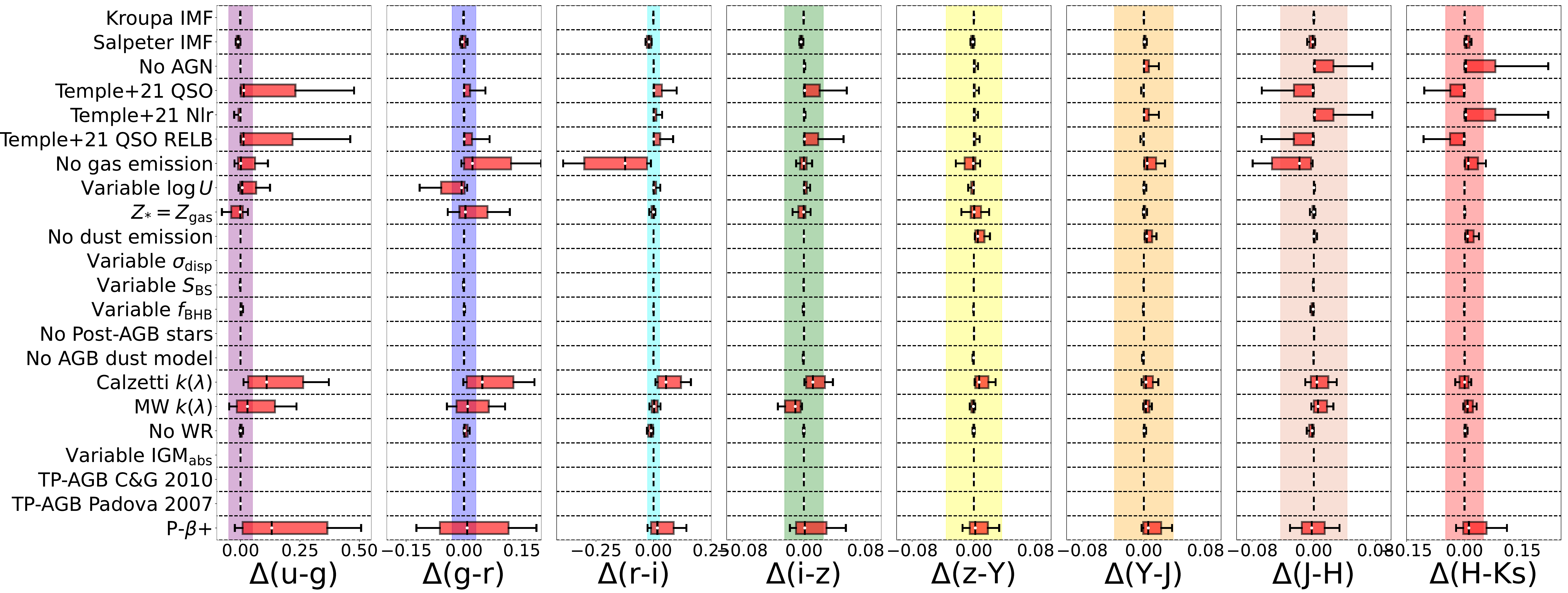}}
      \caption{Rest-frame colour differences between the fiducial galaxy sample (\textsc{Prospector-}$\beta$) and those obtained modifying one stellar population component at the time (y-axis labels). The coloured widths are the sum in quadrature of COSMOS median photometric errors in the two respective bands to indicate distinguishability on a single galaxy basis. The vertical black bars and the white dots on top of them represent the median values of each rest-frame colour difference. The light red boxes show the interquartile range (75th-25th percentile), while the whiskers represent the 84th-16th percentile ranges.}
         \label{fig:tortorelli_2024_figure5}
\end{figure*}

The \textsc{FSPS} code implements the IGM absorption using the \cite{Madau1995} prescription, with a parameter that scales the IGM optical depth. The absorption impacts only the rest-frame wavelengths below the $\mathrm{Ly\alpha}$ wavelength, which becomes visible in optical broad-bands, like the $u$-band, only from $z\sim 1.5$ onwards. The default parameters of \textsc{FSPS} do not activate the \cite{Madau1995} IGM absorption prescription. Therefore, we generated a sample of galaxies where we include the IGM absorption keeping the factor that scales the IGM optical depth fixed to 1. We labelled this sample as `Variable $\mathrm{IGM_{abs}}$'.

\subsection{Combining the various prescriptions}
\label{sect:combinig_prescriptions}

The samples of SEDs detailed in the previous sections have been created varying one single stellar population component at a time with respect to the fiducial \textsc{Prospector}-$\beta$ model. However, the observed population of galaxies in the Universe in reality spans all these properties ranges at the same time. Therefore, we created an additional sample of galaxies,  aimed at evaluating the net effect on colours and redshift distributions of varying simultaneously all the stellar population components according to the prescriptions detailed in each subsection. We labelled this sample as `P-$\beta$+'.

In particular, we randomly generated galaxy SEDs using an IMF among the Chabrier, Kroupa and Salpeter ones, a dust attenuation law among the \cite{Calzetti2000}, \cite{Cardelli1989} and \cite{Charlot2000} ones, and TP-AGB stars prescription among the \cite{Villaume2015}, \cite{Conroy2010}, and PADOVA ones. We replaced the \cite{Nenkova2008,Nenkova2008b} templates with the \cite{Temple2021} ones obtained by modelling SDSS DR16 quasars. The \cite{Temple2021} templates are controlled by a single parameter, $f_{\mathrm{AGN}}$, which  has the same meaning and values range as the one in the fiducial \textsc{Prospector}-$\beta$ model. The gas ionisation is uniformly drawn from the \cite{Byler2017} range, while the gas metallicity is for some objects equal to the stellar metallicity and for others decoupled and uniformly drawn from the \textsc{Prospector}-$\beta$ range. The dust emission, the AGB dust, the post-AGB stars, the WR stars and the IGM absorption are randomly turned on and off. The velocity dispersion follows the \cite{Zahid2016} prescription, while the fraction of blue HB stars and blue stragglers is uniformly drawn from the observational ranges in \cite{Conroy2009}. 

Figure \ref{fig:tortorelli_2024_figure3} shows, for an exemplary galaxy, the directions in colour-magnitude space towards which a galaxy can move when we vary one stellar population component at the time. The fiducial $r$-band magnitude and $g-r$ colour are represented as a black point, while the arrows point towards the set of magnitudes and colours generated with a different stellar population component. The P-$\beta +$ component (magenta arrow) represents the net movement in colour-magnitude space due to all the varied components.

\section{Impact of stellar population components on galaxy colours}
\label{sec:impact_on_colours}

In this section, we evaluated the impact that the different stellar population components have on both the rest-frame and the observed-frame galaxy colours. For each of the $10^6$ galaxies, we computed the difference between the rest-frame and the observed-frame colours estimated with the fiducial \textsc{Prospector-}$\beta$ model and those estimated varying each stellar population model component, $\Delta(\mathrm{colour}) = \mathrm{colour}_{\mathrm{fiducial}} - \mathrm{colour}_{\mathrm{component}}$. We compute this difference for the $u-g$, $g-r$, $r-i$, $i-Z$, $Z-Y$, $Y-J$, $J-H$ and $H-K_s$ colours. We selected galaxies having observed-frame $i \le 25.3$ to reproduce the Rubin-LSST Y10 lens gold sample cut \citep{LSSTDESC2018}. The use of the rest-frame colours allows us to understand the impact of the individual spectral features on the broad-band integration of galaxy spectra, while the observed-frame colours have the aim of reproducing the colour variation that we would observe in a survey that has a similar photometric precision as that expected from Rubin-LSST. The convolution of the colour difference with the redshift distribution of our sample of galaxies has the effect of smearing the colour difference across the wavelength space. The cut is applied on the $i$-band magnitude in the fiducial sample and this selection of galaxies is then held fixed for all variants. This ensures that we evaluated the colour differences on a per-galaxy basis. 

\begin{figure*}
   \resizebox{\hsize}{!}{\includegraphics{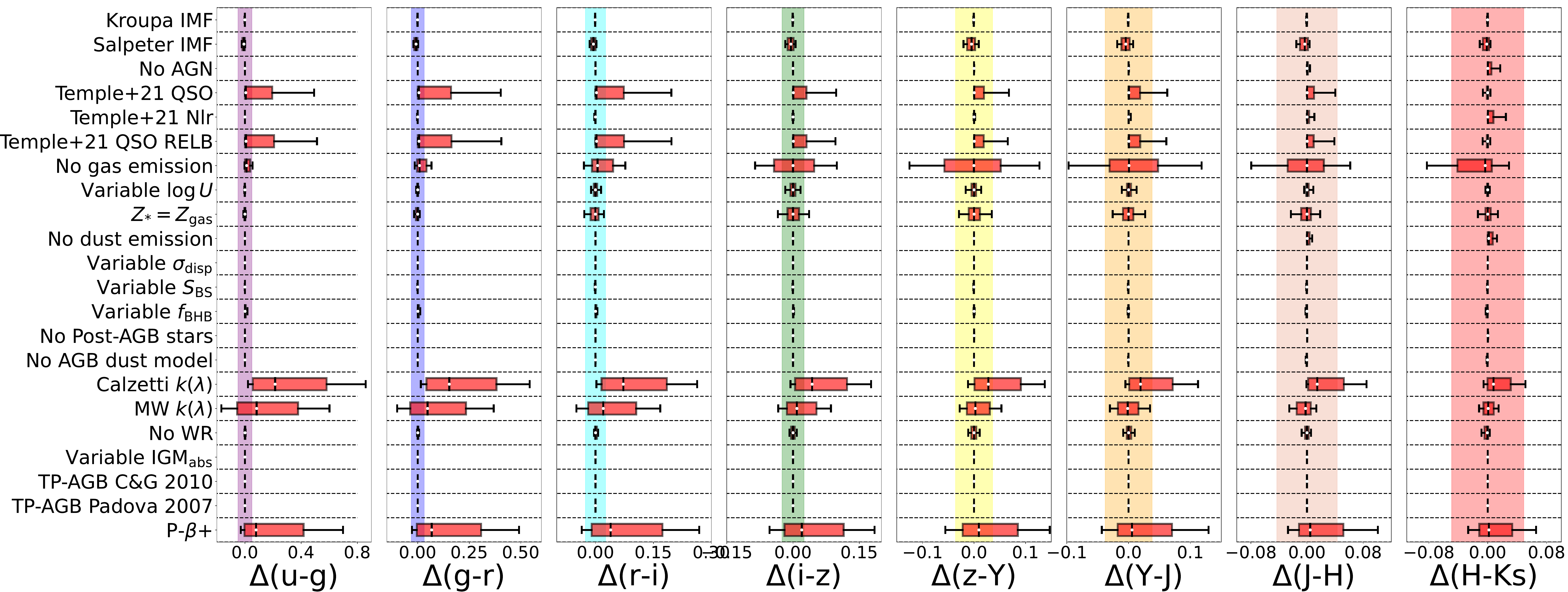}}
      \caption{Observed-frame colour differences between the fiducial galaxy sample (\textsc{Prospector-}$\beta$) and the ones obtained by modifying one stellar population component at a time. The coloured widths are the sum in quadrature of COSMOS median photometric errors in the two respective bands to indicate distinguishability on a single galaxy basis. The vertical black bars and white dots on top of them represent the median values of each observed-frame colour difference. The light red boxes show the interquartile ranges (75th-25th percentiles), while the whiskers represent the 84th-16th percentile ranges.}
         \label{fig:tortorelli_2024_figure6}
\end{figure*}

\subsection{Impact on an example galaxy SED}
\label{sect:impact_on_example_sed}

Figure \ref{fig:tortorelli_2024_figure4} shows the comparison between \textsc{FSPS} generated observed-frame spectra for a subset of the different choices on the stellar population components presented in Sect. \ref{sect:stellar_pop_components_description}. The spectra belong to the same galaxy at $z=0.04$. Different choices have a dramatic impact on the spectral appearance of the galaxy which, in turn, reflect on the galaxy magnitudes and colours.  In this figure, we compare the effects of varying the IMF (red curve), the attenuation law (green curve),  the AGN prescription (magenta curve) and the presence of gas emission (blue curve) with respect to the fiducial \textsc{Prospector-}$\beta$ model (black curve),  leaving all other parameters fixed. The Salpeter IMF produces a galaxy spectrum that overall is fainter with respect to the fiducial case that uses a Chabrier IMF. This leads to fainter magnitudes that might impact the detection of the galaxy in a flux-limited survey, and hence may impact the mix of physical properties of galaxies detected at a given colour. Changing the attenuation law from the fiducial prescription in \textsc{Prospector-}$\beta$ to the \cite{Calzetti2000} model instead induces a wavelength-dependent modification of the galaxy spectrum. The latter is particularly evident towards bluer wavelengths, where the effect of the attenuation is more prominent. Changing the AGN prescription to the one of \cite{Temple2021}, instead, gives rise to an overall brighter spectrum due to the continuum increase, but most notably leads to the appearance of the characteristic broad-line emission of AGN (see e.g. the $\mathrm{[NII]}$ and $\mathrm{H{\alpha}}$ complex in the light yellow bar).  Removing completely the modelling of the galaxy gas emission leads to an overall decrease in flux due to the lack of nebular continuum and the disappearance of the emission lines that reveal the underlying absorption lines of the stellar component. The  galaxy chosen as an example has a high value of the $f_{\mathrm{AGN}}$ parameter to emphasise the dramatic effect of changing the AGN prescription. The units are \emph{not} arbitrarily rescaled, in order to highlight the effect of varying the stellar population components at fixed physical properties on the galaxy's flux.

\subsection{Impact on rest-frame galaxy colours}
\label{sect:impact_on_rest_frame_colours}

Figure \ref{fig:tortorelli_2024_figure5} shows the distribution of the rest-frame colour differences for all the SED modelling components. We report in the figure the median photometric errors of the COSMOS2020 sample \citep{Weaver2022} that has a similar photometric precision to that expected by Rubin-LSST (see Appendix \ref{appendix:comparison_cosmos_rubin_photometry}). The distributions are plotted using a box plot representation to capture the relevant quantile ranges of the distribution of colour differences. The wider the distribution and the more offset its median with respect to the fiducial value, the larger is the effect on colours of the SED model component we vary. For the rest-frame case, the SED components that seem to have an appreciable impact on the colour distributions are those related to the dust attenuation, the AGN prescription, the gas physics and, consequently, the combined P-$\beta$+ sample.

Changing the dust attenuation prescription from the \textsc{Prospector-}$\beta$ fiducial two components \cite{Charlot2000} to the \cite{Calzetti2000} (`Calzetti $k(\lambda)$' label) or \cite{Cardelli1989} (`MW $k(\lambda)$' label) ones induces a wavelength-dependent change in the rest-frame colours, especially in the bluer bands, where the effect of the dust attenuation is more prominent. The distributions of colour differences for the \cite{Calzetti2000} attenuation law are moved towards bluer colours than the fiducial case. The median value of the distributions is beyond the median COSMOS2020 errors for all the optical colours, while it is within the errors for the near-infrared ones. The 84th-16th percentile values also follow the same trend. For the \cite{Cardelli1989} MW dust attenuation prescription, we have a similar, but mitigated behaviour. The median value of the distributions are outside the median COSMOS errors for the $u-g$ and the $g-r$ colours, while they are within the errors for the other redder colours. Interestingly, the $i-z$ rest-frame colour distribution is biased towards redder colours than the fiducial model. The 84th-16th percentile values are instead wider than the errors only for the $u-g$ and the $g-r$ colours, and only marginally for the $i-z$ rest-frame colour.

The AGN prescription effect differs depending on whether we completely remove the AGN component or on whether we use the \cite{Temple2021} model. Removing the AGN component (`No AGN' label) has an appreciable effect only in the rest-frame $J-H$ and $H-K_s$ bands since the AGN prescription in \textsc{FSPS} aims at reproducing the dust torus emission that is stronger in the infrared bands. Replacing the \textsc{FSPS} AGN model with the \cite{Temple2021} model (`Temple+21 QSO' and `Temple+21 QSO RELB' labels) leads to noteworthy changes in the rest-frame near-UV/optical colours and in the $J-H$ and $H-K_s$ colours. The colour difference in the near-UV/optical is mainly due to the fact that the \cite{Temple2021} prescription aims at modelling also the emission coming from the gas in the broad-line region that is subject to the intense ionising radiation field of the accretion disc. The difference in the $J-H$ and $H-K_s$ colours is instead due to a different treatment of the dust torus emission that leads to the \cite{Temple2021} model producing slightly redder colours than the fiducial \textsc{Prospector-}$\beta$ model. By modelling the emission from the ionised gas in the narrow-line region only (`Temple+21 Nlr' label), the rest-frame optical colour differences are mostly within the median COSMOS2020 photometric errors, while the different prescription becomes detectable in the $J-H$ and $H-K_s$ colours. This is due to the lack of dust torus emission when modelling the narrow-line region only, which predicts bluer colours with respect to the \textsc{FSPS} AGN prescription. 

\begin{table*}
\caption{Fraction of galaxies changing SOM cell assignment as function of the SED component.}
        \centering
        \begin{tabular}{p{7cm}p{5cm}p{5cm}} 
                \hline\hline
                SED component & Fraction of cell change for the full sample & Fraction of cell change for the Rubin-LSSY Y10 sample \\
                \hline
                Kroupa IMF & 1.4\% &2.3\% \\
                Salpeter IMF & 14.9\% &23.4\% \\
                No AGN & 6.6\% &6.5\%\\
            Temple+21 QSO & 57.4\% &43.6\% \\
            Temple+21 Nlr & 11.2\% &12.5\%\\
            Temple+21 QSO RELB & 57.3\% &43.7\% \\
            No gas emission & 43.6\% &63.1\%\\
            Variable $\log{U}$ & 17.2\% &25.0\%\\
            $Z_* = Z_{\mathrm{gas}}$ & 25.2\% &36.5\%\\
            No dust emission & 1.5\% &3.1\%\\
            Variable $\sigma_{\mathrm{disp}}$ & 0.1\% &0.1\%\\
            Variable $S_{\mathrm{BS}}$ & 0.8\% &1.8\%\\
            Variable $f_{\mathrm{BHB}}$ & 5.4\%  & 11.1\% \\
            No post-AGB stars & 0.5\% &0.5\%\\
            No AGB dust model & 0.5\% &0.9\%\\
            Calzetti $k(\lambda)$ & 89.6\% &80.9\%\\
            MW $k(\lambda)$ & 88.6\% &76.3\%\\
            No WR & 11.3\% &12.2\%\\
            Variable $\mathrm{IGM_{abs}}$ & 16.4\% &6.0\% \\
            TP-AGB C\&G 2010 & 0.0\% &0.0\%\\
            TP-AGB Padova2007 & 0.0\% &0.0\%\\
            P-$\beta$+ & 84.2\% &82.2\%\\
                \hline
        \end{tabular}
 \tablefoot{For each SED modelling component that we vary (left column), the table shows the fraction of galaxies whose assigned cell in the SOM changes with respect to the one assigned with the fiducial SED model. The central column refers to the case where we consider all the $10^6$ galaxies in the sample. The right-hand column refers, instead, to the case where galaxies are selected using the Rubin-LSST Y10 cut of $i\le 25.3$  (right-hand column). We apply the same cut to each component based on the $i$-band magnitudes of the fiducial sample in order to compare always the same set of galaxies.}
        \label{tab:cell_assignment_variation}
\end{table*}

The different gas physics prescriptions induce changes in the intensity or presence of emission lines which in turn reflect on the galaxy rest-frame colours. Removing completely the modelling of emission lines (`No gas emission' label) results in colour differences that are larger than the median photometric errors for the $u-g$, $g-r$ and $r-i$ optical colours. In particular, the $u-g$ and $g-r$ distributions are biased towards bluer colours, while the $r-i$ towards redder colours. The disagreement with the fiducial model is particularly prominent for the $r-i$ rest-frame colour, because this is the wavelength range where the $\mathrm{H{\alpha}}$, $\mathrm{[NII],}$ and $\mathrm{[SII]}$ lines fall.  The effects in the $u-g$ and $g-r$ colours are instead mainly driven by the complete absence in modelling the $\mathrm{[OII]}$, $\mathrm{H{\beta}}$ and $\mathrm{[OIII]}$ lines. Changing the ionisation state of the gas results in a variable relative intensity of the emission lines, which is particularly evident for the $(\mathrm{[O III]} \ \lambda \lambda 4959, 5007 \ \AA \ / \ \mathrm{[O II]} \ \lambda \lambda 3726,3729 \ \AA)$ and the $(\mathrm{[O III]} \ \lambda 5007 \ \AA \ / \ \mathrm{H \beta})$ emission line ratios. These ratios fall into the $u$ and $g$ bands as testified by the 84th-16th percentile distribution values in the $u-g$ and $g-r$ rest-frame colours being outside of the median photometric errors and leading to bluer and redder colours, respectively, if compared to the fiducial model. The use of a different prescription to assign the gas-phase metallicity to galaxies leads in a change of the overall intensity of the nebular continuum, as well as the intensity and presence of certain complexes of emission lines. The $u-g$ and $g-r$ colours are the ones most affected by the change in the gas prescription, with distributions that are biased in the opposite sense with respect to the `Variable $\log{U}$' case. Besides the lines already mentioned falling into these bands, a visual inspection of the rest-frame spectra shows that other lines contribute to the variation in rest-frame colours, for instance the presence or absence of the $\mathrm{[O III]} \ \lambda 4363 \ \AA$ line.

The distribution in the rest-frame colours for the P-$\beta$+ case can be considered as an average of the effects of all the variable SED components that we implement when creating this sample of galaxies. The optical colours are dominated by the combined effect of the dust attenuation law and AGN prescription that bias the P-$\beta$+ colours towards bluer ones than the fiducial model. The $Z-Y$, $Y-J$ and $J-H$ colour distributions are within the median photometric erros, while the $H-K_s$ infrared colour is biased towards bluer colours being an interplay between the \cite{Temple2021} prescription for the dust torus emission and the lack of dust torus prescription in the narrow-line AGN modelling when compared to the fiducial model.

\subsection{Impact on observed-frame galaxy colours}
\label{sect:impact_on_obs_frame_colours}

Figure \ref{fig:tortorelli_2024_figure6} shows the same set of SED modelling components and colour differences, but in the observed-frame, thus mimicking the effects we would observe in a survey that has a similar photometric precision than that expected from Rubin-LSST. The figure shows that the AGN prescription and the dust attenuation law are the main drivers of the change in observed-frame colours also in this case. Their impact leads to a large fraction of galaxies showing colour variations that exceeds the median photometric errors, implying detectability on a single galaxy basis. Furthermore, the redshift leads to dilution of the colour differences across the wavelength space. This leads, on one hand, to the mitigation of the effects of some of the components, for instance the prescription for the gas ionisation and the gas metallicity that have now most of the galaxies within the photometric errors, while on the other hand, a wider distribution of colour difference values for the AGN prescriptions and dust attenuation laws. The P-$\beta$+ observed-frame colours follow the trends of the main drivers of the colour changes, namely the dust attenuation and the AGN prescription. In particular, all the  distributions are biased towards bluer colours with respect to the fiducial case.

Concerning the dust attenuation, the distribution of colour differences for the \cite{Calzetti2000} attenuation law is moved towards bluer colours than the fiducial case. The median value of the distributions is beyond the median photometric errors for all the optical colours, while it is within the errors for the near-infrared ones. The 84th-16th percentile values are instead all beyond the errors except for the $H-K_s$ colour. For the \cite{Cardelli1989} MW dust attenuation prescription, we have a similar, but mitigated behaviour. The median value of the distributions are outside the median COSMOS2020 errors for the $u-g$ and the $g-r$ colours, while they are within the errors for the other redder colours. The 84th-16th percentile values are instead wider than the errors only up to the $Z-Y$ colour.

Removing the AGN modelling does not lead to changes in the observed colours that are beyond the photometric errors, while replacing the \textsc{FSPS} AGN model with the \cite{Temple2021} model leads to noteworthy changes in the observed-frame colours from the $u-g$ down to the $Y-J$ colour. In this case, the median of the colour difference distributions is only marginally offset from zero, but the 84th-16th percentile distributions are heavily biased towards bluer colours. By modelling the emission from the ionised gas in the narrow-line region only, the observed-frame colour differences are within the photometric errors. Therefore they contribute only marginally to the colour variation with respect to the fiducial model. The other SED modelling choices contribute little to the colour variation, with the case of the modelling of the TP-AGB phase that does not impact at all the studied bands. Therefore, we will not further discuss these two modelling components in the rest of this work.

Figure \ref{fig:tortorelli_2024_figure6} is interesting because it provides us with a quantitative estimate of how much the colours of a population of galaxies change when we vary the SED modelling components, thus guiding the evaluation of which SED components can be left at the fiducial values and which instead require more careful modelling and constraints from the observations. In turn, this is tightly linked to the fact that not accurately knowing the colour distributions makes the modelling of the colour-redshift relation highly uncertain. Indeed, even if the effect of varying the SED component cannot be detected on a colour for a single galaxy when using median COSMOS photometric errors, it could however be of significance on the ensemble colour distribution of a survey and thus on its redshift distributions.

\section{Impact of SED modelling on the SOM-based colour-redshift relation}
\label{sec:impact_on_colour_redshift}

In this work, we modelled the colour-redshift relation using the \cite{Masters2015,Masters2017} SOM, which maps the empirical distribution of galaxies in the multi-dimensional colour space spanned by the COSMOS $u$ band imaging from CFHT, the $griz$ imaging from Subaru Suprime Cam, and the $YJH$ imaging from the UltraVista survey \citep{McCracken2012}. In its original implementation, the SOM was trained on the COSMOS 2015 \citep{Laigle2016} $ugrizYJHK_s$ photometry. We used the KiDS-VIKING $ugriZYJHK_s$ remapped version \citep{McCullough2023} of the \cite{Masters2015,Masters2017} SOM.  In the previous section, we have assessed the dramatic impact that the different stellar population modelling choices have on colours. We show in this section that this directly affects the colour-redshift relation as modelled by the SOM, by evaluating the impact that different stellar population modelling choices have on a galaxy SOM cell assignment and on the mean redshift estimation per SOM cell.

\begin{figure*}
   \resizebox{\hsize}{!}{\includegraphics{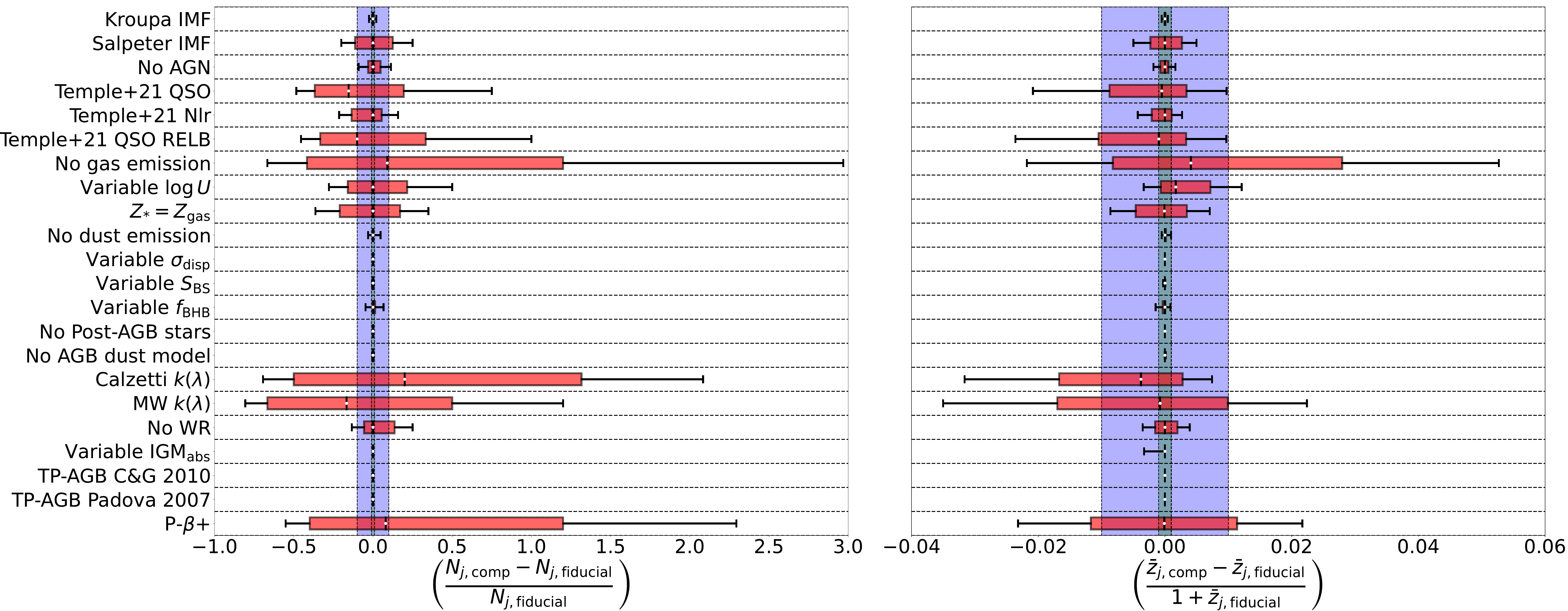}}
      \caption{Distributions of the relative difference in the cell occupation number (left panel) and in the cell mean redshift (right panel) with respect to the fiducial case.  For each pair of fiducial and variable component model, we consider only the cells that are populated by both galaxy samples. The relative differences in the cell occupation number are shown as $ \left( N_{j,\mathrm{comp}}-N_{j,\mathrm{fiducial}}\right)/N_{j,\mathrm{fiducial}}$ for  clarity. The black vertical bars and the white dots on top of them represent the median value of $ \left( N_{j,\mathrm{comp}}-N_{j,\mathrm{fiducial}}\right)/N_{j,\mathrm{fiducial}}$, the boxes represent the 75th - 25th percentile ranges and the whiskers represent the 84th - 16th percentile ranges. The light green and light blue areas show the regions where the relative difference is within $ -0.1 \le \left( N_{j,\mathrm{comp}}-N_{j,\mathrm{fiducial}}\right)/N_{j,\mathrm{fiducial}} \le 0.1$ and $ -0.01 \le \left( N_{j,\mathrm{comp}}-N_{j,\mathrm{fiducial}}\right)/N_{j,\mathrm{fiducial}} \le 0.01$, respectively. The right panel shows  the distributions of the differences in the mean redshift per cell between the fiducial case and the SED components variation cases normalised by $1+z$. The elements on the plot have the same meaning, but the light green and light blue areas show instead the regions where the relative difference of cell mean redshift is within $-0.001 \le \left(\bar{z}_{j,\mathrm{comp}}-\bar{z}_{j,\mathrm{fiducial}}\right) / (1 + \bar{z}_{j,\mathrm{fiducial}}) \le 0.001$ and $-0.01 \le \left(\bar{z}_{j,\mathrm{comp}}-\bar{z}_{j,\mathrm{fiducial}}\right) / (1 + \bar{z}_{j,\mathrm{fiducial}}) \le 0.01)$, respectively.}
         \label{fig:tortorelli_2024_figure7}
\end{figure*}

For each galaxy belonging to the different samples that we built by varying the stellar population components, we performed the SOM cell assignment procedure using the observed-frame model colours, without adding photometric noise and assuming equally sized errors in all colours, which is equivalent to a perfect cell assignment. We show in Appendix \ref{appendix:noise_impact_on_som} that the use of COSMOS2020 photometric errors has an impact on the SOM cell assignment of a galaxy that is negligible compared to its variation in colours due to stellar population choices. 

\begin{figure*}
   \resizebox{\hsize}{!}{\includegraphics{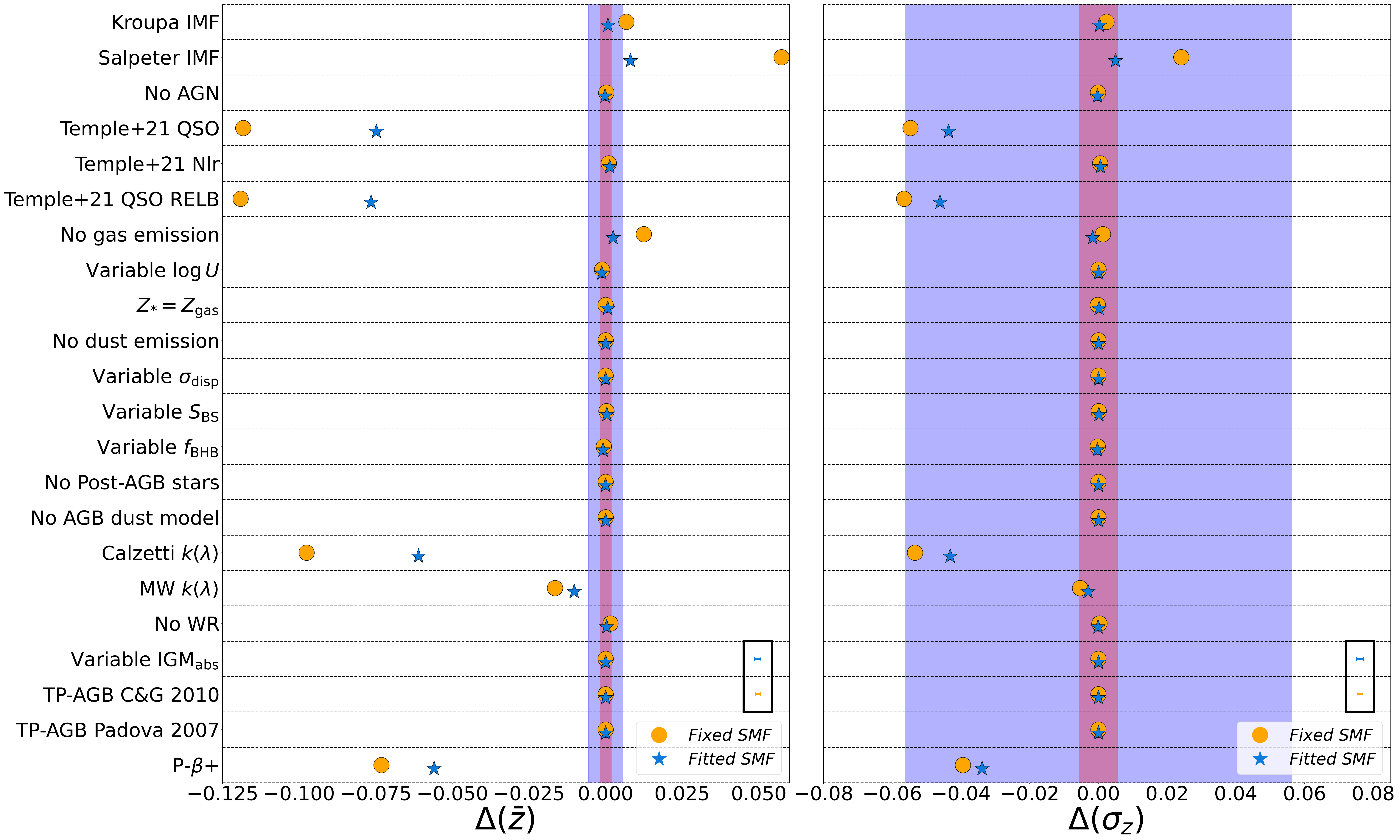}}
      \caption{Differences $\Delta (\bar{z})_{\mathrm{comp}}$ between the mean of the fiducial magnitude-limited ($i\le 25.3$) redshift distribution and those obtained from the magnitude-limited redshift distributions for each varied SED modelling component (left panel). The right panel instead shows the differences $\Delta (\sigma_z)_{\mathrm{comp}}$ on the redshift distributions scatter. The light blue stars refer to the case where the total number of galaxies for each magnitude-selected sample is matched to the fiducial number of objects via an overall magnitude offset (`Fitted SMF'), while the orange circles refer to the unmatched case (`Fixed SMF'). We report the mean errors on $\Delta (\bar{z})_{\mathrm{comp}}$ and $\Delta (\sigma_z)_{\mathrm{comp}}$ estimated via bootstrap resampling in the black boxes in the bottom right-hand corner of each subplot. The blue and red bars show the systematic uncertainty requirements on the redshift distribution mean and scatter for the galaxy clustering and weak lensing 3x2 pt analysis of Rubin-LSST Y10, as detailed in \cite{LSSTDESC2018}.}
         \label{fig:tortorelli_2024_figure8}
\end{figure*}

Table \ref{tab:cell_assignment_variation} shows the fraction of galaxies whose assigned cell in the SOM change with respect to the one assigned with the fiducial model. The left-hand column shows the fraction when considering the full sample of $10^6$ galaxies, while the right-hand column shows the case of galaxies belonging to the Rubin-LSST Y10 sample, $i \le 25.3$. We applied the Y10 cut based on the $i$-band magnitudes of the fiducial sample in order to always compare the same set of galaxies. The Y10 cut selects 191025 galaxies. The fractions track the variation in the colour distribution reported in Sect. \ref{sec:impact_on_colours}, with the largest fraction of cell change in the full sample arises from the attenuation laws. Indeed, the \cite{Calzetti2000} and \cite{Cardelli1989} attenuation laws lead to a change of $89.6 \%$ and $88.6 \%$, respectively, in the assigned cells with respect to the fiducial case for the full sample, while for the Y10 cut the values drop to $80.9\%$ and $76.3\%$. The AGN prescription also leads to a SOM cell assignment variation, which is less dramatic than the attenuation law case, but still leading to a fraction of cell assignment change of roughly $57\%$ ($\sim 43\%$) variation in the full sample (Y10 sample) when adopting the \cite{Temple2021} AGN templates. The fractions are instead lower ($\sim 11\%$) when using only the narrow-line region template. The P-$\beta$+ case leads instead to a fraction of cell change of $84.2\%$ for the full sample and $82.2\%$ for the Y10 sample. Another SED component that generates a sensible change in the SOM cell assignment is that provided by the removal of the gas component emission, `No gas emission'. Although the effect on the galaxy colours distribution of this component is smaller than other components, such as AGN prescription and dust attenuation laws, the colour change they lead to causes a variation in the SOM cell assignment with respect to the fiducial case of $43.6\%$ for the full sample and $63.1 \%$ for the Y10 sample. This is due to the unrealistic galaxy population that this case models, which is consistent with real galaxies only in the case of quiescent objects.  It is also interesting to note that the variation in the SOM cell assignment is greater for the Y10 sample than for the full sample.  The reason resides in the $i$-band cut preferentially selecting star forming objects.  In the full sample there are passive galaxies of which the `No gas emission' actually constitute a good representation of their SEDs,  but in the Y10 sample most of them are not included in the magnitude cut. Therefore in this latter case we are comparing galaxies with no gas emission against mostly star forming objects who could not be properly model in colours unless gas is added to their SEDs, hence the larger fractional change.

Other stellar population components that give a large ($\gtrsim 10\%$) SOM assignment change are the `Salpeter IMF' ($14.9\%$ for the full sample and $23.4\%$ for the Y10 sample), the `Variable $\log{U}$' ($17.2\%$ for the full sample and $25.0\%$ for the Y10 sample), the `$Z_* = Z_{\mathrm{gas}}$' ($17.2\%$ for the full sample and $25.0\%$ for the Y10 sample), and the `No WR' ($11.3\%$ for the full sample and $12.2\%$ for the Y10 sample) components, with the latter inducing a change due to its impact on the rest-frame u-band magnitudes. The other stellar population components lead to variations which are $\lesssim 10\%$.

Having assigned each galaxy to a SOM cell, we aim at characterising how much the occupation number and the mean redshift of each cell change with respect to the fiducial case when using samples generated with different stellar population components. The left-hand panel of Fig. \ref{fig:tortorelli_2024_figure7} shows the distribution of the relative difference in the occupation numbers of the SOM cells. For each pair of fiducial and variable component model, we consider only the cells that are populated by both galaxy samples. We report the relative difference as $ \left( N_{j,\mathrm{comp}}-N_{j,\mathrm{fiducial}}\right)/N_{j,\mathrm{fiducial}}$, with the light green and light blue areas in Fig. \ref{fig:tortorelli_2024_figure7} showing the regions where the relative difference is within $\pm 1\%$ and $\pm 10\%$, respectively. From this figure, we see that the only components that do not lead to a substantial variation in the cell occupation number are those related to the advanced stages of stellar evolution (blue stragglers, blue HB stars, AGB, post-AGB, TP-AGB), as well as the `Kroupa IMF', `No dust emission', `Variable $\sigma_{\mathrm{disp}}$' and `Variable $\mathrm{IGM_{abs}}$' components.

The right-hand panel of Fig. \ref{fig:tortorelli_2024_figure7} shows instead how the mean of the redshift distribution in each cell change when one varies the stellar population components with which galaxy spectra are built. In particular, we report the  difference in the mean redshift between the fiducial model and those obtained by varying one stellar population component at the time normalised by $1+z$, $\left(\bar{z}_{j,\mathrm{comp}}-\bar{z}_{j,\mathrm{fiducial}}\right) / (1 + \bar{z}_{j,\mathrm{fiducial}})$. The light green and light blue areas in Fig. \ref{fig:tortorelli_2024_figure7} show the regions where the relative difference is within $\pm 0.1\%$ and $\pm 1\%$, respectively. The same components that have percent level variation in the occupation number are those that have permille level variation in the mean redshift of the SOM cells, while the other components induce relative variations in the mean redshift that are in some cases even greater than the percent level.

The results shown in Fig. \ref{fig:tortorelli_2024_figure7} highlight the direct impact that varying a single stellar population component has on the colour-redshift relation. For the same galaxy, varying the spectrum generates a variation in colours which in turn associates the galaxy to a different SOM cell. The combined effect of the population of galaxies results in a change of cell occupation numbers and mean redshift per cell that at the cell level is much greater than the percent and permille nominal precision requirements by Stage III and IV surveys, respectively.

\section{SED modelling impact on the redshift distribution means and scatters}
\label{sect:impact_on_tomo_bins}

In this section we evaluated the impact of varying the SED modelling components on the redshift distribution of our sample of simulated galaxies. First, we evaluated how the entire redshift distribution of the sample of galaxies changes between SED model variants when we apply a pure magnitude selection. Then, we used these magnitude-selected samples to create tomographic bins following the prescription in \cite{McCullough2023}. This procedure allows us to evaluate the impact on the mean and the scatter of the tomographic redshift bins induced by the variation in the SED modelling components.

In forward modelling the galaxy population, any mis-specification of the SED model would result in a partial compensation of this impact by the fitted distribution of physical properties of galaxies, particularly the stellar mass function (SMF). What effect this has in practice depends on the galaxy data being used as a constraint and on the degrees of freedom that are given to the galaxy population model. We introduced three limiting cases, which we used to try to evaluate the extent to which this may compensate for the impact of SED mis-modelling on the colour-redshift relation.  We assume that only the SMF that is being fitted is used to compensate for the SED mis-modelling, while other physical properties such as metallicity and star formation history are kept as-is. The SMF is indeed the most flexible element of the galaxy population model for fitting the observed abundance of galaxies, which is the primary constraint in the forward-modelling of photometric surveys. Degrees of freedom on other galaxy physical property distributions may also have an interplay with SED modelling, which however is beyond the scope of this work to investigate. The three SMF cases we consider are:
\begin{itemize}
    \item `Fixed SMF': the different variants keep the distribution of all physical properties of galaxies fixed and we apply the $i$-band selection separately on each variant. This leads to different galaxies, and different total numbers of galaxies, to be selected in each variant.
    \item `Fitted SMF': we apply an overall multiplicative correction of stellar mass that compensates the SED modelling choice dependent stellar mass-to-light ratio. This is done in such a way that the overall abundance of our apparent $i$-band magnitude limited sample is kept fixed.
    \item `Fitted SMF of (type,z)': we assume a highly flexible fit of the SMF as a function of galaxy type and redshift. We realise this practically such that the exact same set of galaxies is selected into the magnitude limited sample, regardless of which variant of the SED modelling is being considered.
\end{itemize}
 
\subsection{The impact on the magnitude-limited sample}
\label{section:mag-lim-case}

A magnitude-limited selection of target galaxies is generally performed using a sharp cut in colours or magnitudes. Since the results presented in Sect. \ref{sec:impact_on_colours} highlight the dramatic effect that varying the SED components has on the galaxy magnitudes and colours, we expect a similar effect to be reflected on the redshift distributions resulting from a magnitude-limited selection. In the case of a Stage IV galaxy survey like Rubin-LSST, the LSST DESC requirements \citep{LSSTDESC2018} for the weak lensing and galaxy clustering analysis on the systematic uncertainty in the mean and redshift scatter of each source tomographic bin refer to a sample of galaxies, the Y10 lens gold sample, that has a limiting magnitude of $i_{\mathrm{lim}} = 25.3$. We used this value to select magnitude-limited samples of objects from each of the 22 sets of galaxies generated by varying the SED modelling components. 

\begin{figure*}
   \resizebox{\hsize}{!}{\includegraphics{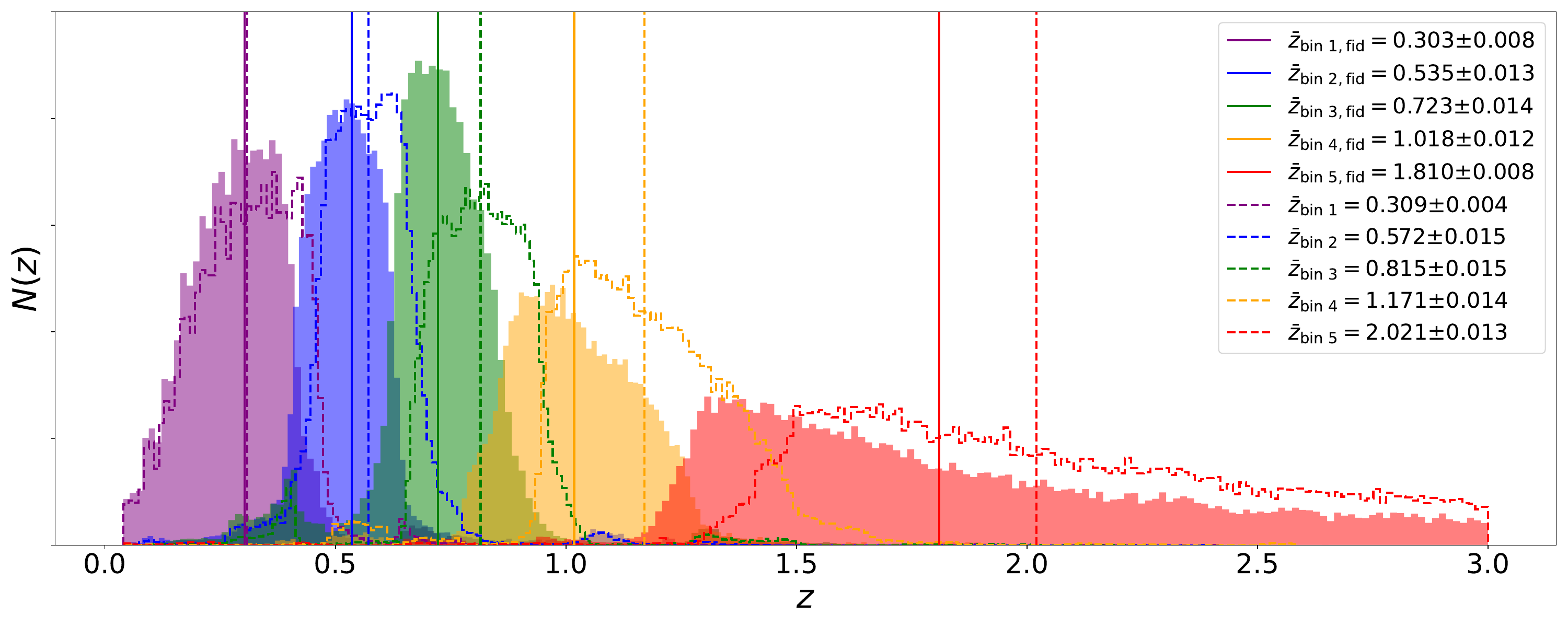}}
      \caption{Redshift distribution $N(z)$ of two sets of five tomographic bins built with the prescription highlighted in Sect. \ref{sect:impact_on_tomo_bins}. The filled histograms refer to the $N(z)$s built with the fiducial sample, while the histograms with the dashed outlines represent the $N(z)$s built using the `Calzetti $k(\lambda)$' component. The solid and dashed vertical lines represent the mean of the tomographic bins for the fiducial and the `Calzetti $k(\lambda)$' components, respectively. The legend shows the tomographic bin means and $1\sigma$ errors on the means. We use the `Calzetti $k(\lambda)$' component to demonstrate our analysis methodology as it induces the largest shift in the $N(z)$. The shift is caused by the brighter rest-frame UV emission that the `Calzetti $k(\lambda)$' component induces (see Fig. \ref{fig:tortorelli_2024_figure4}). This leaves the low-redshift galaxy observables unaffected, while it brightens high-redshift galaxies in the observed optical bands, hence making them more numerous in a magnitude-limited sample.}
         \label{fig:tortorelli_2024_figure9}
\end{figure*}

We computed the mean $\bar{z}_{\mathrm{comp}}$ and the standard deviation $\sigma_{z,\mathrm{comp}}$ for each of the 22 resulting magnitude-limited redshift distributions $N(z)_{\mathrm{comp}}$ and compared those against the mean $\bar{z}_{\mathrm{fiducial}}$ and the standard deviation $\sigma_{z,\mathrm{fiducial}}$ obtained for the magnitude-limited fiducial sample $N(z)_{\mathrm{fiducial}}$,
\begin{equation}
    \begin{split}
        \Delta (\bar{z})_{\mathrm{comp}} &= \bar{z}_{\mathrm{fiducial}} - \bar{z}_{\mathrm{comp}} \ , \\
        \Delta (\sigma_z)_{\mathrm{comp}} &= \sigma_{z,\mathrm{fiducial}} - \sigma_{z,\mathrm{comp}}.
    \end{split}
\end{equation}
The errors on these estimates are obtained using bootstrap resampling. The procedure involves repeatedly sampling with replacement from the original set of galaxies to create multiple bootstrap samples, applying the magnitude selection, computing the mean for each bootstrap sample and then the standard deviation of these bootstrap means. We generated a total of $10^3$ bootstrap samples.

For the `Fixed SMF' sample, the variation in the stellar population components generates samples of galaxies that contain a different number of objects brighter than the $i$-band magnitude cut. The number of galaxies at $i \le 25.3$ ranges from 148447 galaxies in the case of the `Salpeter IMF' to 243329 galaxies in the case of `Temple+21 QSO'. However, in modelling a galaxy population, an important property that is fitted to the observations is the total number of galaxies that the model produces. To account for this, the `Fitted SMF' sample matches the $\mathrm{d}n/\mathrm{d}m_i$ of each of the 22 components to the fiducial one by applying a constant magnitude shift to the $i$-band. This corresponds to an overall multiplicative correction of stellar mass that compensates the SED modelling choice dependent stellar mass-to-light ratio. We estimated the $i$-band magnitude shift for each component variant such that the total number of galaxies having $i \le 25.3$ is equal to that of the fiducial sample (191025). The shifted $i$-band magnitude distributions for the various SED modelling components are then used to select a sample and compute the mean and the standard deviation of the magnitude-limited redshift distributions. We computed the errors on these estimates as in the previous case (i.e. via bootstrap resampling).

Figure \ref{fig:tortorelli_2024_figure8} shows the differences $\Delta (\bar{z})_{\mathrm{comp}}$ and $\Delta (\sigma_z)_{\mathrm{comp}}$ between the mean (left-hand panel) and the scatter (right-hand panel) of the magnitude-limited fiducial redshift distributions and those obtained for each varied SED modelling component. The light blue stars refer to the $\mathrm{d}n/\mathrm{d}m_i$ matched case (`Fitted SMF'), while the orange circles to the unmatched case where the $i$-band selection produces samples of different numbers of galaxies (`Fixed SMF'). The light blue and orange error bars in the black boxes in the bottom right corners show the bootstrap-estimated errors on the mean and the scatter of each $\Delta (\bar{z})_{\mathrm{comp}}$ and $\Delta (\sigma_z)_{\mathrm{comp}}$. In particular, we report the error averaged across all the evaluated components. The blue and the red bars show the systematic uncertainty requirements on the redshift distribution mean and scatter for the galaxy clustering and weak lensing 3x2 pt analysis of Rubin-LSST Y10, as detailed in \cite{LSSTDESC2018}. The requirements state that the systematic uncertainty in the mean redshift of each tomographic bin should not exceed $0.001(1+z)$ for the weak lensing 3x2 pt analysis and $0.003(1+z)$ for the galaxy clustering analysis, while the systematic uncertainty in the photometric redshift scatter should not exceed $0.003(1+z)$ for the weak lensing 3x2 pt analysis and $0.03(1+z)$ for the galaxy clustering analysis.

The `Fixed SMF' case (orange circles) shows the dramatic impact that a sharp magnitude cut has on the mean and standard deviation of the magnitude-selected galaxy redshift distributions.  We find that a magnitude selection leads to changes in the mean of the redshift distributions up to $\Delta z \sim 0.1$. The `Kroupa IMF',  `Salpeter IMF', and `No gas emission' samples lead to redshift distributions that are biased low with respect to the fiducial case, since they tend to shift the $\mathrm{d}n/\mathrm{d}m_i$ towards fainter galaxies that are not included in the magnitude cut.  Keeping the SMF constant while changing the latter modelling components will inevitably lead to a loss of galaxies with fainter $i$-band magnitudes compared to the fiducial case,  as it is visible from the behaviour in Figs. \ref{fig:tortorelli_2024_figure4},   \ref{fig:tortorelli_2024_figure5} and  \ref{fig:tortorelli_2024_figure6}. On the other hand, the `Temple+21 QSO', `Temple+21 QSO RELB',  `Calzetti $k(\lambda)$',  `MW $k(\lambda)$', and  `P-$\beta$+' samples tend to shift the $\mathrm{d}n/\mathrm{d}m_i$ towards brighter galaxies than the fiducial case, leading to redshift distributions that are biased high. 

The mentioned components bias the mean of the redshift distribution with respect to the fiducial case by an amount that exceeds the requirements for the Y10 weak lensing and galaxy clustering analysis. The other components instead cause biases that are within the Y10 requirements. In the case of the redshift scatter, all biases are always within the requirements for the galaxy clustering Y10 analysis, while for the weak lensing analysis the components mentioned above bias the redshift scatter by an amount that is beyond the Y10 requirements.

Matching the total number of galaxies to that of the fiducial case only partially mitigates the bias on the mean and the scatter of the redshift distributions. In particular, the only variants that are made compatible by the matching with both the weak lensing and the galaxy clustering Y10 requirements on the redshift distribution mean are the `Kroupa IMF' and `No gas emission'. Matching the total number of galaxies also leads the `Kroupa IMF' variant to become compatible with both the weak lensing and the galaxy clustering Y10 requirements on the redshift distribution scatter.

\subsection{Impact on the tomographic redshift distributions}
\label{sect:impact_tomo_bins_nz}

To evaluate whether varying the stellar population modelling components introduce biases in the mean and scatter of the tomographic redshift bins, we used the galaxies assigned to the SOM cells to infer the redshift distributions for five tomographic bins built following the prescription in \cite{McCullough2023}. 

\subsubsection{Tomographic redshift distributions construction}
\label{sect:tomo_bins_construction}

After selecting the magnitude-limited sample of galaxies, we performed the SOM cell assignment algorithm, such that the $i$-th cell $c_i$ in the SOM is populated by an associated number of galaxies, $n_i$, with a redshift probability distribution, $P(z|c_i)$. We used the histogram of redshifts of galaxies in a cell as the model for the $P(z|c_i)$. We then sorted the cells according to their median redshift and created five equally populated tomographic bins by placing galaxies in each tomographic bin according to the median redshift of the cell they reside in. The redshift distribution of galaxies in each tomographic bin is then given by the redshift probability distribution in each cell weighted by the cell occupation,
\begin{equation}
    N(z|\mathrm{bin}) = \sum_{i \ \in \ \mathcal{C}_i} n_i P(z|c_i) \ ,
\end{equation}
where $\mathcal{C}_i$ is the set of cells used in the tomographic bin. We populated the lowest redshift bin first and then we moved to the consecutive one once the total number of galaxies in each bin $N_\mathrm{tot}$ is roughly equal to one fifth of the full sample of magnitude-selected objects populating the SOM cells. The mean of the galaxy redshift distribution in each tomographic bin depends on the mean redshift in each colour cell $\bar{z}_{c_i}$ weighted by the number of galaxies per cell $n_i$,
\begin{equation}
    \begin{split}
        \bar{z}_{\mathrm{bin}} &= \frac{\int z \ N(z|\mathrm{bin}) \ \mathrm{d}z}{N_{\mathrm{tot}}} \\
        &= \frac{1}{N_{\mathrm{tot}}} \left [ n_1 \bar{z}_{c_1} + ... + n_{\mathcal{N}} \bar{z}_{c_{\mathcal{N}}} \right] \ ,
    \end{split}
\end{equation}
where $N_\mathrm{tot}$ is the total number of galaxies in the tomographic bin. The variance of each tomographic bin is given by 
\begin{equation}
    \sigma^2_z = \frac{\sum_{i=1}^{\mathcal{N}} n_i \sigma_{z,c_i}^2}{N_{\mathrm{tot}}} + \frac{\sum_{i=1}^{\mathcal{N}} n_i (\bar{z}_{c_i} - \bar{z}_{\mathrm{bin}})^2}{N_{\mathrm{tot}}} \ ,
\end{equation}
where $\sigma_{z,c_i}^2$ is the redshift distribution variance per cell. In Fig. \ref{fig:tortorelli_2024_figure9}, we show the five tomographic bins built for the fiducial sample and for the `Calzetti $k(\lambda)$' sample using the prescription highlighted above. We choose to plot this stellar population variant against the fiducial to highlight the strong impact that stellar population components might have on the tomographic redshift distributions. Indeed, the `Calzetti $k(\lambda)$' is one of the components providing the largest impact on the tomographic redshift distributions. This is due to the brighter rest-frame UV emission that the `Calzetti $k(\lambda)$' variant induces (see Fig. \ref{fig:tortorelli_2024_figure4}). This leaves the low-redshift galaxy observables unaffected, while it brightens high-redshift galaxies in the observed optical bands, hence making them more numerous in a magnitude-limited sample.

\begin{figure*}
   \resizebox{\hsize}{!}{\includegraphics{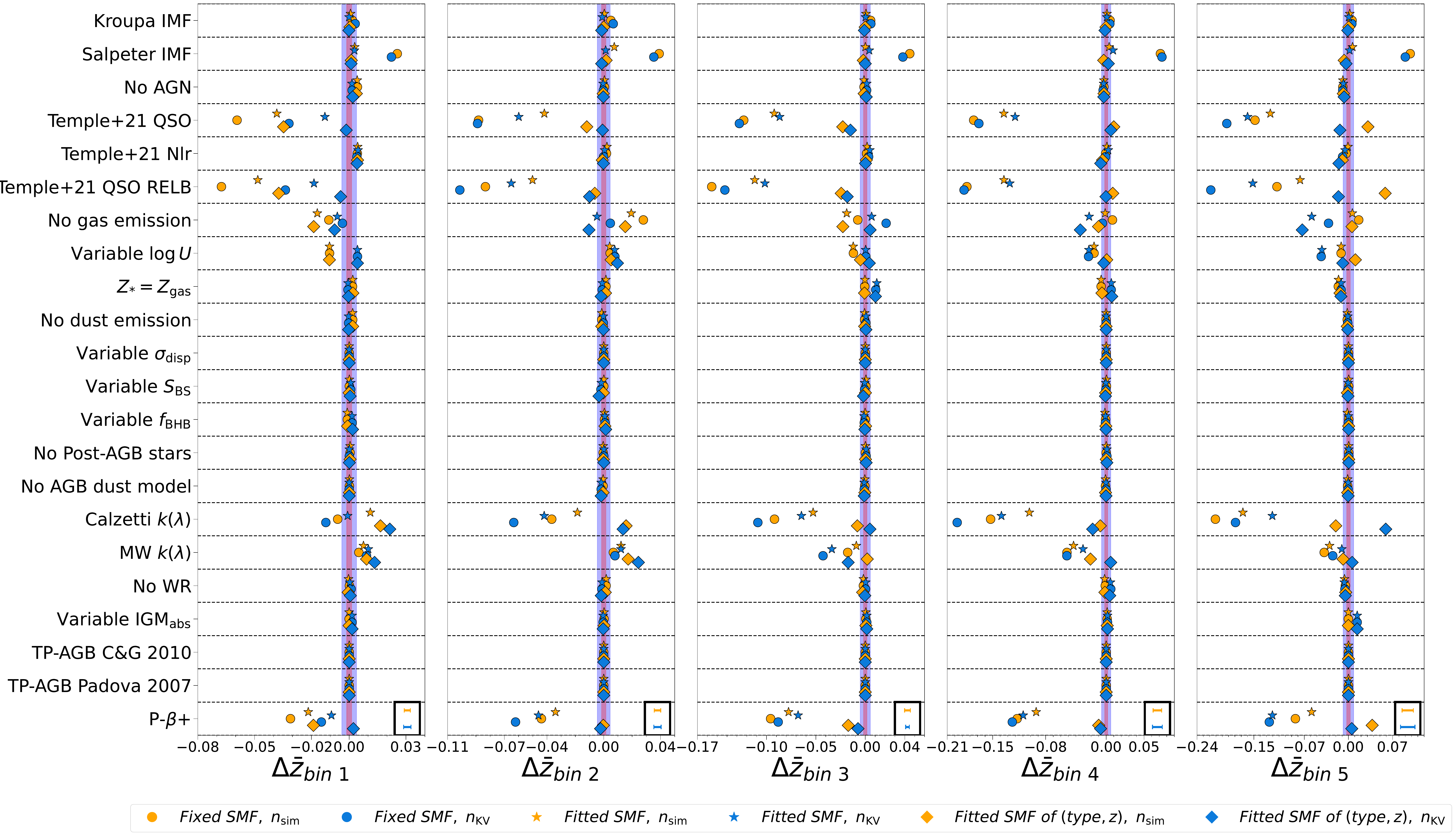}}
      \caption{Impact of varying the choices of the SED modelling component on the mean of the tomographic redshift distribution bins. Each panel shows a different tomographic bin, from the lowest (left-most) to the highest (right-most) redshift one. Each row represents a different SED modelling component we vary. The scatter points represent the difference between the mean redshift per bin obtained with the fiducial \textsc{Prospector-}$\beta$ model and that obtained by varying one SED modelling component at the time. Orange scatter points are obtained by weighting the $P(z|c_i)$ for the number of simulated galaxies per SOM cell, while light blue points are obtained by weighting the $P(z|c_i)$ for the observed abundances in the \cite{Masters2015,McCullough2023} SOM cells. Circle-shaped, star-shaped and diamond-shaped scatter points refer to the cases where we perform an $i$-band magnitude-selection of the sample of galaxies that is different for each component, matched for the total number of galaxies and based on the fiducial $i$-band distribution, respectively. We report the mean errors on the difference of the tomographic bin means estimated via bootstrap resampling in the black boxes in the bottom right-hand corner of each subplot. The blue and red coloured bands refer to the Rubin-LSST Y10  galaxy clustering and weak lensing requirements, respectively.}
         \label{fig:tortorelli_2024_figure10}
\end{figure*}

Figures \ref{fig:tortorelli_2024_figure10} and \ref{fig:tortorelli_2024_figure11} represent the main result of this work. Figure \ref{fig:tortorelli_2024_figure10} shows the bias induced on the tomographic redshift distribution bin means when modelling the galaxy SEDs with different stellar population component choices, while Fig. \ref{fig:tortorelli_2024_figure11} shows the bias induced on the tomographic bins' scatter. The blue and red vertical bars represent the requirements for the Rubin-LSST Y10 galaxy clustering and weak lensing analysis, respectively, as detailed in Sect. \ref{section:mag-lim-case}. The error bars in the black boxes are obtained similarly to Sect. \ref{section:mag-lim-case} (i.e. via bootstrap resampling).

The different points for a given stellar population variant represent different ways in which the redshift probability distribution in each cell is weighted towards the bin average. In particular, the orange scatter points are obtained when weighting the redshift probability distributions $P(z|c_i)$ in each cell by the number of simulated galaxies $n_{\mathrm{sim}}$ that occupy that cell in the same variant. The light blue scatter points, instead, are obtained when weighting the $P(z|c_i)$ by the observed KiDS-VIKING abundance $n_{\mathrm{KV}}$ in the remapped SOM \citep{Masters2015,McCullough2023} cells, similar to what would be the case in a scheme like that of \cite{Myles2021}.

At fixed variation of stellar population variant and scatter point colour, we generated three different samples of magnitude-limited objects with which we construct the tomographic bins. The circles represent the case where, for each different stellar population component, we selected the magnitude-limited sample of galaxies using the $i$-band magnitude distribution from that specific component (`Fixed SMF'). This process selects different numbers of galaxies among the various stellar population components, as pointed out in Sect. \ref{section:mag-lim-case}. The star-shaped scatter points represent instead the case where we matched the $\mathrm{d}n/\mathrm{d}m_i$ of each of the 22 components to the fiducial one by applying a constant magnitude shift to the $i$-band magnitude distributions (`Fitted SMF'). This process aims at matching the total number of magnitude-selected galaxies with that of the fiducial case by matching a model to the observed abundances. The diamond-shaped scatter points instead represent the case where, for each variation of the stellar population components, we selected the magnitude-limited sample of galaxies using the $i$-band magnitude distribution of the fiducial case (`Fitted SMF of (type,z)'). This process effectively employs always the same population of galaxies, but with a colour-redshift relation that is established by the SOM and therefore changes for every stellar population component. 

Figures \ref{fig:tortorelli_2024_figure10} and \ref{fig:tortorelli_2024_figure11} show that the variation of the stellar population component choices with respect to those adopted in the fiducial \textsc{Prospector-}$\beta$ model affects the derived colour-redshift relation. This, in turn, induces biases in both the mean and the scatter of the redshift distributions of tomographic bins that are beyond the requirements set by Stage IV galaxy surveys for several of the variants. As already shown throughout the work, the components that make the greatest contribution to the bias are those related to the IMF, AGN, gas, and dust attenuation modelling, including the P-$\beta$+ component that encapsulates the net effect of all the varied SED modelling components. These variants mostly bias the redshift to higher values. 

In the following sections, when presenting the biases induced by each SED modelling component, we tested the consistency with the weak lensing and clustering requirements using errors estimated via bootstrap resampling. These errors are somewhat different for every SED component however, in Figs. \ref{fig:tortorelli_2024_figure10} and \ref{fig:tortorelli_2024_figure11} we show only the errors averaged across all the components for plot readability.

\subsubsection{Impact of the initial mass function}
\label{sect:impact_imf}

The first SED modelling component of which we evaluated the impact on the tomographic redshift distributions is the IMF. In our work, we tested two different widely used IMFs, the `Kroupa IMF' and `Salpeter IMF' that are compared against the Chabrier IMF adopted in the fiducial model. The `Kroupa IMF' component (first row of Figs. \ref{fig:tortorelli_2024_figure10} and \ref{fig:tortorelli_2024_figure11}) does not induce a dramatic shift in the mean redshift and scatter of the tomographic bins given its similarity with the Chabrier IMF. When considering the case of different magnitude-limited samples of galaxies (`Fixed SMF', circle-shaped points), the shift in the means induced by this component is consistent with the fiducial model within the galaxy clustering requirements in the second and third tomographic bin for $n_{\mathrm{sim}}$ and in the first and third tomographic bin for $n_{\mathrm{kv}}$. It is instead consistent with the weak lensing requirements in the first, fourth and fith tomographic bin for $n_{\mathrm{sim}}$ and in the fourth and fith tomographic bin for $n_{\mathrm{KV}}$. If we account for SED mis-modelling and for the observed abundances, namely, the `Fitted SMF' (star-shaped points) and `Fitted SMF of (type,z)' (diamond-shaped points), the `Kroupa IMF' component becomes consistent with the fiducial model within the Stage IV weak lensing requirements on the mean redshift for every tomographic bin. As for the requirements on the redshift scatter, the `Kroupa IMF' component is consistent with the fiducial model within the weak lensing requirements for every tomographic bin, with the only exception of the `Fixed SMF' weighted by $n_{\mathrm{sim}}$ case, which is consistent with the clustering requirements only in the first tomographic bin.

The `Salpeter IMF' component (second row of Figs. \ref{fig:tortorelli_2024_figure10} and \ref{fig:tortorelli_2024_figure11}) has instead a slightly different behaviour. Since we have seen in Figs. \ref{fig:tortorelli_2024_figure4} and \ref{fig:tortorelli_2024_figure8} that the `Salpeter IMF' has a strong impact on the $i$-band magnitude and thus selection, this component in the `Fixed SMF' case (circle-shaped points) selects a very different sample of objects with respect to the fiducial ones, thereby biasing the mean redshift of the tomographic bins well beyond the Stage IV requirements. Upon matching the total number of galaxies selected to the fiducial model (`Fitted SMF' and `Fitted SMF of (type,z)' cases), the impact is strongly mitigated in every tomographic bin in such a way that for the highest degree of SED mis-modelling compensation (`Fitted SMF of (type,z)' case) we have consistency with the fiducial model within the weak lensing requirements on the redshift mean in every tomographic bin. The impact of the `Salpeter IMF' component on the redshift scatter with respect to the fiducial model is always at least within the clustering requirements. Also in this case, the highest degree of SED mis-modelling compensation (`Fitted SMF of (type,z)' case) makes this component consistent with the fiducial model within the weak lensing requirements.

The results for the IMF show that fitting the SMF in order to match the observed abundance of galaxies to compensate for any possible IMF-related SED mis-modelling is crucial to obtain redshift distribution estimates that are not biased beyond the requirements set by Stage IV surveys.  This is motivated by the fact that changing the IMF mostly impacts the abundance of low mass stars, but it is well known that low mass stars dominate the stellar mass and the number of stars in a galaxy,  while contributing only a few percent to the bolometric light \citep{Conroy2013}.  Fitting the SMF and matching the abundances effectively removes the difference in stellar mass distribution within a sample of galaxies caused by the different adopted IMF.

\subsubsection{Impact of AGNs}
\label{sect:impact_agn}

\begin{figure*}
   \resizebox{\hsize}{!}{\includegraphics{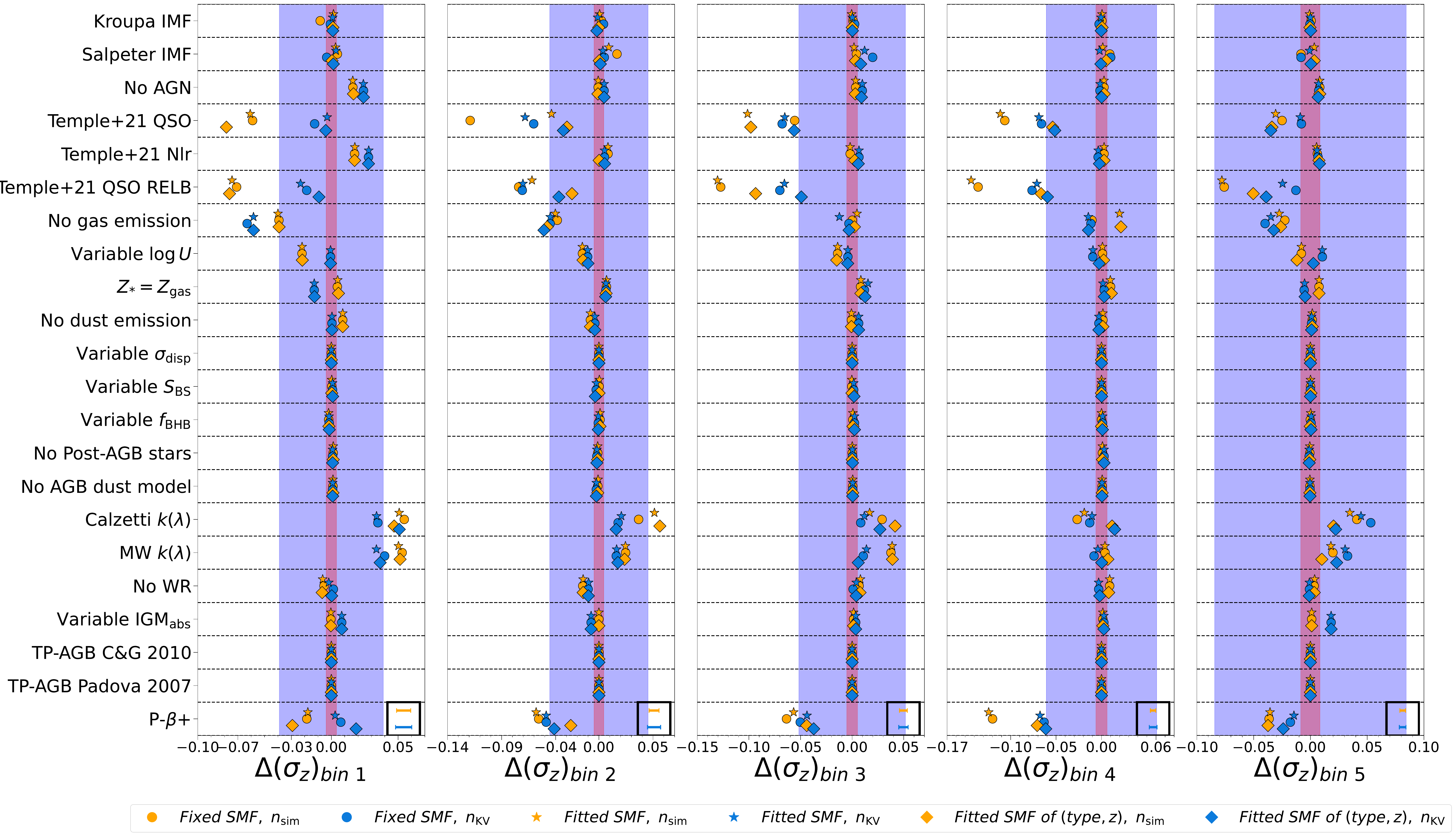}}
      \caption{Impact of varying the choices of the SED modelling component on the scatter of the tomographic redshift distribution bins. Each panel shows a different tomographic bin, from the lowest (left-most) to the highest (right-most) redshift one. Each row represents a different SED modelling component we vary. The scatter points represent the difference between the redshift scatter per bin obtained with the fiducial \textsc{Prospector-}$\beta$ model and that obtained by varying one SED modelling component at the time. The orange scatter points are obtained by weighting the $P(z|c_i)$ for the number of simulated galaxies per SOM cell, while the light blue points are obtained by weighting the $P(z|c_i)$ for the observed abundances in the \cite{Masters2015,McCullough2023} SOM cells. Circle-shaped, star-shaped and diamond-shaped scatter points refer to the cases where we perform an $i$-band magnitude-selection of the sample of galaxies that is different for each component, matched for the total number of galaxies and based on the fiducial $i$-band distribution, respectively. We report the mean errors on the difference of the tomographic bin scatters estimated via bootstrap resampling in the black boxes in the bottom right-hand corner of each subplot. The blue and red coloured bands refer to the Rubin-LSST Y10  galaxy clustering and weak lensing requirements, respectively.}
         \label{fig:tortorelli_2024_figure11}
\end{figure*}

Active galactic nuclei comprise the SED modelling component that induces the largest bias on the mean and scatter of the tomographic redshift bins. In our work, we tested the impact of AGN by replacing the \cite{Nenkova2008,Nenkova2008b} templates implemented in \textsc{FSPS} with the more realistic model presented in \cite{Temple2021}. The `Temple+21 QSO' and the `Temple+21 QSO RELB' components (fourth and sixth rows of Figs. \ref{fig:tortorelli_2024_figure10} and \ref{fig:tortorelli_2024_figure11}) produce redshift distribution estimates that have mean redshifts biased high with respect to the fiducial ones. In the majority of cases they are not consistent with the Stage IV requirements. For the `Fixed SMF' (circle-shaped points) and the `Fitted SMF' (star-shaped points) cases, the `Temple+21 QSO' and the `Temple+21 QSO RELB' components are consistent with the fiducial model neither within the clustering nor within the weak lensing requirements on the mean redshifts, independently of how we weight the $P(z|c_i)$ by the cell occupation. In the `Fitted SMF of (type,z)' case (diamond-shaped points), weighting by $n_{\mathrm{sim}}$ (orange points) leads the `Temple+21 QSO' component to be consistent with the clustering requirements in the fourth tomographic bin and not consistent with any of the requirements in the other four bins, while the `Temple+21 QSO RELB' component becomes consistent with the clustering requirements in the second and fourth bins, and not consistent in the remaining three. Weighting the `Fitted SMF of (type,z)' case by $n_{\mathrm{KV}}$ (light blue diamond-shaped points) leads the `Temple+21 QSO' to be not consistent with any of the requirements in the third bin, consistent with the clustering requirements in the fifth bin and consistent with the weak lensing requirements in the remaining three. The `Temple+21 QSO RELB' component, instead, becomes consistent with the clustering requirements in the first and fifth bins, with the weak lensing requirements in the fourth, and not consistent with any of the requirements in the second and third bin. 

The redshift scatter of the tomographic redshift distributions is also heavily impacted by the AGN modelling choices. For both the `Temple+21 QSO' and `Temple+21 QSO RELB' components, in the first, third and fourth tomographic bins the `Fixed SMF',  `Fitted SMF',  and `Fitted SMF of (type,z)' cases are not consistent with the fiducial redshift scatter within any of the requirements when weighting the $P(z|c_i)$ by $n_{\mathrm{sim}}$ (orange points). When weighting by $n_{\mathrm{KV}}$ (light blue points), instead, the bias is only partially mitigated in the first and fourth bins, while still being not consistent with the requirements in third bin. In the second tomographic bin, only the `Fitted SMF of (type,z)' cases are consistent with the clustering requirements. The only cases where there is consistency with the fiducial model within the weak lensing requirements are the first and fifth bins. More precisely, in the first bin the `Fitted SMF' and `Fitted SMF of (type,z)' cases weighted by $n_{\mathrm{KV}}$ are within the weak lensing requirements, while in the fifth bin this happens for the `Fixed SMF' and `Fitted SMF' cases weighted by $n_{\mathrm{KV}}$.

These results  once again highlight the dramatic impact that the AGN has on the forward modelling-based redshift distributions. Fitting the observed abundance of galaxies to compensate for this SED mis-modelling is not sufficient as it only partially, and not consistently across the weighting schemes, mitigates the effect on the bias of the tomographic bin means and scatters. There is no single re-weighting strategy that mitigates all of the biases in every single tomographic bin, implying that AGN change the colour of a galaxy in a different way with respect to what stars do. 

The modelling of the narrow-line region of AGNs (`Temple+21 Nlr' component, fifth row of Figs. \ref{fig:tortorelli_2024_figure10} and \ref{fig:tortorelli_2024_figure11}) contributes to biasing the redshift mean and scatter with respect to the fiducial model mostly in the first tomographic bin. However, the impact is always within at least the clustering requirements. In all the other bins, independently of the $i$-band selection choices, even at fixed $P(z|c_i)$ weighting, the `Temple+21 Nlr' component leads to tomographic redshift means and scatters that are consistent within the weak lensing requirements with those estimated from the fiducial model.

Excluding completely the modelling of AGN (`No AGN' component, third row of Figs. \ref{fig:tortorelli_2024_figure10} and \ref{fig:tortorelli_2024_figure11}) seems to bias the tomographic bin means and scatters by only a small amount, with a behaviour that is similar to that of the `Temple+21 Nlr' component. In particular, for the tomographic bin means, the three SMF cases weighted by $n_{\mathrm{sim}}$ are consistent with the clustering requirements in the first bin, while all the other cases in all the remaining bins are consistent with the weak lensing requirements. As for the redshift scatter, in the first tomographic bin, all the cases are consistent only with the clustering requirements. In the third bin, the three SMF cases weighted by $n_{\mathrm{KV}}$ are consistent with the clustering requirements, while if weighted by $n_{\mathrm{sim}}$ also with the weak lensing requirements. In the remaining three bins, all the cases are consistent with the fiducial redshift scatter within the weak lensing requirements. Even though the impact is small, this result does not imply that ignoring the AGN modelling does not impact the tomographic bins. Rather, it means that the AGN treatment implemented in \textsc{FSPS} is likely underestimating the effect of the AGN emission on the galaxy SEDs, since the specific implementation in \textsc{FSPS}, coupled with the prior on the AGN parameters, provides a contribution that is pronounced only in the $H,K_s$ rest-frame wavebands. 

\subsubsection{The impact of gas emission}
\label{sect:impact_gas_emission}

The gas emission is another SED modelling component that induces substantial shifts in the mean and scatter of the tomographic bins, although smaller in absolute value with respect to those induced by the AGN modelling components. The extreme case of `No gas emission' (seventh row of Figs. \ref{fig:tortorelli_2024_figure10} and \ref{fig:tortorelli_2024_figure11}) has a behaviour that shows a large degree of scatter among the different $i$-band selection choices, even at fixed $P(z|c_i)$ weighting. Every re-weighting strategy contributes to mitigating the bias only in certain tomographic bins, but not in all of them at the same time. The cases weighted by $n_{\mathrm{sim}}$ are not consistent with any of the requirements in the three lowest redshift tomographic bins. The `Fixed SMF' case (orange round points) is consistent with the clustering requirements in the fourth and fifth bins, while the `Fitted SMF' and `Fitted SMF of (type,z)' cases are consistent with the weak lensing requirements in these bins. If we consider the $n_{\mathrm{KV}}$ weighting, there is no consistency with any of the requirements in the fifth tomographic bin and in the first bin for the `Fitted SMF' and `Fitted SMF of (type,z)' cases. In the second bin, the `Fixed SMF' and `Fitted SMF' cases are consistent with the clustering requirements. In the third bin, the `Fitted SMF' is consistent with the clustering requirements, while the `Fitted SMF of (type,z)' with the weak lensing requirements. In the fourth bin, the only case that is consistent with the weak lensing requirements is the `Fixed SMF' case. The redshift scatter is outside of the requirements for every case in the first bin and consistent only with the clustering requirements in the second, fourth and fifth tomographic bins for all the re-weighting cases. In the third bin, every case but for the `Fitted SMF' weighted by $n_{\mathrm{KV}}$ is consistent with the weak lensing requirements.

Allowing the gas ionisation to vary within the range in \cite{Byler2017} (`Variable $\log{U}$' component, eight row of Figs. \ref{fig:tortorelli_2024_figure10} and \ref{fig:tortorelli_2024_figure11}) affects the tomographic bin means and scatters by an amount that is smaller in absolute value than the `No gas emission' case, but still large enough to induce biases that are beyond the Stage IV requirements. Also in this case, there is no single re-weighting scheme that compensates for the biases sufficiently in every tomographic bin. On the contrary, the weighting scheme have opposite effects in different tomographic bins. For instance, weighting the $P(z|c_i)$ by $n_{\mathrm{KV}}$ makes the bias in the redshift mean with respect to the fiducial case consistent within the weak lensing requirements in the third bin, but inconsistent with any of the requirements in the second tomographic bin, while the opposite happens when weighting the $P(z|c_i)$ by $n_{\mathrm{sim}}$. In the first tomographic bin, no re-weighting scheme makes the bias consistent with the weak lensing requirements, while in the second bin weighting the $P(z|c_i)$ by $n_{\mathrm{sim}}$ leads to a redshift mean that is consistent with the fiducial one within the weak lensing requirements. In the fourth tomographic bin, the `Fitted SMF of (type,z)' case is the only one consistent with the weak lensing requirements independently of the $P(z|c_i)$ weighting, while in the fifth bin the same happens for the clustering requirements. As for the redshift scatter, the behaviour is more consistent within the same $P(z|c_i)$ weighting scheme. The `Fixed SMF', `Fitted SMF' and `Fitted SMF of (type,z)' cases weighted by $n_{\mathrm{KV}}$ are consistent with the weak lensing requirements in every tomographic bin, but for the second one where there is only consistency with the clustering requirements. When weighting by $n_{\mathrm{sim}}$, the three SMF cases are consistent with the clustering requirements in the three lowest redshift bins, while consistent with the weak lensing requirements in the two highest redshift ones.

When setting the gas metallicity equal to the stellar metallicity (ninth row of Figs. \ref{fig:tortorelli_2024_figure10} and \ref{fig:tortorelli_2024_figure11}) as opposed to randomly drawing the gas metallicity from a uniform distribution as in the fiducial \textsc{Prospector-}$\beta$ model, the behaviour within the same $P(z|c_i)$ weighting scheme is more regular than what was shown in the `No gas emission' and `Variable $\log{U}$' cases, for both the redshift mean and scatter. All the re-weighting schemes are consistent with the weak lensing requirements for the redshift mean in the two lowest redshift tomographic bins. In the third tomographic bin, the three SMF cases weighted by $n_{\mathrm{KV}}$ are not consistent with any of the requirements, while those weighted by $n_{\mathrm{sim}}$ are consistent with the weak lensing requirements. In the fifth bin, the three SMF cases are consistent with the clustering requirements when weighted by $n_{\mathrm{KV}}$, while not consistent when weighted by $n_{\mathrm{sim}}$. In the fourth bin all the cases are only consistent with the clustering requirements. In the case of the redshift scatter (Fig. \ref{fig:tortorelli_2024_figure11}), all the re-weighting schemes are consistent with the fiducial redshift scatter within the weak lensing requirements in the second and fifth tomographic bins. In the first and third bins, the three SMF cases weighted by $n_{\mathrm{KV}}$ are only consistent with the clustering requirements, while weighting by $n_{\mathrm{sim}}$ makes the SMF cases consistent with the weak lensing requirements as well. The opposite happens in the fourth tomographic bin.

\subsubsection{Impact of dust attenuation}
\label{sect:impact_dust_attenuation}

Introducing a different modelling of the dust attenuation law with respect to the fiducial \textsc{Prospector-}$\beta$ implementation generates SEDs, and therefore colours, that have an impact on the redshift mean that reaches the $\Delta z \sim 0.1$ level in the highest redshift tomographic bins. Among the two dust attenuation laws we tested, the `Calzetti $k(\lambda)$' component is the one that provides the largest bias. In particular, the `Calzetti $k(\lambda)$' component (seventh row of Figs. \ref{fig:tortorelli_2024_figure10} and \ref{fig:tortorelli_2024_figure11} starting from the bottom) leads to biases in the redshift mean with respect to the fiducial case that are never within the weak lensing requirements (but for the `Fitted SMF' weighted by $n_{\mathrm{KV}}$ case in the first bin), independently of the re-weighting scheme and tomographic bin considered. It becomes however consistent with the clustering requirements in the third and fourth bins when considering the `Fitted SMF of (type,z)' case. Similarly, the redshift scatter is never consistent with the fiducial case within the weak lensing requirements, independently of the re-weighting scheme and tomographic bin considered. It is however consistent with the clustering requirements for all cases in the three highest redshift tomographic bins, in the second bin for the SMF cases weighted by $n_{\mathrm{KV}}$ and in the first bin for the `Fixed SMF' and `Fitted SMF' weighted by $n_{\mathrm{KV}}$ cases.

The `MW $k(\lambda)$' component (sixth row of Figs. \ref{fig:tortorelli_2024_figure10} and \ref{fig:tortorelli_2024_figure11} starting from the bottom) effect is smaller in absolute value with respect to the `Calzetti $k(\lambda)$' case, although similarly inconsistent with the weak lensing requirements across the tomographic bins and re-weighting schemes. In the two lowest redshift tomographic bins, all the re-weighting cases are not consistent with the fiducial redshift means within both the clustering and the weak lensing requirements. The only weak lensing requirements consistency happens in the third and fourth bins with the `Fitted SMF of (type,z)' weighted by $n_{\mathrm{sim}}$ and `Fitted SMF of (type,z)' weighted by $n_{\mathrm{KV}}$, respectively, and in the fifth bin for both the aforementioned cases. All the other cases are not consistent with any of the requirements. When comparing the redshift scatters in each tomographic bin against the fiducial ones, we note that there is consistency with the weak lensing requirements of all the re-weighting cases only in the fourth tomographic bin. The `Fitted SMF of (type,z)' weighted by $n_{\mathrm{KV}}$ and `Fitted SMF of (type,z)' weighted by $n_{\mathrm{sim}}$ cases are also consistent with the weak lensing requirements in the third and fifth bins, respectively. In all the other cases, there is only consistency with the clustering requirements, except for the first tomographic bin where the three SMF cases weighted by $n_{\mathrm{sim}}$ are not consistent with any of the requirements.

The results presented in this section show that modelling the dust attenuation accurately is critical for accurate tomographic redshift distributions as there is no single SMF-based re-weighting strategy that consistently mitigates all of the biases in every tomographic bin. Most importantly, the dust attenuation is the SED modelling component that is less mitigated by the various re-weighting schemes we tested in our work due to the fact that it dramatically changes the colours, not just the overall luminosity.

\subsubsection{Impact of the `P-$\beta$+' component}
\label{sect:impact_pbeta}

Lastly, we consider the case of the `P-$\beta$+' component that encapsulates the net effect of all the variable SED modelling choices applied to the population of galaxies within observationally meaningful parameter values. In the last row of Fig. \ref{fig:tortorelli_2024_figure10}, we see that the `Fixed SMF' and `Fitted SMF' redshift means are not consistent with the fiducial ones within any of the requirements for all the tomographic bins, independently of the $P(z|c_i)$ weighting. The `Fitted SMF of (type,z)' weighted by $n_{\mathrm{KV}}$ case is instead consistent with the clustering requirements in the third tomographic bin and consistent with the weak lensing requirements in the other four bins. When weighted by $n_{\mathrm{sim}}$, the `Fitted SMF of (type,z)' case is consistent with the clustering requirements in the fourth bin and with the weak lensing requirements in the second tomographic bin. In the case of the `P-$\beta$+' component, the highest degree of SED mis-modelling compensation leads the redshift means to be consistent with the fiducial ones for all the tomographic bins, except for the third one where it is consistent with the clustering requirements only.

When considering the redshift scatter (Fig. \ref{fig:tortorelli_2024_figure11}), the only cases that are consistent with the fiducial model within the weak lensing requirements are `Fixed SMF' and `Fitted SMF' weighted by $n_{KV}$ cases in the first tomographic bin. The remaining cases in this bin are instead consistent with the clustering requirements only. In the fourth bin no case lies within the requirements, while in the fifth bin all cases are consistent with the fiducial model within the clustering requirements. In the second and third bin the behaviours are mixed. In the second bin, only the `Fitted SMF of (type,z)' case is consistent with the clustering requirements, while in the third bin the only cases outside of the clustering requirements are the `Fixed SMF' and `Fitted SMF' weighted by $n_{sim}$.

\section{Discussion}
\label{sect:discussion}

The results presented in this work highlight the dramatic impact that variations in the choices and in the parameters of the SED modelling components of SPS-based spectra have on the predicted redshift distributions. We also show that there is no single stellar mass function based re-weighting strategy that consistently mitigates all of the biases in every tomographic bin. This implies that no matter how flexible the galaxy population model is, it is not going to be accurate enough for SPS-based forward modelling -- unless it includes an either a priori accurate or flexible enough model for the stellar population.

The implementation in \textsc{FSPS} of the SED modelling components related to later stages of stellar evolution,  such as blue HB stars, TP-AGB, and Post-AGB, leads to changes in colours that impact the mean redshifts and scatter of the tomographic redshift distributions by an amount that is within the requirements set by Stage IV galaxy surveys. This is (in part) due to the fact that, as short-lived stellar phases, their effect is averaged out over the entire  population. Another possible explanation could be the uncertainty surrounding these phases, both in terms of their modelling and their loosely constrained parameters from observations (see e.g. \citealt{Shiying2024} for a recent detection of TP-AGB features in distant quiescent galaxies using JWST/NIRSpec), which might underestimate the effect that these phases have on galaxy SEDs, even in the extreme variants we tested here. 

The families of SED components that instead make a significant contribution to galaxy colours and lead to inconsistency with Stage IV requirements are those related to the modelling of the IMF, AGN, gas physics, and dust attenuation law. The uncertainty in the true shape of the IMF or at least in the distribution and diversity of IMF parameters needs to be reduced using existing data or new observations, because the results presented in this paper have shown that varying the IMF from a Chabrier to a more bottom-heavy one (Salpeter) leads to variations in the means of the tomographic bins that, if not properly compensated, are greatly inconsistent with the Stage IV requirements. In this regard, more galaxy evolution oriented spectroscopic surveys such as WEAVE-StePS \citep{Iovino2023},  4MOST-StePS \citep{Iovino2023b},  and 4MOST-WAVES \citep{Driver2019} for the local/intermediate redshift Universe and JWST \citep{Cameron2023} for higher redshift objects might help to pin down the diversity of IMF functional forms in the population of galaxies.

The modelling of gas physics is another crucial point that arises from this study. The accurate characterisation of emission lines is an important aspect of modelling spectra because the spectroscopic datasets that are used to calibrate photometric redshifts need galaxies for which we are able to get high confidence in redshift estimation. This aspect biases the samples towards objects with strong spectral features such as emission lines, leading to underrepresented colour-space regions where these features are lacking \citep{Newman2022}. This problem is exacerbated at greater depths where the fraction of highly secure redshifts is small \citep{Hartley2020}. The use of templates, for instance those of \cite{Blanton2007} or \cite{Brown2014}, despite providing good results, is able to take into account neither the diversity of emission lines strength and ratios, nor the relationship that exists between line strengths and physical properties of the galaxies. These templates have been created with training sets that are heavily biased towards local Universe objects, but it is well known that gas-phase metallicities and gas ionisation parameters change at high redshift,  a known result that has been recently strenghten by the observations conducted with JWST \citep{Bunker2023a,Bunker2023b,Calabro2024,Roberts-Borsani2024,Tripodi2024}.  Furthermore,  the creation of galaxy spectra  with some linear combination of templates is very hard to link to the physical properties of galaxies, such as their star formation histories, which is particularly problematic because sSFR is a specific driver of the ionisation parameter of galaxies \citep{Kaasinen2018}. SPS-based spectra offer the most promising avenue in accurately modelling emission lines, however line intensities and ratios are predicted using photoionisation codes,  such as \textsc{CLOUDY} \citep{Ferland2017}, that make strong simplifying assumptions about the geometry and the composition of the gas to reduce the number of free parameters \citep{Byler2017}. One possible solution is to employ methods that empirically map the emission lines strengths to the spectral continuum flux \citep{Beck2016,Khederlarian2024} or develop tools for interpreting nebular emission across a wide range of gas abundances and ionising conditions at different redshifts that do not require a specific ionising spectrum as a source \citep{Li2024}. Alternatively, one could use well characterised set of galaxy spectra and measured gas properties (e.g. \citealt{Bellstedt2020,Thorne2022}) to empirically associate the strength of emission lines to the measured gas ionisation and metallicity.

In a recent work by \cite{Apellaniz2024}, dust extinction has been referred to as the 'elephant in the room' of astronomical observations, since extinction changes between lines of sight within the same galaxy and not all the families of extinction laws are generic enough to be applied to the galaxy population as a whole, but rather the use of a dust attenuation law should be tightly linked to the spectral type and physical properties of the individual galaxies \citep{Kriek2013,Nagaraj2022,Alsing2023}. Our work has shown that this statement is definitely true, since varying the dust attenuation law leads to biases in the mean redshift of the tomographic bins of up to $\Delta z \sim 0.1$, two orders of magnitude greater than Stage IV requirements. A recent work by \cite{Salim2018} has measured the dust attenuation curves of a sample of 230,000 galaxies in the local Universe, thereby providing very stringent constraints on the attenuation curve slopes and UV bump strengths. The comprehensive nature of this work, the sample of low-redshift analogues to high-redshift galaxies, and the functional forms that they provide are a strong initial help in constraining the distribution of dust attenuation curves and parameters that need to be used in SPS-based forward modelling. Furthermore, this work can be complemented with the latest measurements of the $R_V$ distribution in the Milky-Way by LAMOST \citep{Zhang2023}, which can be used for analogues of our own galaxy. Very little information is however available about whether the dust attenuation evolves with redshift. Recent results obtained with the use of JWST data \citep{Markov2024} show that the power-law slope and the bump strength with which $A_{\lambda}$ is usually modelled decrease towards high redshift, possibly due to a higher abundance of large dust grains produced in supernova ejecta. A more refined approach would also consist in modelling each galaxy as a spatially distributed set of stellar populations, where each line of sight has its own values for the dust attenuation parameters.

Concerning the AGN component, we have already discussed how the templates currently implemented in \textsc{FSPS} lack in predicting power towards specific AGN spectral features, such as the broad-line region emission. SDSS \citep{Lyke2020} has helped to build a rather complete census of the quasars and AGN populating the Universe, while Stage IV surveys are either already greatly expanding this census (e.g. DESI, \citealt{Chaussidon2023}) or aiming at probing a whole new regime of lower luminosity objects \citep{Ivezic2017}. If we aim for a forward modelling of the full galaxy population in these surveys it is therefore important to correctly reproduce the colours and spectral characteristics of the population of AGN, including the broad and narrow-line regions emission, the UV continuum, and the $Ly \alpha$ line. In this regard, the use of the \cite{Temple2021} templates can correctly reproduce the observed SDSS-UKIDSS-WIDE quasar colours in a wide range of redshifts and luminosities, as shown in \cite{Temple2021},  and also, in Appendix~\ref{appendix:agn_comparison},  where we compare the AGN colours against SDSS DR16 quasars \citep{Lyke2020} and against AGN from the \textsc{FSPS} \cite{Nenkova2008,Nenkova2008b} templates. Furthermore, the use of DESI data \citep{Chaussidon2023}, with its expected sample of $\sim 3$ million quasars, will help in refining these templates further.

To summarise, the main aim of this paper is to provide a sensitivity study to find which SED components need to be more carefully modelled and better constrained than presently  in order to estimate redshift distributions at the required accuracy for Stage IV galaxy surveys. Although SPS-based models potentially allow for a complete and thorough description of the galaxy population, the stellar population components used to model galaxy SEDs must be carefully chosen; most importantly, these parameters must be constrained using existing or upcoming observations (e.g. \citealt{Davies2018,Driver2019,Costantin2019,Iovino2023,Hahn2023}). Any effort that aims at using SPS-based forward modelling of galaxy surveys as a method to obtain precise redshift distributions for Stage IV cosmological parameters estimation must also consider  the required constraints on galaxy SEDs from observations. 

\section{Conclusions}
\label{sect:conclusions}

The forward modelling of galaxy data is a promising alternative approach to the problem of accurately characterising the tomographic redshift distributions of Stage IV galaxy surveys. It relies on a realistic model for the galaxy population and on a mapping of physical properties of galaxies to their SEDs. The latter is generally obtained using either a set of empirically derived templates or by employing stellar population synthesis (SPS) codes, such as \textsc{FSPS} \citep{Conroy2009}. SPS codes generate realistic galaxy spectra if their parameters are appropriately chosen. Otherwise, SPS model mismatches may cause dramatic changes in the galaxy colours and, thus, on the predicted colour-redshift relation.

In this work, we investigated the impact that choices of the stellar population components and of their parameters have on forward modelling-based tomographic redshift distributions. Specifically, we compared the potential bias induced by these choices with the requirements set by Stage IV cosmological galaxy surveys. We find that the spectral energy distribution (SED) modelling components related to active galactic nuclei (AGNs), the attenuation law, gas physics, and the initial mass function (IMF) lead to biases in the mean and scatter of the tomographic redshift distributions that exceed these requirements. Furthermore, when trying to compensate any SED mis-modelling by fitting the stellar mass function (SMF), we find that there is no single re-weighting strategy that consistently mitigates all of the biases in every single tomographic bin. This highlights the fact that for a galaxy population model to be accurate enough for SPS-based forward modelling, it is necessary to include an either a priori accurate or flexible enough model for the distribution of parameter values for the aforementioned stellar population components, which seems impossible based on purely theoretical considerations. It is therefore of paramount importance to constrain the distribution of stellar population parameters by either designing new experiments or fully exploiting existing observations, together with implementing the appropriate set of comparison metrics between observations and simulations.  

To perform this work, we used the \textsc{Prospector-}$\beta$ galaxy population model \citep{Wang2023} to obtain a set of realistic physical properties of galaxies. We generated a sample of $10^6$ galaxy spectra and magnitudes using these physical properties and the \textsc{FSPS} parameters adopted in \cite{Wang2023}. We labelled this sample as `fiducial' and we compared it against 22 sets of $10^6$ galaxies obtained by keeping the stellar mass, the star formation history, and the stellar metallicity fixed while varying the SED modelling components and their parameters one at the time.

We first evaluated the impact that the stellar population choices have on the rest-frame and the observed-frame galaxy colours. We considered a wide range of wavebands, covering the optical and the near-infrared regime. The bands we employ, $ugriZYJHK_s$, are those used in KiDS-VIKING. We find that the components related to the active galactic nuclei (AGN), the attenuation law and the gas physics modelling modify the distribution of colours with respect to the fiducial case by an amount that is larger than the median photometric errors of COSMOS, used as a proxy of the expected precision of Rubin-LSST photometry (see Appendix \ref{appendix:comparison_cosmos_rubin_photometry}). This implies that the SED components choices lead to an  effect on the simulated galaxy colours that is detectable even on a single galaxy basis.

To then evaluate how the variation in optical and near-infrared colours impact the colour-redshift relation for lensing and clustering, we modelled it using the 
KiDS-VIKING remapped \cite{Masters2015} SOM \citep{McCullough2023}. We assigned a cell membership to each galaxy based on the observed-frame colours and evaluated how the distribution of cell mean redshifts and occupation numbers changes as a result of the different SED modelling choices. We find that there is a large impact of the stellar population component choices on the colour-redshift relation. In particular, the IMF, the AGN, the attenuation law, and the gas physics modelling leads to relative differences in the occupation numbers
of the SOM cells with respect to the fiducial sample of more than $10\%$, with even more than $100\%$ in the case of the attenuation laws. The mean redshifts per cell change as well, with attenuation laws, AGN and gas physics leading to a relative difference with respect to the fiducial sample whose 84th-16th percentile values that exceeds $1\%$, the nominal Stage III survey requirement.

We then proceeded to build a magnitude-limited sample of galaxies using the Rubin-LSST Y10 gold lens sample cut of $i \le 25.3$. The magnitude-limited sample is used to evaluate how much the full redshift distribution of the selected galaxies changes, and to select the sample of objects that are used to build the tomographic bins. We find that the SED modelling choices dramatically impact the mean and the scatter of the full redshift distribution of the magnitude-selected galaxies. This is due to the fact that the SED modelling choices change the $i$-band magnitude distribution and therefore a pure magnitude cut selects different population of galaxies when varying the stellar population components. Even if we account for this by matching the total number of magnitude-selected galaxies to the fiducial case, the difference in the mean redshift and scatter with respect to the fiducial sample is still not consistent with the Stage IV requirements for the variants related to the AGN, dust attenuation, and IMF prescriptions.

For each SED modelling component, the SOM cells are used to create a set of equally populated tomographic bins. We evaluated the mean redshift and the scatter of the five tomographic bins and compared them against those estimated with the fiducial sample. We consider three limiting cases with which we try to evaluate how the impact of the SED mis-modelling on the colour-redshift relation might be compensated by fitting the distribution of physical properties of galaxies, particularly the SMF. 

We find that the IMF, AGN, gas, and attenuation law modelling can bias the mean and the scatter of the tomographic redshift distributions beyond the requirements set by Stage IV surveys. We also find that fitting the SMF is not enough to consistently mitigate all the redshift mean and scatter biases in every single tomographic bin for the AGN, gas, and attenuation law components. This points to the conclusion that if we want to adopt a forward modelling approach to redshift distribution estimates based on SPS models, the choices and the parameters of the stellar population components used to model galaxy SEDs must be carefully evaluated and constrained using existing or upcoming observations.

\begin{acknowledgements}
      LT is grateful to Bingjie Wang for the very useful and insightful discussions on the use of the \textsc{Prospector-}$\beta$ model. LT is also grateful to Dominika Durovcikova, Matthew Temple and Daniel Masters for the helpful discussions on AGN, for the help in the use of \cite{Temple2021} QSO templates and for the clarification on the used SOM wavebands, respectively. LT is grateful to John Franklin Crenshaw for the useful discussion on the comparison between COSMOS and Rubin-LSST photometric errors. This work was funded by the Deutsche Forschungsgemeinschaft (DFG, German Research Foundation) under Germany's Excellence Strategy – EXC-2094 – 390783311. It has also benefitted from support by the Bavaria California Technology Center (BaCaTec).
\end{acknowledgements}

\bibliographystyle{aa}
\bibliography{bibliography}

\begin{thebibliography}{218}
\expandafter\ifx\csname natexlab\endcsname\relax\def\natexlab#1{#1}\fi

\bibitem[{{Abbott} {et~al.}(2022){Abbott}, {Aguena}, {Alarcon}, {Allam},
  {Alves}, {Amon}, {Andrade-Oliveira}, {Annis}, {Avila}, {Bacon}, {Baxter},
  {Bechtol}, {Becker}, {Bernstein}, {Bhargava}, {Birrer}, {Blazek},
  {Brandao-Souza}, {Bridle}, {Brooks}, {Buckley-Geer}, {Burke}, {Camacho},
  {Campos}, {Carnero Rosell}, {Carrasco Kind}, {Carretero}, {Castander},
  {Cawthon}, {Chang}, {Chen}, {Chen}, {Choi}, {Conselice}, {Cordero},
  {Costanzi}, {Crocce}, {da Costa}, {da Silva Pereira}, {Davis}, {Davis}, {De
  Vicente}, {DeRose}, {Desai}, {Di Valentino}, {Diehl}, {Dietrich}, {Dodelson},
  {Doel}, {Doux}, {Drlica-Wagner}, {Eckert}, {Eifler}, {Elsner}, {Elvin-Poole},
  {Everett}, {Evrard}, {Fang}, {Farahi}, {Fernandez}, {Ferrero}, {Fert{\'e}},
  {Fosalba}, {Friedrich}, {Frieman}, {Garc{\'\i}a-Bellido}, {Gatti},
  {Gaztanaga}, {Gerdes}, {Giannantonio}, {Giannini}, {Gruen}, {Gruendl},
  {Gschwend}, {Gutierrez}, {Harrison}, {Hartley}, {Herner}, {Hinton},
  {Hollowood}, {Honscheid}, {Hoyle}, {Huff}, {Huterer}, {Jain}, {James},
  {Jarvis}, {Jeffrey}, {Jeltema}, {Kovacs}, {Krause}, {Kron}, {Kuehn},
  {Kuropatkin}, {Lahav}, {Leget}, {Lemos}, {Liddle}, {Lidman}, {Lima}, {Lin},
  {MacCrann}, {Maia}, {Marshall}, {Martini}, {McCullough}, {Melchior},
  {Mena-Fern{\'a}ndez}, {Menanteau}, {Miquel}, {Mohr}, {Morgan}, {Muir},
  {Myles}, {Nadathur}, {Navarro-Alsina}, {Nichol}, {Ogando}, {Omori},
  {Palmese}, {Pandey}, {Park}, {Paz-Chinch{\'o}n}, {Petravick}, {Pieres},
  {Plazas Malag{\'o}n}, {Porredon}, {Prat}, {Raveri}, {Rodriguez-Monroy},
  {Rollins}, {Romer}, {Roodman}, {Rosenfeld}, {Ross}, {Rykoff}, {Samuroff},
  {S{\'a}nchez}, {Sanchez}, {Sanchez}, {Sanchez Cid}, {Scarpine}, {Schubnell},
  {Scolnic}, {Secco}, {Serrano}, {Sevilla-Noarbe}, {Sheldon}, {Shin}, {Smith},
  {Soares-Santos}, {Suchyta}, {Swanson}, {Tabbutt}, {Tarle}, {Thomas}, {To},
  {Troja}, {Troxel}, {Tucker}, {Tutusaus}, {Varga}, {Walker}, {Weaverdyck},
  {Wechsler}, {Weller}, {Yanny}, {Yin}, {Zhang}, {Zuntz}, \& {DES
  Collaboration}}]{Abbott2022}
{Abbott}, T.~M.~C., {Aguena}, M., {Alarcon}, A., {et~al.} 2022, \prd, 105,
  023520

\bibitem[{{Ahumada} \& {Lapasset}(1995)}]{Ahumada1995}
{Ahumada}, J. \& {Lapasset}, E. 1995, \aaps, 109, 375

\bibitem[{{Ahumada} \& {Lapasset}(2007)}]{Ahumada2007}
{Ahumada}, J.~A. \& {Lapasset}, E. 2007, \aap, 463, 789

\bibitem[{{Albrecht} {et~al.}(2006){Albrecht}, {Bernstein}, {Cahn}, {Freedman},
  {Hewitt}, {Hu}, {Huth}, {Kamionkowski}, {Kolb}, {Knox}, {Mather}, {Staggs},
  \& {Suntzeff}}]{Albrecht2006}
{Albrecht}, A., {Bernstein}, G., {Cahn}, R., {et~al.} 2006, arXiv e-prints,
  arXiv:0609591

\bibitem[{{Alsing} {et~al.}(2020){Alsing}, {Peiris}, {Leja}, {Hahn}, {Tojeiro},
  {Mortlock}, {Leistedt}, {Johnson}, \& {Conroy}}]{Alsing2020}
{Alsing}, J., {Peiris}, H., {Leja}, J., {et~al.} 2020, \apjs, 249, 5

\bibitem[{{Alsing} {et~al.}(2023){Alsing}, {Peiris}, {Mortlock}, {Leja}, \&
  {Leistedt}}]{Alsing2023}
{Alsing}, J., {Peiris}, H., {Mortlock}, D., {Leja}, J., \& {Leistedt}, B. 2023,
  \apjs, 264, 29

\bibitem[{{Alsing} {et~al.}(2024){Alsing}, {Thorp}, {Deger}, {Peiris},
  {Leistedt}, {Mortlock}, \& {Leja}}]{Alsing2024}
{Alsing}, J., {Thorp}, S., {Deger}, S., {et~al.} 2024, arXiv e-prints,
  arXiv:2402.00935

\bibitem[{{Amon} {et~al.}(2022){Amon}, {Gruen}, {Troxel}, {MacCrann},
  {Dodelson}, {Choi}, {Doux}, {Secco}, {Samuroff}, {Krause}, {Cordero},
  {Myles}, {DeRose}, {Wechsler}, {Gatti}, {Navarro-Alsina}, {Bernstein},
  {Jain}, {Blazek}, {Alarcon}, {Fert{\'e}}, {Lemos}, {Raveri}, {Campos},
  {Prat}, {S{\'a}nchez}, {Jarvis}, {Alves}, {Andrade-Oliveira}, {Baxter},
  {Bechtol}, {Becker}, {Bridle}, {Camacho}, {Carnero Rosell}, {Carrasco Kind},
  {Cawthon}, {Chang}, {Chen}, {Chintalapati}, {Crocce}, {Davis}, {Diehl},
  {Drlica-Wagner}, {Eckert}, {Eifler}, {Elvin-Poole}, {Everett}, {Fang},
  {Fosalba}, {Friedrich}, {Gaztanaga}, {Giannini}, {Gruendl}, {Harrison},
  {Hartley}, {Herner}, {Huang}, {Huff}, {Huterer}, {Kuropatkin}, {Leget},
  {Liddle}, {McCullough}, {Muir}, {Pandey}, {Park}, {Porredon}, {Refregier},
  {Rollins}, {Roodman}, {Rosenfeld}, {Ross}, {Rykoff}, {Sanchez},
  {Sevilla-Noarbe}, {Sheldon}, {Shin}, {Troja}, {Tutusaus}, {Tutusaus},
  {Varga}, {Weaverdyck}, {Yanny}, {Yin}, {Zhang}, {Zuntz}, {Aguena}, {Allam},
  {Annis}, {Bacon}, {Bertin}, {Bhargava}, {Brooks}, {Buckley-Geer}, {Burke},
  {Carretero}, {Costanzi}, {da Costa}, {Pereira}, {De Vicente}, {Desai},
  {Dietrich}, {Doel}, {Ferrero}, {Flaugher}, {Frieman}, {Garc{\'\i}a-Bellido},
  {Gaztanaga}, {Gerdes}, {Giannantonio}, {Gschwend}, {Gutierrez}, {Hinton},
  {Hollowood}, {Honscheid}, {Hoyle}, {James}, {Kron}, {Kuehn}, {Lahav}, {Lima},
  {Lin}, {Maia}, {Marshall}, {Martini}, {Melchior}, {Menanteau}, {Miquel},
  {Mohr}, {Morgan}, {Ogando}, {Palmese}, {Paz-Chinch{\'o}n}, {Petravick},
  {Pieres}, {Romer}, {Sanchez}, {Scarpine}, {Schubnell}, {Serrano}, {Smith},
  {Soares-Santos}, {Tarle}, {Thomas}, {To}, {Weller}, \& {DES
  Collaboration}}]{Amon2022}
{Amon}, A., {Gruen}, D., {Troxel}, M.~A., {et~al.} 2022, \prd, 105, 023514

\bibitem[{{Anders} \& {Fritze-v. Alvensleben}(2003)}]{Anders2003}
{Anders}, P. \& {Fritze-v. Alvensleben}, U. 2003, \aap, 401, 1063

\bibitem[{{Baldry} {et~al.}(2012){Baldry}, {Driver}, {Loveday}, {Taylor},
  {Kelvin}, {Liske}, {Norberg}, {Robotham}, {Brough}, {Hopkins}, {Bamford},
  {Peacock}, {Bland-Hawthorn}, {Conselice}, {Croom}, {Jones}, {Parkinson},
  {Popescu}, {Prescott}, {Sharp}, \& {Tuffs}}]{Baldry2012}
{Baldry}, I.~K., {Driver}, S.~P., {Loveday}, J., {et~al.} 2012, \mnras, 421,
  621

\bibitem[{{Baldwin}(1977)}]{Baldwin1977}
{Baldwin}, J.~A. 1977, \apj, 214, 679

\bibitem[{{Beck} {et~al.}(2016){Beck}, {Dobos}, {Yip}, {Szalay}, \&
  {Csabai}}]{Beck2016}
{Beck}, R., {Dobos}, L., {Yip}, C.-W., {Szalay}, A.~S., \& {Csabai}, I. 2016,
  \mnras, 457, 362

\bibitem[{{Bedijn}(1987)}]{Bedijn1987}
{Bedijn}, P.~J. 1987, \aap, 186, 136

\bibitem[{{Behroozi} {et~al.}(2019){Behroozi}, {Wechsler}, {Hearin}, \&
  {Conroy}}]{Behroozi2019}
{Behroozi}, P., {Wechsler}, R.~H., {Hearin}, A.~P., \& {Conroy}, C. 2019,
  \mnras, 488, 3143

\bibitem[{{Belfiore} {et~al.}(2016){Belfiore}, {Maiolino}, {Maraston},
  {Emsellem}, {Bershady}, {Masters}, {Yan}, {Bizyaev}, {Boquien}, {Brownstein},
  {Bundy}, {Drory}, {Heckman}, {Law}, {Roman-Lopes}, {Pan}, {Stanghellini},
  {Thomas}, {Weijmans}, \& {Westfall}}]{Belfiore2016}
{Belfiore}, F., {Maiolino}, R., {Maraston}, C., {et~al.} 2016, \mnras, 461,
  3111

\bibitem[{{Bellstedt} {et~al.}(2021){Bellstedt}, {Robotham}, {Driver},
  {Thorne}, {Davies}, {Holwerda}, {Hopkins}, {Lara-Lopez},
  {L{\'o}pez-S{\'a}nchez}, \& {Phillipps}}]{Bellstedt2021}
{Bellstedt}, S., {Robotham}, A. S.~G., {Driver}, S.~P., {et~al.} 2021, \mnras,
  503, 3309

\bibitem[{{Bellstedt} {et~al.}(2020){Bellstedt}, {Robotham}, {Driver},
  {Thorne}, {Davies}, {Lagos}, {Stevens}, {Taylor}, {Baldry}, {Moffett},
  {Hopkins}, \& {Phillipps}}]{Bellstedt2020}
{Bellstedt}, S., {Robotham}, A. S.~G., {Driver}, S.~P., {et~al.} 2020, \mnras,
  498, 5581

\bibitem[{{Bernstein} {et~al.}(2002){Bernstein}, {Freedman}, \&
  {Madore}}]{Bernstein2002}
{Bernstein}, R.~A., {Freedman}, W.~L., \& {Madore}, B.~F. 2002, \apj, 571, 107

\bibitem[{{Bezanson} {et~al.}(2022){Bezanson}, {Labbe}, {Whitaker}, {Leja},
  {Price}, {Franx}, {Brammer}, {Marchesini}, {Zitrin}, {Wang}, {Weaver},
  {Furtak}, {Atek}, {Coe}, {Cutler}, {Dayal}, {van Dokkum}, {Feldmann},
  {Forster Schreiber}, {Fujimoto}, {Geha}, {Glazebrook}, {de Graaff}, {Greene},
  {Juneau}, {Kassin}, {Kriek}, {Khullar}, {Maseda}, {Mowla}, {Muzzin},
  {Nanayakkara}, {Nelson}, {Oesch}, {Pacifici}, {Pan}, {Papovich}, {Setton},
  {Shapley}, {Smit}, {Stefanon}, {Taylor}, \& {Williams}}]{Bezanson2022}
{Bezanson}, R., {Labbe}, I., {Whitaker}, K.~E., {et~al.} 2022, arXiv e-prints,
  arXiv:2212.04026

\bibitem[{{Blanton} \& {Roweis}(2007)}]{Blanton2007}
{Blanton}, M.~R. \& {Roweis}, S. 2007, \aj, 133, 734

\bibitem[{{Blitz} \& {Shu}(1980)}]{Blitz1980}
{Blitz}, L. \& {Shu}, F.~H. 1980, \apj, 238, 148

\bibitem[{{Brown} {et~al.}(2014){Brown}, {Moustakas}, {Smith}, {da Cunha},
  {Jarrett}, {Imanishi}, {Armus}, {Brandl}, \& {Peek}}]{Brown2014}
{Brown}, M. J.~I., {Moustakas}, J., {Smith}, J. D.~T., {et~al.} 2014, \apjs,
  212, 18

\bibitem[{{Bruderer} {et~al.}(2016){Bruderer}, {Chang}, {Refregier}, {Amara},
  {Berg{\'e}}, \& {Gamper}}]{Bruderer2016}
{Bruderer}, C., {Chang}, C., {Refregier}, A., {et~al.} 2016, \apj, 817, 25

\bibitem[{{Buchs} {et~al.}(2019){Buchs}, {Davis}, {Gruen}, {DeRose}, {Alarcon},
  {Bernstein}, {S{\'a}nchez}, {Myles}, {Roodman}, {Allen}, {Amon}, {Choi},
  {Masters}, {Miquel}, {Troxel}, {Wechsler}, {Abbott}, {Annis}, {Avila},
  {Bechtol}, {Bridle}, {Brooks}, {Buckley-Geer}, {Burke}, {Carnero Rosell},
  {Carrasco Kind}, {Carretero}, {Castander}, {Cawthon}, {D'Andrea}, {da Costa},
  {De Vicente}, {Desai}, {Diehl}, {Doel}, {Drlica-Wagner}, {Eifler}, {Evrard},
  {Flaugher}, {Fosalba}, {Frieman}, {Garc{\'\i}a-Bellido}, {Gaztanaga},
  {Gruendl}, {Gschwend}, {Gutierrez}, {Hartley}, {Hollowood}, {Honscheid},
  {James}, {Kuehn}, {Kuropatkin}, {Lima}, {Lin}, {Maia}, {March}, {Marshall},
  {Melchior}, {Menanteau}, {Ogando}, {Plazas}, {Rykoff}, {Sanchez}, {Scarpine},
  {Serrano}, {Sevilla-Noarbe}, {Smith}, {Soares-Santos}, {Sobreira}, {Suchyta},
  {Swanson}, {Tarle}, {Thomas}, {Vikram}, \& {DES Collaboration}}]{Buchs2019}
{Buchs}, R., {Davis}, C., {Gruen}, D., {et~al.} 2019, \mnras, 489, 820

\bibitem[{{Bunker} {et~al.}(2023{\natexlab{a}}){Bunker}, {Cameron},
  {Curtis-Lake}, {Jakobsen}, {Carniani}, {Curti}, {Witstok}, {Maiolino},
  {D'Eugenio}, {Looser}, {Willott}, {Bonaventura}, {Hainline}, {Uebler},
  {Willmer}, {Saxena}, {Smit}, {Alberts}, {Arribas}, {Baker}, {Baum},
  {Bhatawdekar}, {Bowler}, {Boyett}, {Charlot}, {Chen}, {Chevallard},
  {Circosta}, {DeCoursey}, {de Graaff}, {Egami}, {Eisenstein}, {Endsley},
  {Ferruit}, {Giardino}, {Hausen}, {Helton}, {Hviding}, {Ji}, {Johnson},
  {Jones}, {Kumari}, {Laseter}, {Luetzgendorf}, {Maseda}, {Nelson}, {Parlanti},
  {Perna}, {Rauscher}, {Rawle}, {Rix}, {Rieke}, {Robertson}, {Rodriguez Del
  Pino}, {Sandles}, {Scholtz}, {Sharpe}, {Skarbinski}, {Stark}, {Sun},
  {Tacchella}, {Topping}, {Villanueva}, {Wallace}, {Williams}, \&
  {Woodrum}}]{Bunker2023a}
{Bunker}, A.~J., {Cameron}, A.~J., {Curtis-Lake}, E., {et~al.}
  2023{\natexlab{a}}, arXiv e-prints, arXiv:2306.02467

\bibitem[{{Bunker} {et~al.}(2023{\natexlab{b}}){Bunker}, {Saxena}, {Cameron},
  {Willott}, {Curtis-Lake}, {Jakobsen}, {Carniani}, {Smit}, {Maiolino},
  {Witstok}, {Curti}, {D'Eugenio}, {Jones}, {Ferruit}, {Arribas}, {Charlot},
  {Chevallard}, {Giardino}, {de Graaff}, {Looser}, {L{\"u}tzgendorf}, {Maseda},
  {Rawle}, {Rix}, {Del Pino}, {Alberts}, {Egami}, {Eisenstein}, {Endsley},
  {Hainline}, {Hausen}, {Johnson}, {Rieke}, {Rieke}, {Robertson}, {Shivaei},
  {Stark}, {Sun}, {Tacchella}, {Tang}, {Williams}, {Willmer}, {Baker}, {Baum},
  {Bhatawdekar}, {Bowler}, {Boyett}, {Chen}, {Circosta}, {Helton}, {Ji},
  {Kumari}, {Lyu}, {Nelson}, {Parlanti}, {Perna}, {Sandles}, {Scholtz},
  {Suess}, {Topping}, {{\"U}bler}, {Wallace}, \& {Whitler}}]{Bunker2023b}
{Bunker}, A.~J., {Saxena}, A., {Cameron}, A.~J., {et~al.} 2023{\natexlab{b}},
  \aap, 677, A88

\bibitem[{{Byler} {et~al.}(2017){Byler}, {Dalcanton}, {Conroy}, \&
  {Johnson}}]{Byler2017}
{Byler}, N., {Dalcanton}, J.~J., {Conroy}, C., \& {Johnson}, B.~D. 2017, \apj,
  840, 44

\bibitem[{{Byler} {et~al.}(2019){Byler}, {Dalcanton}, {Conroy}, {Johnson},
  {Choi}, {Dotter}, \& {Rosenfield}}]{Byler2019}
{Byler}, N., {Dalcanton}, J.~J., {Conroy}, C., {et~al.} 2019, \aj, 158, 2

\bibitem[{{Calabro} {et~al.}(2024){Calabro}, {Castellano}, {Zavala},
  {Pentericci}, {Arrabal Haro}, {Bakx}, {Burgarella}, {Casey}, {Dickinson},
  {Finkelstein}, {Fontana}, {Llerena}, {Mascia}, {Merlin}, {Mitsuhashi},
  {Napolitano}, {Paris}, {Perez-Gonzalez}, {Roberts-Borsani}, {Santini},
  {Treu}, \& {Vanzella}}]{Calabro2024}
{Calabro}, A., {Castellano}, M., {Zavala}, J.~A., {et~al.} 2024, arXiv
  e-prints, arXiv:2403.12683

\bibitem[{{Calzetti} {et~al.}(2000){Calzetti}, {Armus}, {Bohlin}, {Kinney},
  {Koornneef}, \& {Storchi-Bergmann}}]{Calzetti2000}
{Calzetti}, D., {Armus}, L., {Bohlin}, R.~C., {et~al.} 2000, \apj, 533, 682

\bibitem[{{Cameron} {et~al.}(2023){Cameron}, {Katz}, {Witten}, {Saxena},
  {Laporte}, \& {Bunker}}]{Cameron2023}
{Cameron}, A.~J., {Katz}, H., {Witten}, C., {et~al.} 2023, arXiv e-prints,
  arXiv:2311.02051

\bibitem[{{Cardelli} {et~al.}(1989){Cardelli}, {Clayton}, \&
  {Mathis}}]{Cardelli1989}
{Cardelli}, J.~A., {Clayton}, G.~C., \& {Mathis}, J.~S. 1989, \apj, 345, 245

\bibitem[{{Chabrier}(2003)}]{Chabrier2003}
{Chabrier}, G. 2003, \pasp, 115, 763

\bibitem[{{Charlot} \& {Fall}(2000)}]{Charlot2000}
{Charlot}, S. \& {Fall}, S.~M. 2000, \apj, 539, 718

\bibitem[{{Chaussidon} {et~al.}(2023){Chaussidon}, {Y{\`e}che},
  {Palanque-Delabrouille}, {Alexander}, {Yang}, {Ahlen}, {Bailey}, {Brooks},
  {Cai}, {Chabanier}, {Davis}, {Dawson}, {de laMacorra}, {Dey}, {Dey},
  {Eftekharzadeh}, {Eisenstein}, {Fanning}, {Font-Ribera}, {Gazta{\~n}aga}, {A
  Gontcho}, {Gonzalez-Morales}, {Guy}, {Herrera-Alcantar}, {Honscheid},
  {Ishak}, {Jiang}, {Juneau}, {Kehoe}, {Kisner}, {Kov{\'a}cs}, {Kremin}, {Lan},
  {Landriau}, {Le Guillou}, {Levi}, {Magneville}, {Martini}, {Meisner},
  {Moustakas}, {Mu{\~n}oz-Guti{\'e}rrez}, {Myers}, {Newman}, {Nie}, {Percival},
  {Poppett}, {Prada}, {Raichoor}, {Ravoux}, {Ross}, {Schlafly}, {Schlegel},
  {Tan}, {Tarl{\'e}}, {Zhou}, {Zhou}, \& {Zou}}]{Chaussidon2023}
{Chaussidon}, E., {Y{\`e}che}, C., {Palanque-Delabrouille}, N., {et~al.} 2023,
  \apj, 944, 107

\bibitem[{{Chisari} \& {Kelson}(2012)}]{Chisari2012}
{Chisari}, N.~E. \& {Kelson}, D.~D. 2012, \apj, 753, 94

\bibitem[{{Choi} {et~al.}(2016){Choi}, {Dotter}, {Conroy}, {Cantiello},
  {Paxton}, \& {Johnson}}]{Choi2016}
{Choi}, J., {Dotter}, A., {Conroy}, C., {et~al.} 2016, \apj, 823, 102

\bibitem[{{Chon} {et~al.}(2021){Chon}, {Omukai}, \& {Schneider}}]{Chon2021}
{Chon}, S., {Omukai}, K., \& {Schneider}, R. 2021, \mnras, 508, 4175

\bibitem[{{Cohn} {et~al.}(2018){Cohn}, {Leja}, {Tran}, {Forrest}, {Johnson},
  {Tillman}, {Alcorn}, {Conroy}, {Glazebrook}, {Kacprzak}, {Kelson},
  {Nanayakkara}, {Papovich}, {van Dokkum}, \& {Yuan}}]{Cohn2018}
{Cohn}, J.~H., {Leja}, J., {Tran}, K.-V.~H., {et~al.} 2018, \apj, 869, 141

\bibitem[{{Conroy}(2013)}]{Conroy2013}
{Conroy}, C. 2013, \araa, 51, 393

\bibitem[{{Conroy} \& {Gunn}(2010)}]{Conroy2010}
{Conroy}, C. \& {Gunn}, J.~E. 2010, \apj, 712, 833

\bibitem[{{Conroy} {et~al.}(2009){Conroy}, {Gunn}, \& {White}}]{Conroy2009}
{Conroy}, C., {Gunn}, J.~E., \& {White}, M. 2009, \apj, 699, 486

\bibitem[{{Costantin} {et~al.}(2019){Costantin}, {Iovino}, {Zibetti},
  {Longhetti}, {Gallazzi}, {Mercurio}, {Lonoce}, {Balcells}, {Bolzonella},
  {Busarello}, {Dalton}, {Ferr{\'e}-Mateu}, {Garc{\'\i}a-Benito}, {Gargiulo},
  {Haines}, {Jin}, {La Barbera}, {McGee}, {Merluzzi}, {Morelli}, {Murphy},
  {Peralta de Arriba}, {Pizzella}, {Poggianti}, {Pozzetti},
  {S{\'a}nchez-Bl{\'a}zquez}, {Talia}, {Tortora}, {Trager}, {Vazdekis},
  {Vergani}, \& {Vulcani}}]{Costantin2019}
{Costantin}, L., {Iovino}, A., {Zibetti}, S., {et~al.} 2019, \aap, 632, A9

\bibitem[{{Crenshaw} {et~al.}(2024){Crenshaw}, {Bryce Kalmbach}, {Gagliano},
  {Yan}, {Connolly}, {Malz}, {Schmidt}, \& {The LSST Dark Energy Science
  Collaboration}}]{Crenshaw2024}
{Crenshaw}, J.~F., {Bryce Kalmbach}, J., {Gagliano}, A., {et~al.} 2024, arXiv
  e-prints, arXiv:2405.04740

\bibitem[{{Crocce} {et~al.}(2016){Crocce}, {Carretero}, {Bauer}, {Ross},
  {Sevilla-Noarbe}, {Giannantonio}, {Sobreira}, {Sanchez}, {Gaztanaga},
  {Carrasco Kind}, {S{\'a}nchez}, {Bonnett}, {Benoit-L{\'e}vy}, {Brunner},
  {Carnero Rosell}, {Cawthon}, {Fosalba}, {Hartley}, {Kim}, {Leistedt},
  {Miquel}, {Peiris}, {Percival}, {Rosenfeld}, {Rykoff}, {S{\'a}nchez},
  {Abbott}, {Abdalla}, {Allam}, {Banerji}, {Bernstein}, {Bertin}, {Brooks},
  {Buckley-Geer}, {Burke}, {Capozzi}, {Castander}, {Cunha}, {D'Andrea}, {da
  Costa}, {Desai}, {Diehl}, {Eifler}, {Evrard}, {Fausti Neto}, {Fernandez},
  {Finley}, {Flaugher}, {Frieman}, {Gerdes}, {Gruen}, {Gruendl}, {Gutierrez},
  {Honscheid}, {James}, {Kuehn}, {Kuropatkin}, {Lahav}, {Li}, {Lima}, {Maia},
  {March}, {Marshall}, {Martini}, {Melchior}, {Miller}, {Neilsen}, {Nichol},
  {Nord}, {Ogando}, {Plazas}, {Romer}, {Sako}, {Santiago}, {Schubnell},
  {Smith}, {Soares-Santos}, {Suchyta}, {Swanson}, {Tarle}, {Thaler}, {Thomas},
  {Vikram}, {Walker}, {Wechsler}, {Weller}, {Zuntz}, \& {DES
  Collaboration}}]{Crocce2016}
{Crocce}, M., {Carretero}, J., {Bauer}, A.~H., {et~al.} 2016, \mnras, 455, 4301

\bibitem[{{Crowther}(2007)}]{Crowther2007}
{Crowther}, P.~A. 2007, \araa, 45, 177

\bibitem[{{Culpan} {et~al.}(2021){Culpan}, {Pelisoli}, \& {Geier}}]{Culpan2021}
{Culpan}, R., {Pelisoli}, I., \& {Geier}, S. 2021, \aap, 654, A107

\bibitem[{{Curti} {et~al.}(2020){Curti}, {Mannucci}, {Cresci}, \&
  {Maiolino}}]{Curti2020}
{Curti}, M., {Mannucci}, F., {Cresci}, G., \& {Maiolino}, R. 2020, \mnras, 491,
  944

\bibitem[{{Dalal} {et~al.}(2023){Dalal}, {Li}, {Nicola}, {Zuntz}, {Strauss},
  {Sugiyama}, {Zhang}, {Rau}, {Mandelbaum}, {Takada}, {More}, {Miyatake},
  {Kannawadi}, {Shirasaki}, {Taniguchi}, {Takahashi}, {Osato}, {Hamana},
  {Oguri}, {Nishizawa}, {Plazas Malag{\'o}n}, {Sunayama}, {Alonso}, {Slosar},
  {Armstrong}, {Bosch}, {Komiyama}, {Lupton}, {Lust}, {MacArthur}, {Miyazaki},
  {Murayama}, {Nishimichi}, {Okura}, {Price}, {Tait}, {Tanaka}, \&
  {Wang}}]{Dalal2023}
{Dalal}, R., {Li}, X., {Nicola}, A., {et~al.} 2023, arXiv e-prints,
  arXiv:2304.00701

\bibitem[{{Davidzon} {et~al.}(2017){Davidzon}, {Ilbert}, {Laigle}, {Coupon},
  {McCracken}, {Delvecchio}, {Masters}, {Capak}, {Hsieh}, {Le F{\`e}vre},
  {Tresse}, {Bethermin}, {Chang}, {Faisst}, {Le Floc'h}, {Steinhardt}, {Toft},
  {Aussel}, {Dubois}, {Hasinger}, {Salvato}, {Sanders}, {Scoville}, \&
  {Silverman}}]{Davidzon2017}
{Davidzon}, I., {Ilbert}, O., {Laigle}, C., {et~al.} 2017, \aap, 605, A70

\bibitem[{{Davies} {et~al.}(2018){Davies}, {Robotham}, {Driver}, {Lagos},
  {Cortese}, {Mannering}, {Foster}, {Lidman}, {Hashemizadeh}, {Koushan},
  {O'Toole}, {Baldry}, {Bilicki}, {Bland-Hawthorn}, {Bremer}, {Brown},
  {Bryant}, {Catinella}, {Croom}, {Grootes}, {Holwerda}, {Jarvis}, {Maddox},
  {Meyer}, {Moffett}, {Phillipps}, {Taylor}, {Windhorst}, \&
  {Wolf}}]{Davies2018}
{Davies}, L.~J.~M., {Robotham}, A.~S.~G., {Driver}, S.~P., {et~al.} 2018,
  \mnras, 480, 768

\bibitem[{{Davies} {et~al.}(1994){Davies}, {Benz}, \& {Hills}}]{Davies1994}
{Davies}, M.~B., {Benz}, W., \& {Hills}, J.~G. 1994, \apj, 424, 870

\bibitem[{{Desjacques} {et~al.}(2018){Desjacques}, {Jeong}, \&
  {Schmidt}}]{Desjacques2018}
{Desjacques}, V., {Jeong}, D., \& {Schmidt}, F. 2018, \physrep, 733, 1

\bibitem[{{Dore} {et~al.}(2019){Dore}, {Hirata}, {Wang}, {Weinberg}, {Eifler},
  {Foley}, {Heinrich}, {Krause}, {Perlmutter}, {Pisani}, {Scolnic}, {Spergel},
  {Suntzeff}, {Aldering}, {Baltay}, {Capak}, {Choi}, {Dvorkin}, {Fall}, {Fang},
  {Fruchter}, {Galbany}, {Ho}, {Hounsell}, {Izard}, {Jain}, {Koekemoer},
  {Kruk}, {Leauthaud}, {Malhotra}, {Mandelbaum}, {Massara}, {Masters},
  {Miyatake}, {Plazas}, {Rhoads}, {Rhodes}, {Rose}, {Rubin}, {Sako},
  {Samushia}, {Shirasaki}, {Simet}, {Takada}, {Troxel}, {Wu}, {Yoshida}, \&
  {Zhai}}]{Dore2019}
{Dore}, O., {Hirata}, C., {Wang}, Y., {et~al.} 2019, \baas, 51, 341

\bibitem[{{Dotter}(2016)}]{Dotter2016}
{Dotter}, A. 2016, \apjs, 222, 8

\bibitem[{{Draine}(2003)}]{Draine2003}
{Draine}, B.~T. 2003, \araa, 41, 241

\bibitem[{{Draine} \& {Lee}(1984)}]{Draine1984}
{Draine}, B.~T. \& {Lee}, H.~M. 1984, \apj, 285, 89

\bibitem[{{Draine} \& {Li}(2007)}]{Draine2007}
{Draine}, B.~T. \& {Li}, A. 2007, \apj, 657, 810

\bibitem[{{Driver} {et~al.}(2019){Driver}, {Liske}, {Davies}, {Robotham},
  {Baldry}, {Brown}, {Cluver}, {Kuijken}, {Loveday}, {McMahon}, {Meyer},
  {Norberg}, {Owers}, {Power}, {Taylor}, \& {WAVES Team}}]{Driver2019}
{Driver}, S.~P., {Liske}, J., {Davies}, L.~J.~M., {et~al.} 2019, The Messenger,
  175, 46

\bibitem[{{Fagioli} {et~al.}(2018){Fagioli}, {Riebartsch}, {Nicola}, {Herbel},
  {Amara}, {Refregier}, {Chang}, {Gamper}, \& {Tortorelli}}]{Fagioli2018}
{Fagioli}, M., {Riebartsch}, J., {Nicola}, A., {et~al.} 2018, \jcap, 2018, 015

\bibitem[{{Fagioli} {et~al.}(2020){Fagioli}, {Tortorelli}, {Herbel},
  {Z{\"u}rcher}, {Refregier}, \& {Amara}}]{Fagioli2020}
{Fagioli}, M., {Tortorelli}, L., {Herbel}, J., {et~al.} 2020, \jcap, 2020, 050

\bibitem[{{Ferland} {et~al.}(2017){Ferland}, {Chatzikos}, {Guzm{\'a}n},
  {Lykins}, {van Hoof}, {Williams}, {Abel}, {Badnell}, {Keenan}, {Porter}, \&
  {Stancil}}]{Ferland2017}
{Ferland}, G.~J., {Chatzikos}, M., {Guzm{\'a}n}, F., {et~al.} 2017, \rmxaa, 53,
  385

\bibitem[{{Fortuni} {et~al.}(2023){Fortuni}, {Merlin}, {Fontana}, {Giocoli},
  {Romelli}, {Graziani}, {Santini}, {Castellano}, {Charlot}, \&
  {Chevallard}}]{Fortuni2023}
{Fortuni}, F., {Merlin}, E., {Fontana}, A., {et~al.} 2023, \aap, 677, A102

\bibitem[{{Fotopoulou} {et~al.}(2016){Fotopoulou}, {Buchner},
  {Georgantopoulos}, {Hasinger}, {Salvato}, {Georgakakis}, {Cappelluti},
  {Ranalli}, {Hsu}, {Brusa}, {Comastri}, {Miyaji}, {Nandra}, {Aird}, \&
  {Paltani}}]{Fotopoulou2016}
{Fotopoulou}, S., {Buchner}, J., {Georgantopoulos}, I., {et~al.} 2016, \aap,
  587, A142

\bibitem[{{Gallazzi} {et~al.}(2005){Gallazzi}, {Charlot}, {Brinchmann},
  {White}, \& {Tremonti}}]{Gallazzi2005}
{Gallazzi}, A., {Charlot}, S., {Brinchmann}, J., {White}, S. D.~M., \&
  {Tremonti}, C.~A. 2005, \mnras, 362, 41

\bibitem[{{Garcia-Lario} {et~al.}(1997){Garcia-Lario}, {Manchado}, {Pych}, \&
  {Pottasch}}]{Garcia1997}
{Garcia-Lario}, P., {Manchado}, A., {Pych}, W., \& {Pottasch}, S.~R. 1997,
  \aaps, 126, 479

\bibitem[{{Georgakakis} {et~al.}(2017){Georgakakis}, {Aird}, {Schulze},
  {Dwelly}, {Salvato}, {Nandra}, {Merloni}, \& {Schneider}}]{Georgakakis2017}
{Georgakakis}, A., {Aird}, J., {Schulze}, A., {et~al.} 2017, \mnras, 471, 1976

\bibitem[{{Girardi} {et~al.}(2000){Girardi}, {Bressan}, {Bertelli}, \&
  {Chiosi}}]{Girardi2000}
{Girardi}, L., {Bressan}, A., {Bertelli}, G., \& {Chiosi}, C. 2000, \aaps, 141,
  371

\bibitem[{{Grazian} {et~al.}(2015){Grazian}, {Fontana}, {Santini}, {Dunlop},
  {Ferguson}, {Castellano}, {Amorin}, {Ashby}, {Barro}, {Behroozi}, {Boutsia},
  {Caputi}, {Chary}, {Dekel}, {Dickinson}, {Faber}, {Fazio}, {Finkelstein},
  {Galametz}, {Giallongo}, {Giavalisco}, {Grogin}, {Guo}, {Kocevski},
  {Koekemoer}, {Koo}, {Lee}, {Lu}, {Merlin}, {Mobasher}, {Nonino}, {Papovich},
  {Paris}, {Pentericci}, {Reddy}, {Renzini}, {Salmon}, {Salvato}, {Sommariva},
  {Song}, \& {Vanzella}}]{Grazian2015}
{Grazian}, A., {Fontana}, A., {Santini}, P., {et~al.} 2015, \aap, 575, A96

\bibitem[{{Hahn} {et~al.}(2023){Hahn}, {Wilson}, {Ruiz-Macias}, {Cole},
  {Weinberg}, {Moustakas}, {Kremin}, {Tinker}, {Smith}, {Wechsler}, {Ahlen},
  {Alam}, {Bailey}, {Brooks}, {Cooper}, {Davis}, {Dawson}, {Dey}, {Dey},
  {Eftekharzadeh}, {Eisenstein}, {Fanning}, {Forero-Romero}, {Frenk},
  {Gazta{\~n}aga}, {Gontcho A Gontcho}, {Guy}, {Honscheid}, {Ishak}, {Juneau},
  {Kehoe}, {Kisner}, {Lan}, {Landriau}, {Le Guillou}, {Levi}, {Magneville},
  {Martini}, {Meisner}, {Myers}, {Nie}, {Norberg}, {Palanque-Delabrouille},
  {Percival}, {Poppett}, {Prada}, {Raichoor}, {Ross}, {Safonova}, {Saulder},
  {Schlafly}, {Schlegel}, {Sierra-Porta}, {Tarle}, {Weaver}, {Y{\`e}che},
  {Zarrouk}, {Zhou}, {Zhou}, \& {Zou}}]{Hahn2023}
{Hahn}, C., {Wilson}, M.~J., {Ruiz-Macias}, O., {et~al.} 2023, \aj, 165, 253

\bibitem[{{Hartley} {et~al.}(2020){Hartley}, {Chang}, {Samani}, {Carnero
  Rosell}, {Davis}, {Hoyle}, {Gruen}, {Asorey}, {Gschwend}, {Lidman}, {Kuehn},
  {King}, {Rau}, {Wechsler}, {DeRose}, {Hinton}, {Whiteway}, {Abbott},
  {Aguena}, {Allam}, {Annis}, {Avila}, {Bernstein}, {Bertin}, {Bridle},
  {Brooks}, {Burke}, {Carrasco Kind}, {Carretero}, {Castander}, {Cawthon},
  {Costanzi}, {da Costa}, {Desai}, {Diehl}, {Dietrich}, {Flaugher}, {Fosalba},
  {Frieman}, {Garc{\'\i}a-Bellido}, {Gaztanaga}, {Gerdes}, {Gruendl},
  {Gutierrez}, {Hollowood}, {Honscheid}, {James}, {Kent}, {Krause},
  {Kuropatkin}, {Lahav}, {Lima}, {Maia}, {Marshall}, {Melchior}, {Menanteau},
  {Miquel}, {Ogando}, {Palmese}, {Paz-Chinch{\'o}n}, {Plazas}, {Roodman},
  {Rykoff}, {Sanchez}, {Scarpine}, {Schubnell}, {Serrano}, {Sevilla-Noarbe},
  {Smith}, {Soares-Santos}, {Suchyta}, {Tarle}, {Troxel}, {Tucker}, {Varga},
  {Weller}, {Wilkinson}, \& {DES Collaboration}}]{Hartley2020}
{Hartley}, W.~G., {Chang}, C., {Samani}, S., {et~al.} 2020, \mnras, 496, 4769

\bibitem[{{Hearin} {et~al.}(2023){Hearin}, {Chaves-Montero}, {Alarcon},
  {Becker}, \& {Benson}}]{Hearin2023}
{Hearin}, A.~P., {Chaves-Montero}, J., {Alarcon}, A., {Becker}, M.~R., \&
  {Benson}, A. 2023, \mnras, 521, 1741

\bibitem[{{Hensley} \& {Draine}(2021)}]{Hensley2021}
{Hensley}, B.~S. \& {Draine}, B.~T. 2021, \apj, 906, 73

\bibitem[{{Herbel} {et~al.}(2017){Herbel}, {Kacprzak}, {Amara}, {Refregier},
  {Bruderer}, \& {Nicola}}]{Herbel2017}
{Herbel}, J., {Kacprzak}, T., {Amara}, A., {et~al.} 2017, \jcap, 2017, 035

\bibitem[{{Heymans} {et~al.}(2021){Heymans}, {Tr{\"o}ster}, {Asgari}, {Blake},
  {Hildebrandt}, {Joachimi}, {Kuijken}, {Lin}, {S{\'a}nchez}, {van den Busch},
  {Wright}, {Amon}, {Bilicki}, {de Jong}, {Crocce}, {Dvornik}, {Erben},
  {Fortuna}, {Getman}, {Giblin}, {Glazebrook}, {Hoekstra}, {Joudaki},
  {Kannawadi}, {K{\"o}hlinger}, {Lidman}, {Miller}, {Napolitano}, {Parkinson},
  {Schneider}, {Shan}, {Valentijn}, {Verdoes Kleijn}, \& {Wolf}}]{Heymans2021}
{Heymans}, C., {Tr{\"o}ster}, T., {Asgari}, M., {et~al.} 2021, \aap, 646, A140

\bibitem[{{Hildebrandt} {et~al.}(2020){Hildebrandt}, {K{\"o}hlinger}, {van den
  Busch}, {Joachimi}, {Heymans}, {Kannawadi}, {Wright}, {Asgari}, {Blake},
  {Hoekstra}, {Joudaki}, {Kuijken}, {Miller}, {Morrison}, {Tr{\"o}ster},
  {Amon}, {Archidiacono}, {Brieden}, {Choi}, {de Jong}, {Erben}, {Giblin},
  {Mead}, {Peacock}, {Radovich}, {Schneider}, {Sif{\'o}n}, \&
  {Tewes}}]{Hildebrandt2020}
{Hildebrandt}, H., {K{\"o}hlinger}, F., {van den Busch}, J.~L., {et~al.} 2020,
  \aap, 633, A69

\bibitem[{{Hills} \& {Day}(1976)}]{Hills1976}
{Hills}, J.~G. \& {Day}, C.~A. 1976, \aplett, 17, 87

\bibitem[{{Inoue} {et~al.}(2014){Inoue}, {Shimizu}, {Iwata}, \&
  {Tanaka}}]{Inoue2014}
{Inoue}, A.~K., {Shimizu}, I., {Iwata}, I., \& {Tanaka}, M. 2014, \mnras, 442,
  1805

\bibitem[{{Iovino} {et~al.}(2023{\natexlab{a}}){Iovino}, {Mercurio},
  {Gallazzi}, {La Barbera}, {Longhetti}, {Tortora}, {Zibetti}, {Belfiore},
  {Bianconi}, {Busarello}, {Corsini}, {Costantin}, {De Lucia}, {De Propris},
  {D'Eugenio}, {Fontanot}, {Garc{\'\i}a-Benito}, {Hirschmann}, {Haines},
  {Mannucci}, {McGee}, {Merluzzi}, {Morelli}, {Moretti}, {Pasquali},
  {Poggianti}, {Pozzetti}, {Rodighiero}, {S{\'a}nchez-Bl{\'a}zquez}, {van der
  Wel}, {Vazdekis}, {Vulcani}, {Zanella}, {Annunziatella}, {Concas},
  {Cassar{\`a}}, {Cresci}, {Curti}, {de Lorenzo-C{\'a}ceres}, {Mateu},
  {Delgado}, {Mancini}, {Pacifici}, {Perez-Montero}, {Pizzella},
  {Perez-Gonzalez}, {Trager}, \& {Vergani}}]{Iovino2023b}
{Iovino}, A., {Mercurio}, A., {Gallazzi}, A.~R., {et~al.} 2023{\natexlab{a}},
  The Messenger, 190, 22

\bibitem[{{Iovino} {et~al.}(2023{\natexlab{b}}){Iovino}, {Poggianti},
  {Mercurio}, {Longhetti}, {Bolzonella}, {Busarello}, {Gullieuszik}, {La
  Barbera}, {Merluzzi}, {Morelli}, {Tortora}, {Vergani}, {Zibetti}, {Haines},
  {Costantin}, {Ditrani}, {Pozzetti}, {Angthopo}, {Balcells}, {Bardelli},
  {Benn}, {Bianconi}, {Cassar{\`a}}, {Corsini}, {Cucciati}, {Dalton},
  {Ferr{\'e}-Mateu}, {Fossati}, {Gallazzi}, {Garc{\'\i}a-Benito}, {Granett},
  {Gonz{\'a}lez Delgado}, {Ikhsanova}, {Iodice}, {Jin}, {Knapen}, {McGee},
  {Moretti}, {Murphy}, {Peralta de Arriba}, {Pizzella},
  {S{\'a}nchez-Bl{\'a}zquez}, {Spiniello}, {Talia}, {Trager}, {Vazdekis},
  {Vulcani}, \& {Zucca}}]{Iovino2023}
{Iovino}, A., {Poggianti}, B.~M., {Mercurio}, A., {et~al.} 2023{\natexlab{b}},
  \aap, 672, A87

\bibitem[{{Ivezi{\'c}}(2017)}]{Ivezic2017}
{Ivezi{\'c}}, {\v{Z}}. 2017, in New Frontiers in Black Hole Astrophysics, ed.
  A.~{Gomboc}, Vol. 324, 330--337

\bibitem[{{Ivezi{\'c}} {et~al.}(2019){Ivezi{\'c}}, {Kahn}, {Tyson}, {Abel},
  {Acosta}, {Allsman}, {Alonso}, {AlSayyad}, {Anderson}, {Andrew}, {Angel},
  {Angeli}, {Ansari}, {Antilogus}, {Araujo}, {Armstrong}, {Arndt}, {Astier},
  {Aubourg}, {Auza}, {Axelrod}, {Bard}, {Barr}, {Barrau}, {Bartlett}, {Bauer},
  {Bauman}, {Baumont}, {Bechtol}, {Bechtol}, {Becker}, {Becla}, {Beldica},
  {Bellavia}, {Bianco}, {Biswas}, {Blanc}, {Blazek}, {Blandford}, {Bloom},
  {Bogart}, {Bond}, {Booth}, {Borgland}, {Borne}, {Bosch}, {Boutigny},
  {Brackett}, {Bradshaw}, {Brandt}, {Brown}, {Bullock}, {Burchat}, {Burke},
  {Cagnoli}, {Calabrese}, {Callahan}, {Callen}, {Carlin}, {Carlson},
  {Chandrasekharan}, {Charles-Emerson}, {Chesley}, {Cheu}, {Chiang}, {Chiang},
  {Chirino}, {Chow}, {Ciardi}, {Claver}, {Cohen-Tanugi}, {Cockrum}, {Coles},
  {Connolly}, {Cook}, {Cooray}, {Covey}, {Cribbs}, {Cui}, {Cutri}, {Daly},
  {Daniel}, {Daruich}, {Daubard}, {Daues}, {Dawson}, {Delgado}, {Dellapenna},
  {de Peyster}, {de Val-Borro}, {Digel}, {Doherty}, {Dubois},
  {Dubois-Felsmann}, {Durech}, {Economou}, {Eifler}, {Eracleous}, {Emmons},
  {Fausti Neto}, {Ferguson}, {Figueroa}, {Fisher-Levine}, {Focke}, {Foss},
  {Frank}, {Freemon}, {Gangler}, {Gawiser}, {Geary}, {Gee}, {Geha}, {Gessner},
  {Gibson}, {Gilmore}, {Glanzman}, {Glick}, {Goldina}, {Goldstein}, {Goodenow},
  {Graham}, {Gressler}, {Gris}, {Guy}, {Guyonnet}, {Haller}, {Harris},
  {Hascall}, {Haupt}, {Hernandez}, {Herrmann}, {Hileman}, {Hoblitt}, {Hodgson},
  {Hogan}, {Howard}, {Huang}, {Huffer}, {Ingraham}, {Innes}, {Jacoby}, {Jain},
  {Jammes}, {Jee}, {Jenness}, {Jernigan}, {Jevremovi{\'c}}, {Johns}, {Johnson},
  {Johnson}, {Jones}, {Juramy-Gilles}, {Juri{\'c}}, {Kalirai}, {Kallivayalil},
  {Kalmbach}, {Kantor}, {Karst}, {Kasliwal}, {Kelly}, {Kessler}, {Kinnison},
  {Kirkby}, {Knox}, {Kotov}, {Krabbendam}, {Krughoff}, {Kub{\'a}nek},
  {Kuczewski}, {Kulkarni}, {Ku}, {Kurita}, {Lage}, {Lambert}, {Lange},
  {Langton}, {Le Guillou}, {Levine}, {Liang}, {Lim}, {Lintott}, {Long},
  {Lopez}, {Lotz}, {Lupton}, {Lust}, {MacArthur}, {Mahabal}, {Mandelbaum},
  {Markiewicz}, {Marsh}, {Marshall}, {Marshall}, {May}, {McKercher}, {McQueen},
  {Meyers}, {Migliore}, {Miller}, {Mills}, {Miraval}, {Moeyens}, {Moolekamp},
  {Monet}, {Moniez}, {Monkewitz}, {Montgomery}, {Morrison}, {Mueller},
  {Muller}, {Mu{\~n}oz Arancibia}, {Neill}, {Newbry}, {Nief}, {Nomerotski},
  {Nordby}, {O'Connor}, {Oliver}, {Olivier}, {Olsen}, {O'Mullane}, {Ortiz},
  {Osier}, {Owen}, {Pain}, {Palecek}, {Parejko}, {Parsons}, {Pease},
  {Peterson}, {Peterson}, {Petravick}, {Libby Petrick}, {Petry},
  {Pierfederici}, {Pietrowicz}, {Pike}, {Pinto}, {Plante}, {Plate}, {Plutchak},
  {Price}, {Prouza}, {Radeka}, {Rajagopal}, {Rasmussen}, {Regnault}, {Reil},
  {Reiss}, {Reuter}, {Ridgway}, {Riot}, {Ritz}, {Robinson}, {Roby}, {Roodman},
  {Rosing}, {Roucelle}, {Rumore}, {Russo}, {Saha}, {Sassolas}, {Schalk},
  {Schellart}, {Schindler}, {Schmidt}, {Schneider}, {Schneider}, {Schoening},
  {Schumacher}, {Schwamb}, {Sebag}, {Selvy}, {Sembroski}, {Seppala}, {Serio},
  {Serrano}, {Shaw}, {Shipsey}, {Sick}, {Silvestri}, {Slater}, {Smith},
  {Smith}, {Sobhani}, {Soldahl}, {Storrie-Lombardi}, {Stover}, {Strauss},
  {Street}, {Stubbs}, {Sullivan}, {Sweeney}, {Swinbank}, {Szalay}, {Takacs},
  {Tether}, {Thaler}, {Thayer}, {Thomas}, {Thornton}, {Thukral}, {Tice},
  {Trilling}, {Turri}, {Van Berg}, {Vanden Berk}, {Vetter}, {Virieux},
  {Vucina}, {Wahl}, {Walkowicz}, {Walsh}, {Walter}, {Wang}, {Wang}, {Warner},
  {Wiecha}, {Willman}, {Winters}, {Wittman}, {Wolff}, {Wood-Vasey}, {Wu},
  {Xin}, {Yoachim}, \& {Zhan}}]{Ivezic2019}
{Ivezi{\'c}}, {\v{Z}}., {Kahn}, S.~M., {Tyson}, J.~A., {et~al.} 2019, \apj,
  873, 111

\bibitem[{{Jimenez} {et~al.}(1998){Jimenez}, {Flynn}, \&
  {Kotoneva}}]{Jimenez1998}
{Jimenez}, R., {Flynn}, C., \& {Kotoneva}, E. 1998, \mnras, 299, 515

\bibitem[{{Kaasinen} {et~al.}(2018){Kaasinen}, {Kewley}, {Bian}, {Groves},
  {Kashino}, {Silverman}, \& {Kartaltepe}}]{Kaasinen2018}
{Kaasinen}, M., {Kewley}, L., {Bian}, F., {et~al.} 2018, \mnras, 477, 5568

\bibitem[{{Kacprzak} {et~al.}(2014){Kacprzak}, {Bridle}, {Rowe}, {Voigt},
  {Zuntz}, {Hirsch}, \& {MacCrann}}]{Kacprzak2014}
{Kacprzak}, T., {Bridle}, S., {Rowe}, B., {et~al.} 2014, \mnras, 441, 2528

\bibitem[{{Kacprzak} {et~al.}(2020){Kacprzak}, {Herbel}, {Nicola}, {Sgier},
  {Tarsitano}, {Bruderer}, {Amara}, {Refregier}, {Bridle}, {Drlica-Wagner},
  {Gruen}, {Hartley}, {Hoyle}, {Secco}, {Zuntz}, {Annis}, {Avila}, {Bertin},
  {Brooks}, {Buckley-Geer}, {Carnero Rosell}, {Carrasco Kind}, {Carretero}, {da
  Costa}, {De Vicente}, {Desai}, {Diehl}, {Doel}, {Garc{\'\i}a-Bellido},
  {Gaztanaga}, {Gruendl}, {Gschwend}, {Gutierrez}, {Hollowood}, {Honscheid},
  {James}, {Jarvis}, {Lima}, {Maia}, {Marshall}, {Melchior}, {Menanteau},
  {Miquel}, {Paz-Chinch{\'o}n}, {Plazas}, {Sanchez}, {Scarpine}, {Serrano},
  {Sevilla-Noarbe}, {Smith}, {Suchyta}, {Swanson}, {Tarle}, {Vikram}, {Weller},
  \& {DES Collaboration}}]{Kacprzak2020}
{Kacprzak}, T., {Herbel}, J., {Nicola}, A., {et~al.} 2020, \prd, 101, 082003

\bibitem[{{Kacprzak} {et~al.}(2012){Kacprzak}, {Zuntz}, {Rowe}, {Bridle},
  {Refregier}, {Amara}, {Voigt}, \& {Hirsch}}]{Kacprzak2012}
{Kacprzak}, T., {Zuntz}, J., {Rowe}, B., {et~al.} 2012, \mnras, 427, 2711

\bibitem[{{Kelson} \& {Holden}(2010)}]{Kelson2010}
{Kelson}, D.~D. \& {Holden}, B.~P. 2010, \apjl, 713, L28

\bibitem[{{Khederlarian} {et~al.}(2024){Khederlarian}, {Newman}, {Andrews},
  {Dey}, {Moustakas}, {Hearin}, {Juneau}, {Tortorelli}, {Gruen}, {Hahn},
  {Canning}, {Aguilar}, {Ahlen}, {Brooks}, {Claybaugh}, {de la Macorra},
  {Doel}, {Fanning}, {Ferraro}, {Forero-Romero}, {Gazta{\~n}aga}, {Gontcho},
  {Kehoe}, {Kisner}, {Kremin}, {Lambert}, {Landriau}, {Manera}, {Meisner},
  {Miquel}, {Mueller}, {Mu{\~n}oz-Guti{\'e}rrez}, {Myers}, {Nie}, {Poppett},
  {Prada}, {Rezaie}, {Rossi}, {Sanchez}, {Schubnell}, {Silber}, {Sprayberry},
  {Tarl{\'e}}, {Weaver}, {Zhou}, \& {Zou}}]{Khederlarian2024}
{Khederlarian}, A., {Newman}, J.~A., {Andrews}, B.~H., {et~al.} 2024, \mnras,
  531, 1454

\bibitem[{{Kohonen}(2001)}]{Kohonen2001}
{Kohonen}, T. 2001, {Self-Organizing Maps}

\bibitem[{{Korytov} {et~al.}(2019){Korytov}, {Hearin}, {Kovacs}, {Larsen},
  {Rangel}, {Hollowed}, {Benson}, {Heitmann}, {Mao}, {Bahmanyar}, {Chang},
  {Campbell}, {DeRose}, {Finkel}, {Frontiere}, {Gawiser}, {Habib}, {Joachimi},
  {Lanusse}, {Li}, {Mandelbaum}, {Morrison}, {Newman}, {Pope}, {Rykoff},
  {Simet}, {To}, {Vikraman}, {Wechsler}, {White}, \& {(The LSST Dark Energy
  Science Collaboration}}]{Korytov2019}
{Korytov}, D., {Hearin}, A., {Kovacs}, E., {et~al.} 2019, \apjs, 245, 26

\bibitem[{{Kriek} \& {Conroy}(2013)}]{Kriek2013}
{Kriek}, M. \& {Conroy}, C. 2013, \apjl, 775, L16

\bibitem[{{Kroupa}(2001)}]{Kroupa2001}
{Kroupa}, P. 2001, \mnras, 322, 231

\bibitem[{{Kulkarni} {et~al.}(2019){Kulkarni}, {Worseck}, \&
  {Hennawi}}]{Kulkarni2019}
{Kulkarni}, G., {Worseck}, G., \& {Hennawi}, J.~F. 2019, \mnras, 488, 1035

\bibitem[{{Laigle} {et~al.}(2016){Laigle}, {McCracken}, {Ilbert}, {Hsieh},
  {Davidzon}, {Capak}, {Hasinger}, {Silverman}, {Pichon}, {Coupon}, {Aussel},
  {Le Borgne}, {Caputi}, {Cassata}, {Chang}, {Civano}, {Dunlop}, {Fynbo},
  {Kartaltepe}, {Koekemoer}, {Le F{\`e}vre}, {Le Floc'h}, {Leauthaud}, {Lilly},
  {Lin}, {Marchesi}, {Milvang-Jensen}, {Salvato}, {Sanders}, {Scoville},
  {Smolcic}, {Stockmann}, {Taniguchi}, {Tasca}, {Toft}, {Vaccari}, \&
  {Zabl}}]{Laigle2016}
{Laigle}, C., {McCracken}, H.~J., {Ilbert}, O., {et~al.} 2016, \apjs, 224, 24

\bibitem[{{Lee} {et~al.}(2002){Lee}, {Lee}, \& {Gibson}}]{Lee2002}
{Lee}, H.-c., {Lee}, Y.-W., \& {Gibson}, B.~K. 2002, \aj, 124, 2664

\bibitem[{{Lee} {et~al.}(1990){Lee}, {Demarque}, \& {Zinn}}]{Lee1990}
{Lee}, Y.-W., {Demarque}, P., \& {Zinn}, R. 1990, \apj, 350, 155

\bibitem[{{Lee} {et~al.}(1994){Lee}, {Demarque}, \& {Zinn}}]{Lee1994}
{Lee}, Y.-W., {Demarque}, P., \& {Zinn}, R. 1994, \apj, 423, 248

\bibitem[{{Leistedt} {et~al.}(2023){Leistedt}, {Alsing}, {Peiris}, {Mortlock},
  \& {Leja}}]{Leistedt2023}
{Leistedt}, B., {Alsing}, J., {Peiris}, H., {Mortlock}, D., \& {Leja}, J. 2023,
  \apjs, 264, 23

\bibitem[{{Leja} {et~al.}(2019{\natexlab{a}}){Leja}, {Carnall}, {Johnson},
  {Conroy}, \& {Speagle}}]{Leja2019b}
{Leja}, J., {Carnall}, A.~C., {Johnson}, B.~D., {Conroy}, C., \& {Speagle},
  J.~S. 2019{\natexlab{a}}, \apj, 876, 3

\bibitem[{{Leja} {et~al.}(2018){Leja}, {Johnson}, {Conroy}, \& {van
  Dokkum}}]{Leja2018}
{Leja}, J., {Johnson}, B.~D., {Conroy}, C., \& {van Dokkum}, P. 2018, \apj,
  854, 62

\bibitem[{{Leja} {et~al.}(2019{\natexlab{b}}){Leja}, {Johnson}, {Conroy}, {van
  Dokkum}, {Speagle}, {Brammer}, {Momcheva}, {Skelton}, {Whitaker}, {Franx}, \&
  {Nelson}}]{Leja2019}
{Leja}, J., {Johnson}, B.~D., {Conroy}, C., {et~al.} 2019{\natexlab{b}}, \apj,
  877, 140

\bibitem[{{Leja} {et~al.}(2017){Leja}, {Johnson}, {Conroy}, {van Dokkum}, \&
  {Byler}}]{Leja2017}
{Leja}, J., {Johnson}, B.~D., {Conroy}, C., {van Dokkum}, P.~G., \& {Byler}, N.
  2017, \apj, 837, 170

\bibitem[{{Leja} {et~al.}(2020){Leja}, {Speagle}, {Johnson}, {Conroy}, {van
  Dokkum}, \& {Franx}}]{Leja2020}
{Leja}, J., {Speagle}, J.~S., {Johnson}, B.~D., {et~al.} 2020, \apj, 893, 111

\bibitem[{{Leja} {et~al.}(2022){Leja}, {Speagle}, {Ting}, {Johnson}, {Conroy},
  {Whitaker}, {Nelson}, {van Dokkum}, \& {Franx}}]{Leja2022}
{Leja}, J., {Speagle}, J.~S., {Ting}, Y.-S., {et~al.} 2022, \apj, 936, 165

\bibitem[{{Li} {et~al.}(2023){Li}, {Zhang}, {Sugiyama}, {Dalal}, {Rau},
  {Mandelbaum}, {Takada}, {More}, {Strauss}, {Miyatake}, {Shirasaki}, {Hamana},
  {Oguri}, {Luo}, {Nishizawa}, {Takahashi}, {Nicola}, {Osato}, {Kannawadi},
  {Sunayama}, {Armstrong}, {Komiyama}, {Lupton}, {Lust}, {Miyazaki},
  {Murayama}, {Nishimichi}, {Okura}, {Price}, {Tait}, {Tanaka}, \&
  {Wang}}]{Li2023}
{Li}, X., {Zhang}, T., {Sugiyama}, S., {et~al.} 2023, arXiv e-prints,
  arXiv:2304.00702

\bibitem[{{Li} {et~al.}(2024){Li}, {Leja}, {Johnson}, {Tacchella}, {Davies},
  {Belli}, {Park}, \& {Emami}}]{Li2024}
{Li}, Y., {Leja}, J., {Johnson}, B.~D., {et~al.} 2024, arXiv e-prints,
  arXiv:2405.04598

\bibitem[{{Li} \& {Han}(2007)}]{Li2007}
{Li}, Z. \& {Han}, Z. 2007, arXiv e-prints, arXiv:0712.1859

\bibitem[{{Lu} {et~al.}(2024){Lu}, {Daddi}, {Maraston}, {Dickinson}, {Arrabal
  Haro}, {Gobat}, {Renzini}, {Giavalisco}, {Bagley}, {Calabr{\`o}}, {Cheng},
  {de la Vega}, {D'Eugenio}, {Elbaz}, {Finkelstein}, {G{\'o}mez-Guijarro},
  {Gu}, {Hathi}, {Huertas-Company}, {Kartaltepe}, {Koekemoer}, {Le Bail},
  {Lyu}, {Magnelli}, {Mobasher}, {Papovich}, {Pirzkal}, {Rich}, {Tacchella}, \&
  {Yung}}]{Shiying2024}
{Lu}, S., {Daddi}, E., {Maraston}, C., {et~al.} 2024, arXiv e-prints,
  arXiv:2403.07414

\bibitem[{{Lyke} {et~al.}(2020){Lyke}, {Higley}, {McLane}, {Schurhammer},
  {Myers}, {Ross}, {Dawson}, {Chabanier}, {Martini}, {Busca}, {Mas des
  Bourboux}, {Salvato}, {Streblyanska}, {Zarrouk}, {Burtin}, {Anderson},
  {Bautista}, {Bizyaev}, {Brandt}, {Brinkmann}, {Brownstein}, {Comparat},
  {Green}, {de la Macorra}, {Mu{\~n}oz Guti{\'e}rrez}, {Hou}, {Newman},
  {Palanque-Delabrouille}, {P{\^a}ris}, {Percival}, {Petitjean}, {Rich},
  {Rossi}, {Schneider}, {Smith}, {Vivek}, \& {Weaver}}]{Lyke2020}
{Lyke}, B.~W., {Higley}, A.~N., {McLane}, J.~N., {et~al.} 2020, \apjs, 250, 8

\bibitem[{{Madau}(1995)}]{Madau1995}
{Madau}, P. 1995, \apj, 441, 18

\bibitem[{{Madau} \& {Dickinson}(2014)}]{Madau2014}
{Madau}, P. \& {Dickinson}, M. 2014, \araa, 52, 415

\bibitem[{{Magrini} {et~al.}(2012){Magrini}, {Hunt}, {Galli}, {Schneider},
  {Bianchi}, {Maiolino}, {Romano}, {Tosi}, \& {Valiante}}]{Magrini2012}
{Magrini}, L., {Hunt}, L., {Galli}, D., {et~al.} 2012, \mnras, 427, 1075

\bibitem[{{Ma{\'\i}z Apell{\'a}niz}(2024)}]{Apellaniz2024}
{Ma{\'\i}z Apell{\'a}niz}, J. 2024, arXiv e-prints, arXiv:2401.01116

\bibitem[{{Ma{\'\i}z Apell{\'a}niz} {et~al.}(2017){Ma{\'\i}z Apell{\'a}niz},
  {Trigueros P{\'a}ez}, {Bostroem}, {Barb{\'a}}, \& {Evans}}]{Apellaniz2017}
{Ma{\'\i}z Apell{\'a}niz}, J., {Trigueros P{\'a}ez}, E., {Bostroem}, A.~K.,
  {Barb{\'a}}, R.~H., \& {Evans}, C.~J. 2017, in Highlights on Spanish
  Astrophysics IX, ed. S.~{Arribas}, A.~{Alonso-Herrero}, F.~{Figueras},
  C.~{Hern{\'a}ndez-Monteagudo}, A.~{S{\'a}nchez-Lavega}, \&
  S.~{P{\'e}rez-Hoyos}, 510--510

\bibitem[{{Mandelbaum}(2015)}]{Mandelbaum2015}
{Mandelbaum}, R. 2015, Journal of Instrumentation, 10, C05017

\bibitem[{{Mandelbaum}(2018)}]{Mandelbaum2018}
{Mandelbaum}, R. 2018, \araa, 56, 393

\bibitem[{{Mandelbaum} {et~al.}(2018){Mandelbaum}, {Eifler}, {Hlo{\v{z}}ek},
  {Collett}, {Gawiser}, {Scolnic}, {Alonso}, {Awan}, {Biswas}, {Blazek},
  {Burchat}, {Chisari}, {Dell'Antonio}, {Digel}, {Frieman}, {Goldstein},
  {Hook}, {Ivezi{\'c}}, {Kahn}, {Kamath}, {Kirkby}, {Kitching}, {Krause},
  {Leget}, {Marshall}, {Meyers}, {Miyatake}, {Newman}, {Nichol}, {Rykoff},
  {Sanchez}, {Slosar}, {Sullivan}, \& {Troxel}}]{LSSTDESC2018}
{Mandelbaum}, R., {Eifler}, T., {Hlo{\v{z}}ek}, R., {et~al.} 2018, arXiv
  e-prints, arXiv:1809.01669

\bibitem[{{Maraston} {et~al.}(2006){Maraston}, {Daddi}, {Renzini}, {Cimatti},
  {Dickinson}, {Papovich}, {Pasquali}, \& {Pirzkal}}]{Maraston2006}
{Maraston}, C., {Daddi}, E., {Renzini}, A., {et~al.} 2006, \apj, 652, 85

\bibitem[{{Marchesini} {et~al.}(2009){Marchesini}, {van Dokkum}, {F{\"o}rster
  Schreiber}, {Franx}, {Labb{\'e}}, \& {Wuyts}}]{Marchesini2009}
{Marchesini}, D., {van Dokkum}, P.~G., {F{\"o}rster Schreiber}, N.~M., {et~al.}
  2009, \apj, 701, 1765

\bibitem[{{Marigo} \& {Girardi}(2007)}]{Marigo2007}
{Marigo}, P. \& {Girardi}, L. 2007, \aap, 469, 239

\bibitem[{{Marigo} {et~al.}(2008){Marigo}, {Girardi}, {Bressan}, {Groenewegen},
  {Silva}, \& {Granato}}]{Marigo2008}
{Marigo}, P., {Girardi}, L., {Bressan}, A., {et~al.} 2008, \aap, 482, 883

\bibitem[{{Markov} {et~al.}(2024){Markov}, {Gallerani}, {Ferrara},
  {Pallottini}, {Parlanti}, {Di Mascia}, {Sommovigo}, \&
  {Kohandel}}]{Markov2024}
{Markov}, V., {Gallerani}, S., {Ferrara}, A., {et~al.} 2024, arXiv e-prints,
  arXiv:2402.05996

\bibitem[{{Martini} {et~al.}(2013){Martini}, {Dicken}, \&
  {Storchi-Bergmann}}]{Martini2013}
{Martini}, P., {Dicken}, D., \& {Storchi-Bergmann}, T. 2013, \apj, 766, 121

\bibitem[{{Massey} {et~al.}(2013){Massey}, {Hoekstra}, {Kitching}, {Rhodes},
  {Cropper}, {Amiaux}, {Harvey}, {Mellier}, {Meneghetti}, {Miller},
  {Paulin-Henriksson}, {Pires}, {Scaramella}, \& {Schrabback}}]{Massey2013}
{Massey}, R., {Hoekstra}, H., {Kitching}, T., {et~al.} 2013, \mnras, 429, 661

\bibitem[{{Masters} {et~al.}(2015){Masters}, {Capak}, {Stern}, {Ilbert},
  {Salvato}, {Schmidt}, {Longo}, {Rhodes}, {Paltani}, {Mobasher}, {Hoekstra},
  {Hildebrandt}, {Coupon}, {Steinhardt}, {Speagle}, {Faisst}, {Kalinich},
  {Brodwin}, {Brescia}, \& {Cavuoti}}]{Masters2015}
{Masters}, D., {Capak}, P., {Stern}, D., {et~al.} 2015, \apj, 813, 53

\bibitem[{{Masters} {et~al.}(2017){Masters}, {Stern}, {Cohen}, {Capak},
  {Rhodes}, {Castander}, \& {Paltani}}]{Masters2017}
{Masters}, D.~C., {Stern}, D.~K., {Cohen}, J.~G., {et~al.} 2017, \apj, 841, 111

\bibitem[{{Masters} {et~al.}(2019){Masters}, {Stern}, {Cohen}, {Capak},
  {Stanford}, {Hernitschek}, {Galametz}, {Davidzon}, {Rhodes}, {Sanders},
  {Mobasher}, {Castander}, {Pruett}, \& {Fotopoulou}}]{Masters2019}
{Masters}, D.~C., {Stern}, D.~K., {Cohen}, J.~G., {et~al.} 2019, \apj, 877, 81

\bibitem[{{Mathieu} \& {Geller}(2015)}]{Mathieu2015}
{Mathieu}, R.~D. \& {Geller}, A.~M. 2015, in Astrophysics and Space Science
  Library, Vol. 413, Astrophysics and Space Science Library, ed. H.~M.~J.
  {Boffin}, G.~{Carraro}, \& G.~{Beccari}, 29

\bibitem[{{Mathis} {et~al.}(1977){Mathis}, {Rumpl}, \&
  {Nordsieck}}]{Mathis1977}
{Mathis}, J.~S., {Rumpl}, W., \& {Nordsieck}, K.~H. 1977, \apj, 217, 425

\bibitem[{{McCracken} {et~al.}(2012){McCracken}, {Milvang-Jensen}, {Dunlop},
  {Franx}, {Fynbo}, {Le F{\`e}vre}, {Holt}, {Caputi}, {Goranova}, {Buitrago},
  {Emerson}, {Freudling}, {Hudelot}, {L{\'o}pez-Sanjuan}, {Magnard}, {Mellier},
  {M{\o}ller}, {Nilsson}, {Sutherland}, {Tasca}, \& {Zabl}}]{McCracken2012}
{McCracken}, H.~J., {Milvang-Jensen}, B., {Dunlop}, J., {et~al.} 2012, \aap,
  544, A156

\bibitem[{{McCrea}(1964)}]{McCrea1964}
{McCrea}, W.~H. 1964, \mnras, 128, 147

\bibitem[{{McCullough} {et~al.}(2023){McCullough}, {Gruen}, {Amon}, {Roodman},
  {Masters}, {Raichoor}, {Schlegel}, {Canning}, {Castander}, {DeRose},
  {Miquel}, {Myles}, {Newman}, {Slosar}, {Speagle}, {Wilson}, {Aguilar},
  {Ahlen}, {Bailey}, {Brooks}, {Claybaugh}, {Cole}, {Dawson}, {de la Macorra},
  {Doel}, {Forero-Romero}, {Gontcho}, {Guy}, {Kehoe}, {Kremin}, {Landriau}, {Le
  Guillou}, {Levi}, {Manera}, {Martini}, {Meisner}, {Moustakas}, {Nie},
  {Percival}, {Poppett}, {Prada}, {Rezaie}, {Rossi}, {Sanchez}, {Seo},
  {Tarl{\'e}}, {Weaver}, {Zhou}, \& {Zou}}]{McCullough2023}
{McCullough}, J., {Gruen}, D., {Amon}, A., {et~al.} 2023, arXiv e-prints,
  arXiv:2309.13109

\bibitem[{{Meiksin}(2006)}]{Meiksin2006}
{Meiksin}, A. 2006, \mnras, 365, 807

\bibitem[{{Melbourne} \& {Boyer}(2013)}]{Melbourne2013}
{Melbourne}, J. \& {Boyer}, M.~L. 2013, \apj, 764, 30

\bibitem[{{Melbourne} {et~al.}(2012){Melbourne}, {Williams}, {Dalcanton},
  {Rosenfield}, {Girardi}, {Marigo}, {Weisz}, {Dolphin}, {Boyer}, {Olsen},
  {Skillman}, \& {Seth}}]{Melbourne2012}
{Melbourne}, J., {Williams}, B.~F., {Dalcanton}, J.~J., {et~al.} 2012, \apj,
  748, 47

\bibitem[{{Melchior} {et~al.}(2023){Melchior}, {Liang}, {Hahn}, \&
  {Goulding}}]{Melchior2023}
{Melchior}, P., {Liang}, Y., {Hahn}, C., \& {Goulding}, A. 2023, \aj, 166, 74

\bibitem[{{Milone} \& {Latham}(1994)}]{Milone1994}
{Milone}, A.~A.~E. \& {Latham}, D.~W. 1994, \aj, 108, 1828

\bibitem[{{Mor} {et~al.}(2009){Mor}, {Netzer}, \& {Elitzur}}]{Mor2009}
{Mor}, R., {Netzer}, H., \& {Elitzur}, M. 2009, \apj, 705, 298

\bibitem[{{Moser} {et~al.}(2024){Moser}, {Kacprzak}, {Fischbacher},
  {Refregier}, {Grimm}, \& {Tortorelli}}]{Moser2024}
{Moser}, B., {Kacprzak}, T., {Fischbacher}, S., {et~al.} 2024, \jcap, 2024, 049

\bibitem[{{Moustakas} {et~al.}(2013){Moustakas}, {Coil}, {Aird}, {Blanton},
  {Cool}, {Eisenstein}, {Mendez}, {Wong}, {Zhu}, \& {Arnouts}}]{Moustakas2013}
{Moustakas}, J., {Coil}, A.~L., {Aird}, J., {et~al.} 2013, \apj, 767, 50

\bibitem[{{Muzzin} {et~al.}(2013){Muzzin}, {Marchesini}, {Stefanon}, {Franx},
  {McCracken}, {Milvang-Jensen}, {Dunlop}, {Fynbo}, {Brammer}, {Labb{\'e}}, \&
  {van Dokkum}}]{Muzzin2013}
{Muzzin}, A., {Marchesini}, D., {Stefanon}, M., {et~al.} 2013, \apj, 777, 18

\bibitem[{{Myles} {et~al.}(2021){Myles}, {Alarcon}, {Amon}, {S{\'a}nchez},
  {Everett}, {DeRose}, {McCullough}, {Gruen}, {Bernstein}, {Troxel},
  {Dodelson}, {Campos}, {MacCrann}, {Yin}, {Raveri}, {Amara}, {Becker}, {Choi},
  {Cordero}, {Eckert}, {Gatti}, {Giannini}, {Gschwend}, {Gruendl}, {Harrison},
  {Hartley}, {Huff}, {Kuropatkin}, {Lin}, {Masters}, {Miquel}, {Prat},
  {Roodman}, {Rykoff}, {Sevilla-Noarbe}, {Sheldon}, {Wechsler}, {Yanny},
  {Abbott}, {Aguena}, {Allam}, {Annis}, {Bacon}, {Bertin}, {Bhargava},
  {Bridle}, {Brooks}, {Burke}, {Carnero Rosell}, {Carrasco Kind}, {Carretero},
  {Castander}, {Conselice}, {Costanzi}, {Crocce}, {da Costa}, {Pereira},
  {Desai}, {Diehl}, {Eifler}, {Elvin-Poole}, {Evrard}, {Ferrero}, {Fert{\'e}},
  {Flaugher}, {Fosalba}, {Frieman}, {Garc{\'\i}a-Bellido}, {Gaztanaga},
  {Giannantonio}, {Hinton}, {Hollowood}, {Honscheid}, {Hoyle}, {Huterer},
  {James}, {Krause}, {Kuehn}, {Lahav}, {Lima}, {Maia}, {Marshall}, {Martini},
  {Melchior}, {Menanteau}, {Mohr}, {Morgan}, {Muir}, {Ogando}, {Palmese},
  {Paz-Chinch{\'o}n}, {Plazas}, {Rodriguez-Monroy}, {Samuroff}, {Sanchez},
  {Scarpine}, {Secco}, {Serrano}, {Smith}, {Soares-Santos}, {Suchyta},
  {Swanson}, {Tarle}, {Thomas}, {To}, {Varga}, {Weller}, \&
  {Wester}}]{Myles2021}
{Myles}, J., {Alarcon}, A., {Amon}, A., {et~al.} 2021, \mnras, 505, 4249

\bibitem[{{Nagaraj} {et~al.}(2022){Nagaraj}, {Forbes}, {Leja},
  {Foreman-Mackey}, \& {Hayward}}]{Nagaraj2022}
{Nagaraj}, G., {Forbes}, J.~C., {Leja}, J., {Foreman-Mackey}, D., \& {Hayward},
  C.~C. 2022, \apj, 932, 54

\bibitem[{{Nelson} {et~al.}(2019){Nelson}, {Tadaki}, {Tacconi}, {Lutz},
  {F{\"o}rster Schreiber}, {Cibinel}, {Wuyts}, {Lang}, {Leja}, {Montes},
  {Oesch}, {Belli}, {Davies}, {Davies}, {Genzel}, {Lippa}, {Price},
  {{\"U}bler}, \& {Wisnioski}}]{Nelson2019}
{Nelson}, E.~J., {Tadaki}, K.-i., {Tacconi}, L.~J., {et~al.} 2019, \apj, 870,
  130

\bibitem[{{Nenkova} {et~al.}(2008{\natexlab{a}}){Nenkova}, {Sirocky},
  {Ivezi{\'c}}, \& {Elitzur}}]{Nenkova2008}
{Nenkova}, M., {Sirocky}, M.~M., {Ivezi{\'c}}, {\v{Z}}., \& {Elitzur}, M.
  2008{\natexlab{a}}, \apj, 685, 147

\bibitem[{{Nenkova} {et~al.}(2008{\natexlab{b}}){Nenkova}, {Sirocky},
  {Nikutta}, {Ivezi{\'c}}, \& {Elitzur}}]{Nenkova2008b}
{Nenkova}, M., {Sirocky}, M.~M., {Nikutta}, R., {Ivezi{\'c}}, {\v{Z}}., \&
  {Elitzur}, M. 2008{\natexlab{b}}, \apj, 685, 160

\bibitem[{{Newman} \& {Gruen}(2022)}]{Newman2022}
{Newman}, J.~A. \& {Gruen}, D. 2022, \araa, 60, 363

\bibitem[{{Noll} {et~al.}(2009){Noll}, {Burgarella}, {Giovannoli}, {Buat},
  {Marcillac}, \& {Mu{\~n}oz-Mateos}}]{Noll2009}
{Noll}, S., {Burgarella}, D., {Giovannoli}, E., {et~al.} 2009, \aap, 507, 1793

\bibitem[{{Padovani} {et~al.}(2017){Padovani}, {Alexander}, {Assef}, {De
  Marco}, {Giommi}, {Hickox}, {Richards}, {Smol{\v{c}}i{\'c}},
  {Hatziminaoglou}, {Mainieri}, \& {Salvato}}]{Padovani2017}
{Padovani}, P., {Alexander}, D.~M., {Assef}, R.~J., {et~al.} 2017, \aapr, 25, 2

\bibitem[{{Parikh} {et~al.}(2024){Parikh}, {Saglia}, {Thomas}, {Mehrgan},
  {Bender}, \& {Maraston}}]{Parikh2024}
{Parikh}, T., {Saglia}, R., {Thomas}, J., {et~al.} 2024, \mnras, 528, 7338

\bibitem[{{Paxton} {et~al.}(2011){Paxton}, {Bildsten}, {Dotter}, {Herwig},
  {Lesaffre}, \& {Timmes}}]{Paxton2011}
{Paxton}, B., {Bildsten}, L., {Dotter}, A., {et~al.} 2011, \apjs, 192, 3

\bibitem[{{Paxton} {et~al.}(2013){Paxton}, {Cantiello}, {Arras}, {Bildsten},
  {Brown}, {Dotter}, {Mankovich}, {Montgomery}, {Stello}, {Timmes}, \&
  {Townsend}}]{Paxton2013}
{Paxton}, B., {Cantiello}, M., {Arras}, P., {et~al.} 2013, \apjs, 208, 4

\bibitem[{{Paxton} {et~al.}(2015){Paxton}, {Marchant}, {Schwab}, {Bauer},
  {Bildsten}, {Cantiello}, {Dessart}, {Farmer}, {Hu}, {Langer}, {Townsend},
  {Townsley}, \& {Timmes}}]{Paxton2015}
{Paxton}, B., {Marchant}, P., {Schwab}, J., {et~al.} 2015, \apjs, 220, 15

\bibitem[{{Pereira} \& {Miranda}(2007)}]{Pereira2007}
{Pereira}, C.~B. \& {Miranda}, L.~F. 2007, \aap, 462, 231

\bibitem[{{Piotto} {et~al.}(2004){Piotto}, {De Angeli}, {King}, {Djorgovski},
  {Bono}, {Cassisi}, {Meylan}, {Recio-Blanco}, {Rich}, \&
  {Davies}}]{Piotto2004}
{Piotto}, G., {De Angeli}, F., {King}, I.~R., {et~al.} 2004, \apjl, 604, L109

\bibitem[{{Planck Collaboration} {et~al.}(2020){Planck Collaboration},
  {Aghanim}, {Akrami}, {Ashdown}, {Aumont}, {Baccigalupi}, {Ballardini},
  {Banday}, {Barreiro}, {Bartolo}, {Basak}, {Battye}, {Benabed}, {Bernard},
  {Bersanelli}, {Bielewicz}, {Bock}, {Bond}, {Borrill}, {Bouchet}, {Boulanger},
  {Bucher}, {Burigana}, {Butler}, {Calabrese}, {Cardoso}, {Carron},
  {Challinor}, {Chiang}, {Chluba}, {Colombo}, {Combet}, {Contreras}, {Crill},
  {Cuttaia}, {de Bernardis}, {de Zotti}, {Delabrouille}, {Delouis}, {Di
  Valentino}, {Diego}, {Dor{\'e}}, {Douspis}, {Ducout}, {Dupac}, {Dusini},
  {Efstathiou}, {Elsner}, {En{\ss}lin}, {Eriksen}, {Fantaye}, {Farhang},
  {Fergusson}, {Fernandez-Cobos}, {Finelli}, {Forastieri}, {Frailis},
  {Fraisse}, {Franceschi}, {Frolov}, {Galeotta}, {Galli}, {Ganga},
  {G{\'e}nova-Santos}, {Gerbino}, {Ghosh}, {Gonz{\'a}lez-Nuevo}, {G{\'o}rski},
  {Gratton}, {Gruppuso}, {Gudmundsson}, {Hamann}, {Handley}, {Hansen},
  {Herranz}, {Hildebrandt}, {Hivon}, {Huang}, {Jaffe}, {Jones}, {Karakci},
  {Keih{\"a}nen}, {Keskitalo}, {Kiiveri}, {Kim}, {Kisner}, {Knox},
  {Krachmalnicoff}, {Kunz}, {Kurki-Suonio}, {Lagache}, {Lamarre}, {Lasenby},
  {Lattanzi}, {Lawrence}, {Le Jeune}, {Lemos}, {Lesgourgues}, {Levrier},
  {Lewis}, {Liguori}, {Lilje}, {Lilley}, {Lindholm}, {L{\'o}pez-Caniego},
  {Lubin}, {Ma}, {Mac{\'\i}as-P{\'e}rez}, {Maggio}, {Maino}, {Mandolesi},
  {Mangilli}, {Marcos-Caballero}, {Maris}, {Martin}, {Martinelli},
  {Mart{\'\i}nez-Gonz{\'a}lez}, {Matarrese}, {Mauri}, {McEwen}, {Meinhold},
  {Melchiorri}, {Mennella}, {Migliaccio}, {Millea}, {Mitra},
  {Miville-Desch{\^e}nes}, {Molinari}, {Montier}, {Morgante}, {Moss}, {Natoli},
  {N{\o}rgaard-Nielsen}, {Pagano}, {Paoletti}, {Partridge}, {Patanchon},
  {Peiris}, {Perrotta}, {Pettorino}, {Piacentini}, {Polastri}, {Polenta},
  {Puget}, {Rachen}, {Reinecke}, {Remazeilles}, {Renzi}, {Rocha}, {Rosset},
  {Roudier}, {Rubi{\~n}o-Mart{\'\i}n}, {Ruiz-Granados}, {Salvati}, {Sandri},
  {Savelainen}, {Scott}, {Shellard}, {Sirignano}, {Sirri}, {Spencer},
  {Sunyaev}, {Suur-Uski}, {Tauber}, {Tavagnacco}, {Tenti}, {Toffolatti},
  {Tomasi}, {Trombetti}, {Valenziano}, {Valiviita}, {Van Tent}, {Vibert},
  {Vielva}, {Villa}, {Vittorio}, {Wandelt}, {Wehus}, {White}, {White},
  {Zacchei}, \& {Zonca}}]{Planck2020}
{Planck Collaboration}, {Aghanim}, N., {Akrami}, Y., {et~al.} 2020, \aap, 641,
  A6

\bibitem[{{Plazas Malag{\'o}n}(2020)}]{Plazas2020}
{Plazas Malag{\'o}n}, A.~A. 2020, Symmetry, 12, 494

\bibitem[{{Preston} \& {Sneden}(2000)}]{Preston2000}
{Preston}, G.~W. \& {Sneden}, C. 2000, \aj, 120, 1014

\bibitem[{{Pritchard} {et~al.}(2007){Pritchard}, {Furlanetto}, \&
  {Kamionkowski}}]{Pritchard2007}
{Pritchard}, J.~R., {Furlanetto}, S.~R., \& {Kamionkowski}, M. 2007, \mnras,
  374, 159

\bibitem[{{Racca} {et~al.}(2016){Racca}, {Laureijs}, {Stagnaro}, {Salvignol},
  {Lorenzo Alvarez}, {Saavedra Criado}, {Gaspar Venancio}, {Short}, {Strada},
  {B{\"o}nke}, {Colombo}, {Calvi}, {Maiorano}, {Piersanti}, {Prezelus},
  {Rosato}, {Pinel}, {Rozemeijer}, {Lesna}, {Musi}, {Sias}, {Anselmi},
  {Cazaubiel}, {Vaillon}, {Mellier}, {Amiaux}, {Berth{\'e}}, {Sauvage},
  {Azzollini}, {Cropper}, {Pottinger}, {Jahnke}, {Ealet}, {Maciaszek},
  {Pasian}, {Zacchei}, {Scaramella}, {Hoar}, {Kohley}, {Vavrek}, {Rudolph}, \&
  {Schmidt}}]{Racca2016}
{Racca}, G.~D., {Laureijs}, R., {Stagnaro}, L., {et~al.} 2016, in Society of
  Photo-Optical Instrumentation Engineers (SPIE) Conference Series, Vol. 9904,
  Space Telescopes and Instrumentation 2016: Optical, Infrared, and Millimeter
  Wave, ed. H.~A. {MacEwen}, G.~G. {Fazio}, M.~{Lystrup}, N.~{Batalha},
  N.~{Siegler}, \& E.~C. {Tong}, 99040O

\bibitem[{{Rain} {et~al.}(2021){Rain}, {Ahumada}, \& {Carraro}}]{Rain2021}
{Rain}, M.~J., {Ahumada}, J.~A., \& {Carraro}, G. 2021, \aap, 650, A67

\bibitem[{{Reines} {et~al.}(2010){Reines}, {Nidever}, {Whelan}, \&
  {Johnson}}]{Reines2010}
{Reines}, A.~E., {Nidever}, D.~L., {Whelan}, D.~G., \& {Johnson}, K.~E. 2010,
  \apj, 708, 26

\bibitem[{{Richards} {et~al.}(2006){Richards}, {Lacy}, {Storrie-Lombardi},
  {Hall}, {Gallagher}, {Hines}, {Fan}, {Papovich}, {Vanden Berk}, {Trammell},
  {Schneider}, {Vestergaard}, {York}, {Jester}, {Anderson}, {Budav{\'a}ri}, \&
  {Szalay}}]{Richards2006}
{Richards}, G.~T., {Lacy}, M., {Storrie-Lombardi}, L.~J., {et~al.} 2006, \apjs,
  166, 470

\bibitem[{{Rigby} \& {Rieke}(2004)}]{Rigby2004}
{Rigby}, J.~R. \& {Rieke}, G.~H. 2004, \apj, 606, 237

\bibitem[{{Roberts-Borsani} {et~al.}(2024){Roberts-Borsani}, {Treu}, {Shapley},
  {Fontana}, {Pentericci}, {Castellano}, {Morishita}, {Bergamini}, \&
  {Rosati}}]{Roberts-Borsani2024}
{Roberts-Borsani}, G., {Treu}, T., {Shapley}, A., {et~al.} 2024, arXiv
  e-prints, arXiv:2403.07103

\bibitem[{{Rodr{\'\i}guez-Monroy} {et~al.}(2022){Rodr{\'\i}guez-Monroy},
  {Weaverdyck}, {Elvin-Poole}, {Crocce}, {Carnero Rosell}, {Andrade-Oliveira},
  {Avila}, {Bechtol}, {Bernstein}, {Blazek}, {Camacho}, {Cawthon}, {De
  Vicente}, {DeRose}, {Dodelson}, {Everett}, {Fang}, {Ferrero}, {Fert{\'e}},
  {Friedrich}, {Gaztanaga}, {Giannini}, {Gruendl}, {Hartley}, {Herner}, {Huff},
  {Jarvis}, {Krause}, {MacCrann}, {Mena-Fern{\'a}ndez}, {Muir}, {Pandey},
  {Park}, {Porredon}, {Prat}, {Rosenfeld}, {Ross}, {Rozo}, {Rykoff}, {Sanchez},
  {Sanchez Cid}, {Sevilla-Noarbe}, {Tabbutt}, {To}, {Wagoner}, {Wechsler},
  {Aguena}, {Allam}, {Amon}, {Annis}, {Bacon}, {Baxter}, {Bertin}, {Bhargava},
  {Brooks}, {Burke}, {Carrasco Kind}, {Carretero}, {Castander}, {Choi},
  {Conselice}, {Costanzi}, {da Costa}, {Pereira}, {Desai}, {Diehl}, {Flaugher},
  {Fosalba}, {Frieman}, {Garc{\'\i}a-Bellido}, {Giannantonio}, {Gruen},
  {Gschwend}, {Gutierrez}, {Hinton}, {Hollowood}, {Honscheid}, {Huterer},
  {Jain}, {James}, {Kuehn}, {Kuropatkin}, {Lima}, {Maia}, {March}, {Marshall},
  {Melchior}, {Menanteau}, {Miller}, {Miquel}, {Mohr}, {Morgan}, {Palmese},
  {Paz-Chinch{\'o}n}, {Pieres}, {Plazas Malag{\'o}n}, {Roodman}, {Scarpine},
  {Serrano}, {Smith}, {Soares-Santos}, {Suchyta}, {Tarle}, {Thomas}, {Varga},
  \& {DES Collaboration}}]{Rodriguez-Monroy2022}
{Rodr{\'\i}guez-Monroy}, M., {Weaverdyck}, N., {Elvin-Poole}, J., {et~al.}
  2022, \mnras, 511, 2665

\bibitem[{{Salim} {et~al.}(2018){Salim}, {Boquien}, \& {Lee}}]{Salim2018}
{Salim}, S., {Boquien}, M., \& {Lee}, J.~C. 2018, \apj, 859, 11

\bibitem[{{Salim} {et~al.}(2016){Salim}, {Lee}, {Janowiecki}, {da Cunha},
  {Dickinson}, {Boquien}, {Burgarella}, {Salzer}, \& {Charlot}}]{Salim2016}
{Salim}, S., {Lee}, J.~C., {Janowiecki}, S., {et~al.} 2016, \apjs, 227, 2

\bibitem[{{Salim} \& {Narayanan}(2020)}]{Salim2020}
{Salim}, S. \& {Narayanan}, D. 2020, \araa, 58, 529

\bibitem[{{Salinas} {et~al.}(2012){Salinas}, {J{\'\i}lkov{\'a}}, {Carraro},
  {Catelan}, \& {Amigo}}]{Salinas2012}
{Salinas}, R., {J{\'\i}lkov{\'a}}, L., {Carraro}, G., {Catelan}, M., \&
  {Amigo}, P. 2012, \mnras, 421, 960

\bibitem[{{Salpeter}(1955)}]{Salpeter1955}
{Salpeter}, E.~E. 1955, \apj, 121, 161

\bibitem[{{Sandage}(1953)}]{Sandage1953}
{Sandage}, A.~R. 1953, \aj, 58, 61

\bibitem[{{Santucci} {et~al.}(2015){Santucci}, {Placco}, {Rossi}, {Beers},
  {Reggiani}, {Lee}, {Xue}, \& {Carollo}}]{Santucci2015}
{Santucci}, R.~M., {Placco}, V.~M., {Rossi}, S., {et~al.} 2015, \apj, 801, 116

\bibitem[{{Sextl} {et~al.}(2023){Sextl}, {Kudritzki}, {Zahid}, \&
  {Ho}}]{Sextl2023}
{Sextl}, E., {Kudritzki}, R.-P., {Zahid}, H.~J., \& {Ho}, I.~T. 2023, \apj,
  949, 60

\bibitem[{{Shakura} \& {Sunyaev}(1973)}]{Shakura1973}
{Shakura}, N.~I. \& {Sunyaev}, R.~A. 1973, \aap, 24, 337

\bibitem[{{Shapiro} {et~al.}(2013){Shapiro}, {Rowe}, {Goodsall}, {Hirata},
  {Fucik}, {Rhodes}, {Seshadri}, \& {Smith}}]{Shapiro2013}
{Shapiro}, C., {Rowe}, B.~T.~P., {Goodsall}, T., {et~al.} 2013, \pasp, 125,
  1496

\bibitem[{{Shapley} {et~al.}(2015){Shapley}, {Reddy}, {Kriek}, {Freeman},
  {Sanders}, {Siana}, {Coil}, {Mobasher}, {Shivaei}, {Price}, \& {de
  Groot}}]{Shapley2015}
{Shapley}, A.~E., {Reddy}, N.~A., {Kriek}, M., {et~al.} 2015, \apj, 801, 88

\bibitem[{{Silva} {et~al.}(1998){Silva}, {Granato}, {Bressan}, \&
  {Danese}}]{Silva1998}
{Silva}, L., {Granato}, G.~L., {Bressan}, A., \& {Danese}, L. 1998, \apj, 509,
  103

\bibitem[{{Skelton} {et~al.}(2014){Skelton}, {Whitaker}, {Momcheva}, {Brammer},
  {van Dokkum}, {Labb{\'e}}, {Franx}, {van der Wel}, {Bezanson}, {Da Cunha},
  {Fumagalli}, {F{\"o}rster Schreiber}, {Kriek}, {Leja}, {Lundgren}, {Magee},
  {Marchesini}, {Maseda}, {Nelson}, {Oesch}, {Pacifici}, {Patel}, {Price},
  {Rix}, {Tal}, {Wake}, \& {Wuyts}}]{Skelton2014}
{Skelton}, R.~E., {Whitaker}, K.~E., {Momcheva}, I.~G., {et~al.} 2014, \apjs,
  214, 24

\bibitem[{{Smith} {et~al.}(2010){Smith}, {Bailer-Jones}, {Klement}, \&
  {Xue}}]{Smith2010}
{Smith}, K.~W., {Bailer-Jones}, C.~A.~L., {Klement}, R.~J., \& {Xue}, X.~X.
  2010, \aap, 522, A88

\bibitem[{{Smith}(2020)}]{Smith2020}
{Smith}, R.~J. 2020, \araa, 58, 577

\bibitem[{{Sneppen} {et~al.}(2022){Sneppen}, {Steinhardt}, {Hensley}, {Jermyn},
  {Mostafa}, \& {Weaver}}]{Sneppen2022}
{Sneppen}, A., {Steinhardt}, C.~L., {Hensley}, H., {et~al.} 2022, \apj, 931, 57

\bibitem[{{Song} {et~al.}(2016){Song}, {Finkelstein}, {Ashby}, {Grazian}, {Lu},
  {Papovich}, {Salmon}, {Somerville}, {Dickinson}, {Duncan}, {Faber}, {Fazio},
  {Ferguson}, {Fontana}, {Guo}, {Hathi}, {Lee}, {Merlin}, \&
  {Willner}}]{Song2016}
{Song}, M., {Finkelstein}, S.~L., {Ashby}, M. L.~N., {et~al.} 2016, \apj, 825,
  5

\bibitem[{{Stanford} {et~al.}(2021){Stanford}, {Masters}, {Darvish}, {Stern},
  {Cohen}, {Capak}, {Hernitschek}, {Davidzon}, {Rhodes}, {Sanders}, {Mobasher},
  {Castander}, {Paltani}, {Aghanim}, {Amara}, {Auricchio}, {Balestra},
  {Bender}, {Bodendorf}, {Bonino}, {Branchini}, {Brinchmann}, {Capobianco},
  {Carbone}, {Carretero}, {Casas}, {Castellano}, {Cavuoti}, {Cimatti},
  {Cledassou}, {Conselice}, {Corcione}, {Costille}, {Cropper}, {Degaudenzi},
  {Douspis}, {Dubath}, {Dusini}, {Fosalba}, {Frailis}, {Franceschi},
  {Franzetti}, {Fumana}, {Garilli}, {Giocoli}, {Grupp}, {Haugan}, {Hoekstra},
  {Holmes}, {Hormuth}, {Hudelot}, {Jahnke}, {Kiessling}, {Kilbinger},
  {Kitching}, {Kubik}, {K{\"u}mmel}, {Kunz}, {Kurki-Suonio}, {Laureijs},
  {Ligori}, {Lilje}, {Lloro}, {Maiorano}, {Marggraf}, {Markovic}, {Massey},
  {Meneghetti}, {Meylan}, {Moscardini}, {Niemi}, {Padilla}, {Pasian},
  {Pedersen}, {Pettorino}, {Pires}, {Poncet}, {Popa}, {Pozzetti}, {Raison},
  {Roncarelli}, {Rossetti}, {Saglia}, {Scaramella}, {Schneider}, {Secroun},
  {Seidel}, {Serrano}, {Sirignano}, {Sirri}, {Taylor}, {Teplitz}, {Tereno},
  {Toledo-Moreo}, {Valentijn}, {Valenziano}, {Verdoes Kleijn}, {Wang},
  {Zamorani}, {Zoubian}, {Brescia}, {Congedo}, {Conversi}, {Copin}, {Kermiche},
  {Kohley}, {Medinaceli}, {Mei}, {Moresco}, {Morin}, {Munari}, {Polenta},
  {Sureau}, {Tallada Cresp{\'\i}}, {Vassallo}, {Zacchei}, {Andreon}, {Aussel},
  {Baccigalupi}, {Balaguera-Antol{\'\i}nez}, {Baldi}, {Bardelli}, {Biviano},
  {Borsato}, {Bozzo}, {Burigana}, {Cabanac}, {Camera}, {Cappi}, {Carvalho},
  {Casas}, {Castignani}, {Colodro-Conde}, {Coupon}, {Courtois}, {Cuby}, {Da
  Silva}, {de la Torre}, {Di Ferdinando}, {Duncan}, {Dupac}, {Fabricius},
  {Farina}, {Farrens}, {Ferreira}, {Finelli}, {Flose-Reimberg}, {Fotopoulou},
  {Galeotta}, {Ganga}, {Gillard}, {Gozaliasl}, {Graci{\'a}-Carpio}, {Keihanen},
  {Kirkpatrick}, {Lindholm}, {Mainetti}, {Maino}, {Martinet}, {Marulli},
  {Maturi}, {Maurogordato}, {Metcalf}, {Nakajima}, {Neissner}, {Nightingale},
  {Nucita}, {Patrizii}, {Potter}, {Renzi}, {Riccio}, {Romelli}, {S{\'a}nchez},
  {Sapone}, {Schirmer}, {Schultheis}, {Scottez}, {Stanco}, {Tenti}, {Teyssier},
  {Torradeflot}, {Valiviita}, {Viel}, {Whittaker}, {Zucca}, \& {Euclid
  Collaboration}}]{Stanford2021}
{Stanford}, S.~A., {Masters}, D., {Darvish}, B., {et~al.} 2021, \apjs, 256, 9

\bibitem[{{Steidel} {et~al.}(2016){Steidel}, {Strom}, {Pettini}, {Rudie},
  {Reddy}, \& {Trainor}}]{Steidel2016}
{Steidel}, C.~C., {Strom}, A.~L., {Pettini}, M., {et~al.} 2016, \apj, 826, 159

\bibitem[{{Sweigart}(1987)}]{Sweigart1987}
{Sweigart}, A.~V. 1987, \apjs, 65, 95

\bibitem[{{Tacchella} {et~al.}(2018){Tacchella}, {Bose}, {Conroy},
  {Eisenstein}, \& {Johnson}}]{Tacchella2018}
{Tacchella}, S., {Bose}, S., {Conroy}, C., {Eisenstein}, D.~J., \& {Johnson},
  B.~D. 2018, \apj, 868, 92

\bibitem[{{Temple} {et~al.}(2021){Temple}, {Hewett}, \& {Banerji}}]{Temple2021}
{Temple}, M.~J., {Hewett}, P.~C., \& {Banerji}, M. 2021, \mnras, 508, 737

\bibitem[{{Thorne} {et~al.}(2022{\natexlab{a}}){Thorne}, {Robotham},
  {Bellstedt}, {Davies}, {Cook}, {Cortese}, {Holwerda}, {Phillipps}, \&
  {Siudek}}]{Thorne2022}
{Thorne}, J.~E., {Robotham}, A. S.~G., {Bellstedt}, S., {et~al.}
  2022{\natexlab{a}}, \mnras, 517, 6035

\bibitem[{{Thorne} {et~al.}(2022{\natexlab{b}}){Thorne}, {Robotham}, {Davies},
  {Bellstedt}, {Brown}, {Croom}, {Delvecchio}, {Groves}, {Jarvis}, {Shabala},
  {Seymour}, {Whittam}, {Bravo}, {Cook}, {Driver}, {Holwerda}, {Phillipps}, \&
  {Siudek}}]{Thorne2022b}
{Thorne}, J.~E., {Robotham}, A. S.~G., {Davies}, L. J.~M., {et~al.}
  2022{\natexlab{b}}, \mnras, 509, 4940

\bibitem[{{Tomczak} {et~al.}(2014){Tomczak}, {Quadri}, {Tran}, {Labb{\'e}},
  {Straatman}, {Papovich}, {Glazebrook}, {Allen}, {Brammer}, {Kacprzak},
  {Kawinwanichakij}, {Kelson}, {McCarthy}, {Mehrtens}, {Monson}, {Persson},
  {Spitler}, {Tilvi}, \& {van Dokkum}}]{Tomczak2014}
{Tomczak}, A.~R., {Quadri}, R.~F., {Tran}, K.-V.~H., {et~al.} 2014, \apj, 783,
  85

\bibitem[{{Tortorelli} {et~al.}(2018){Tortorelli}, {Della Bruna}, {Herbel},
  {Amara}, {Refregier}, {Alarcon}, {Carretero}, {Castander}, {De Vicente},
  {Eriksen}, {Fernandez}, {Folger}, {Garc{\'\i}a-Bellido}, {Gaztanaga},
  {Miquel}, {Padilla}, {Sanchez}, {Serrano}, {Stothert}, {Tallada}, \&
  {Tonello}}]{Tortorelli2018}
{Tortorelli}, L., {Della Bruna}, L., {Herbel}, J., {et~al.} 2018, \jcap, 2018,
  035

\bibitem[{{Tortorelli} {et~al.}(2020){Tortorelli}, {Fagioli}, {Herbel},
  {Amara}, {Kacprzak}, \& {Refregier}}]{Tortorelli2020}
{Tortorelli}, L., {Fagioli}, M., {Herbel}, J., {et~al.} 2020, \jcap, 2020, 048

\bibitem[{{Tortorelli} {et~al.}(2021){Tortorelli}, {Siudek}, {Moser},
  {Kacprzak}, {Berner}, {Refregier}, {Amara}, {Garc{\'\i}a-Bellido}, {Cabayol},
  {Carretero}, {Castander}, {De Vicente}, {Eriksen}, {Fernandez}, {Gaztanaga},
  {Hildebrandt}, {Joachimi}, {Miquel}, {Sevilla-Noarbe}, {Padilla}, {Renard},
  {Sanchez}, {Serrano}, {Tallada-Cresp{\'\i}}, \& {Wright}}]{Tortorelli2021}
{Tortorelli}, L., {Siudek}, M., {Moser}, B., {et~al.} 2021, \jcap, 2021, 013

\bibitem[{{Tremonti} {et~al.}(2004){Tremonti}, {Heckman}, {Kauffmann},
  {Brinchmann}, {Charlot}, {White}, {Seibert}, {Peng}, {Schlegel}, {Uomoto},
  {Fukugita}, \& {Brinkmann}}]{Tremonti2004}
{Tremonti}, C.~A., {Heckman}, T.~M., {Kauffmann}, G., {et~al.} 2004, \apj, 613,
  898

\bibitem[{{Tripodi} {et~al.}(2024){Tripodi}, {D'Eugenio}, {Maiolino}, {Curti},
  {Scholtz}, {Tacchella}, {Bunker}, {Trussler}, {Cameron}, {Arribas}, {Baker},
  {Brada{\v{c}}}, {Carniani}, {Charlot}, {Ji}, {Ji}, {Robertson}, {{\"U}bler},
  {Venturi}, {Willmer}, \& {Witstok}}]{Tripodi2024}
{Tripodi}, R., {D'Eugenio}, F., {Maiolino}, R., {et~al.} 2024, arXiv e-prints,
  arXiv:2403.08431

\bibitem[{{Urry} \& {Padovani}(1995)}]{Urry1995}
{Urry}, C.~M. \& {Padovani}, P. 1995, \pasp, 107, 803

\bibitem[{{van den Busch} {et~al.}(2022){van den Busch}, {Wright},
  {Hildebrandt}, {Bilicki}, {Asgari}, {Joudaki}, {Blake}, {Heymans},
  {Kannawadi}, {Shan}, \& {Tr{\"o}ster}}]{vandenbusch2022}
{van den Busch}, J.~L., {Wright}, A.~H., {Hildebrandt}, H., {et~al.} 2022,
  \aap, 664, A170

\bibitem[{{van Winckel}(2003)}]{VanWinckel2003}
{van Winckel}, H. 2003, \araa, 41, 391

\bibitem[{{Vassiliadis} \& {Wood}(1994)}]{Vassiliadis1994}
{Vassiliadis}, E. \& {Wood}, P.~R. 1994, \apjs, 92, 125

\bibitem[{{Viel} {et~al.}(2005){Viel}, {Lesgourgues}, {Haehnelt}, {Matarrese},
  \& {Riotto}}]{Viel2005}
{Viel}, M., {Lesgourgues}, J., {Haehnelt}, M.~G., {Matarrese}, S., \& {Riotto},
  A. 2005, \prd, 71, 063534

\bibitem[{{Villaume} {et~al.}(2015){Villaume}, {Conroy}, \&
  {Johnson}}]{Villaume2015}
{Villaume}, A., {Conroy}, C., \& {Johnson}, B.~D. 2015, \apj, 806, 82

\bibitem[{{Wang} {et~al.}(2023{\natexlab{a}}){Wang}, {Leja}, {Atek}, {Labbe},
  {Li}, {Bezanson}, {Brammer}, {Cutler}, {Dayal}, {Furtak}, {Greene},
  {Kokorev}, {Pan}, {Price}, {Suess}, {Weaver}, {Whitaker}, \&
  {Williams}}]{Wang2023b}
{Wang}, B., {Leja}, J., {Atek}, H., {et~al.} 2023{\natexlab{a}}, arXiv
  e-prints, arXiv:2310.06781

\bibitem[{{Wang} {et~al.}(2023{\natexlab{b}}){Wang}, {Leja}, {Bezanson},
  {Johnson}, {Khullar}, {Labb{\'e}}, {Price}, {Weaver}, \&
  {Whitaker}}]{Wang2023}
{Wang}, B., {Leja}, J., {Bezanson}, R., {et~al.} 2023{\natexlab{b}}, \apjl,
  944, L58

\bibitem[{{Wang} {et~al.}(2024){Wang}, {Leja}, {Labb{\'e}}, {Bezanson},
  {Whitaker}, {Brammer}, {Furtak}, {Weaver}, {Price}, {Zitrin}, {Atek}, {Coe},
  {Cutler}, {Dayal}, {van Dokkum}, {Feldmann}, {Marchesini}, {Franx},
  {F{\"o}rster Schreiber}, {Fujimoto}, {Geha}, {Glazebrook}, {de Graaff},
  {Greene}, {Juneau}, {Kassin}, {Kriek}, {Khullar}, {Maseda}, {Mowla},
  {Muzzin}, {Nanayakkara}, {Nelson}, {Oesch}, {Pacifici}, {Pan}, {Papovich},
  {Setton}, {Shapley}, {Smit}, {Stefanon}, {Suess}, {Taylor}, \&
  {Williams}}]{Wang2024}
{Wang}, B., {Leja}, J., {Labb{\'e}}, I., {et~al.} 2024, \apjs, 270, 12

\bibitem[{{Wang} {et~al.}(2023{\natexlab{c}}){Wang}, {Yang}, \&
  {Li}}]{WangYang2023}
{Wang}, Q., {Yang}, X.~J., \& {Li}, A. 2023{\natexlab{c}}, \mnras, 525, 983

\bibitem[{{Weaver} {et~al.}(2023){Weaver}, {Davidzon}, {Toft}, {Ilbert},
  {McCracken}, {Gould}, {Jespersen}, {Steinhardt}, {Lagos}, {Capak}, {Casey},
  {Chartab}, {Faisst}, {Hayward}, {Kartaltepe}, {Kauffmann}, {Koekemoer},
  {Kokorev}, {Laigle}, {Liu}, {Long}, {Magdis}, {McPartland}, {Milvang-Jensen},
  {Mobasher}, {Moneti}, {Peng}, {Sanders}, {Shuntov}, {Sneppen}, {Valentino},
  {Zalesky}, \& {Zamorani}}]{Weaver2023}
{Weaver}, J.~R., {Davidzon}, I., {Toft}, S., {et~al.} 2023, \aap, 677, A184

\bibitem[{{Weaver} {et~al.}(2022){Weaver}, {Kauffmann}, {Ilbert}, {McCracken},
  {Moneti}, {Toft}, {Brammer}, {Shuntov}, {Davidzon}, {Hsieh}, {Laigle},
  {Anastasiou}, {Jespersen}, {Vinther}, {Capak}, {Casey}, {McPartland},
  {Milvang-Jensen}, {Mobasher}, {Sanders}, {Zalesky}, {Arnouts}, {Aussel},
  {Dunlop}, {Faisst}, {Franx}, {Furtak}, {Fynbo}, {Gould}, {Greve}, {Gwyn},
  {Kartaltepe}, {Kashino}, {Koekemoer}, {Kokorev}, {Le F{\`e}vre}, {Lilly},
  {Masters}, {Magdis}, {Mehta}, {Peng}, {Riechers}, {Salvato}, {Sawicki},
  {Scarlata}, {Scoville}, {Shirley}, {Silverman}, {Sneppen}, {Smolc̆i{\'c}},
  {Steinhardt}, {Stern}, {Tanaka}, {Taniguchi}, {Teplitz}, {Vaccari}, {Wang},
  \& {Zamorani}}]{Weaver2022}
{Weaver}, J.~R., {Kauffmann}, O.~B., {Ilbert}, O., {et~al.} 2022, \apjs, 258,
  11

\bibitem[{{Weigel} {et~al.}(2016){Weigel}, {Schawinski}, \&
  {Bruderer}}]{Weigel2016}
{Weigel}, A.~K., {Schawinski}, K., \& {Bruderer}, C. 2016, \mnras, 459, 2150

\bibitem[{{Wright} {et~al.}(2018){Wright}, {Driver}, \&
  {Robotham}}]{Wright2018}
{Wright}, A.~H., {Driver}, S.~P., \& {Robotham}, A.~S.~G. 2018, \mnras, 480,
  3491

\bibitem[{{Wright} {et~al.}(2019){Wright}, {Hildebrandt}, {Kuijken}, {Erben},
  {Blake}, {Buddelmeijer}, {Choi}, {Cross}, {de Jong}, {Edge},
  {Gonzalez-Fernandez}, {Gonz{\'a}lez Solares}, {Grado}, {Heymans}, {Irwin},
  {Kupcu Yoldas}, {Lewis}, {Mann}, {Napolitano}, {Radovich}, {Schneider},
  {Sif{\'o}n}, {Sutherland}, {Sutorius}, \& {Verdoes Kleijn}}]{Wright2019}
{Wright}, A.~H., {Hildebrandt}, H., {Kuijken}, K., {et~al.} 2019, \aap, 632,
  A34

\bibitem[{{Wright} {et~al.}(2020){Wright}, {Hildebrandt}, {van den Busch}, \&
  {Heymans}}]{Wright2020}
{Wright}, A.~H., {Hildebrandt}, H., {van den Busch}, J.~L., \& {Heymans}, C.
  2020, \aap, 637, A100

\bibitem[{{Xin} \& {Deng}(2005)}]{Xin2005}
{Xin}, Y. \& {Deng}, L. 2005, \apj, 619, 824

\bibitem[{{Xue} {et~al.}(2008){Xue}, {Rix}, {Zhao}, {Re Fiorentin}, {Naab},
  {Steinmetz}, {van den Bosch}, {Beers}, {Lee}, {Bell}, {Rockosi}, {Yanny},
  {Newberg}, {Wilhelm}, {Kang}, {Smith}, \& {Schneider}}]{Xue2008}
{Xue}, X.~X., {Rix}, H.~W., {Zhao}, G., {et~al.} 2008, \apj, 684, 1143

\bibitem[{{Zahid} {et~al.}(2016){Zahid}, {Geller}, {Fabricant}, \&
  {Hwang}}]{Zahid2016}
{Zahid}, H.~J., {Geller}, M.~J., {Fabricant}, D.~G., \& {Hwang}, H.~S. 2016,
  \apj, 832, 203

\bibitem[{{Zhang} {et~al.}(2023){Zhang}, {Yuan}, \& {Chen}}]{Zhang2023}
{Zhang}, R., {Yuan}, H., \& {Chen}, B. 2023, \apjs, 269, 6

\bibitem[{{Zhao} {et~al.}(2012){Zhao}, {Zhao}, {Chu}, {Jing}, \&
  {Deng}}]{Zhao2012}
{Zhao}, G., {Zhao}, Y.-H., {Chu}, Y.-Q., {Jing}, Y.-P., \& {Deng}, L.-C. 2012,
  Research in Astronomy and Astrophysics, 12, 723

\end{thebibliography}

\begin{appendix}

\section{AGN template comparison with SDSS DR16}
\label{appendix:agn_comparison}

\begin{figure}[h]
   \centering
   \includegraphics[width=\hsize]{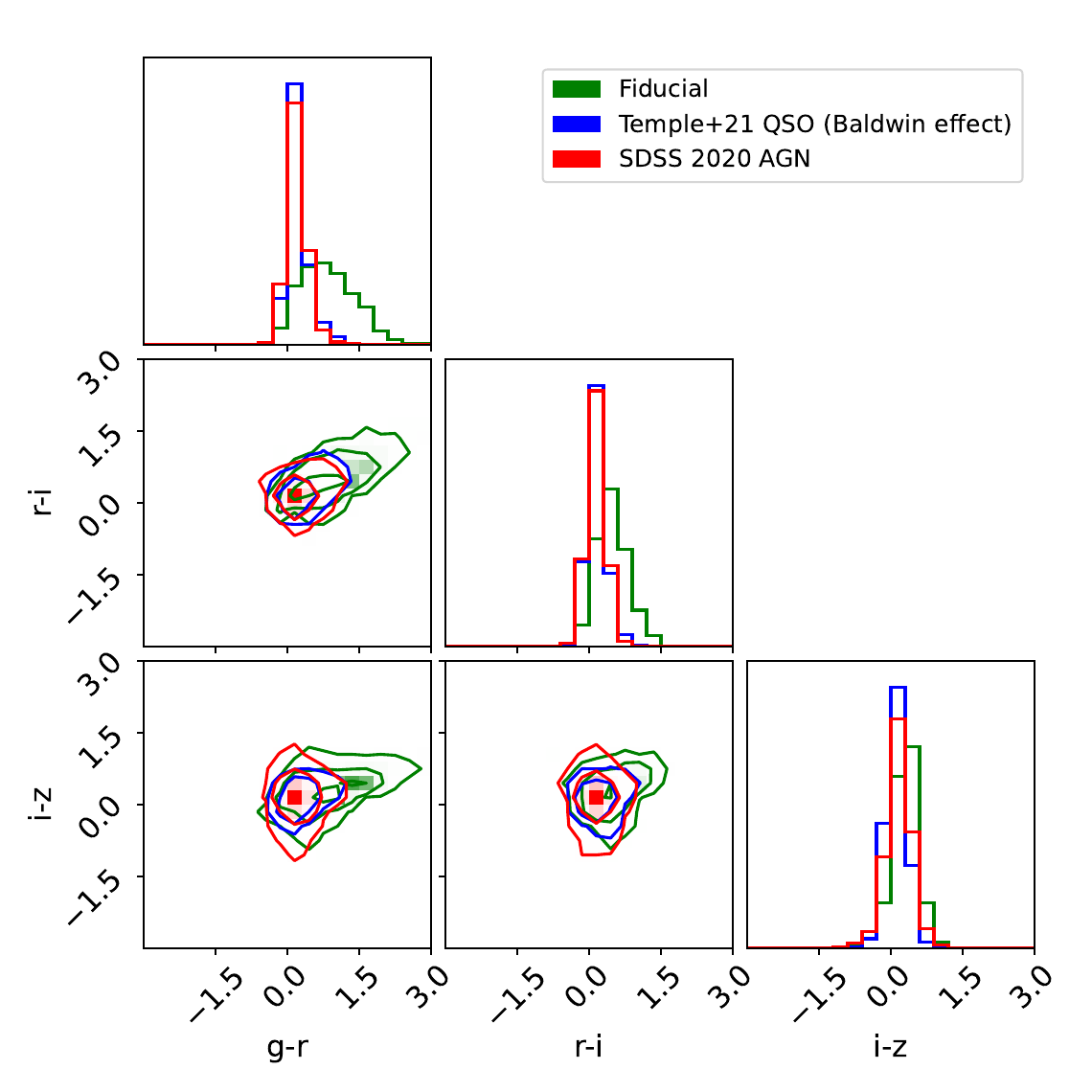}
      \caption{Comparison between the SDSS quasar $g-r$, $r-i$ and $i-z$ colours (red contours), the \textsc{Prospector-}$\beta$ colours with the \citealt{Nenkova2008,Nenkova2008b} templates (green contours), and the \textsc{Prospector-}$\beta$ colours with the \cite{Temple2021} templates (blue contours).
              }
    \label{fig:tortorelli_2024_figureA1}
\end{figure}

In this work, we use the \cite{Temple2021} templates to model the AGN contribution to the galaxy SEDs. These templates were developed using the SDSS DR 16 quasar population presented in \cite{Lyke2020}. The parametric model is able to reproduce the distribution of observed SDSS-UKIDSS-WISE quasar colours over a wide range of redshift ($0 < z < 5$) and luminosity ($-29 < M_i < -22 $) with the aim of providing predictions for the Rubin-LSST and Euclid surveys. We adopt these templates because the \cite{Nenkova2008,Nenkova2008b} templates aim at reproducing only the dust torus emission from the central AGN of a galaxy, while the \cite{Temple2021} templates provide a representation of the complete spectral features that an AGN imprints on a galaxy SED. 

Figure \ref{fig:tortorelli_2024_figureA1} shows the comparison between the SDSS $g-r$, $r-i$ and $i-z$ optical colours from the DR16 quasar catalogue (red contours) and those generated with the fiducial \textsc{Prospector-}$\beta$ model (green contours) and with the addition of the \cite{Temple2021} templates (blue contours). To compare these samples and highlight the AGN contribution, we selected the objects for which the bolometric luminosity of the AGN is greater than $10^{-1}$ that of the hosting galaxy \citep{Leja2018}. We also cut the samples at the $i$-band limiting magnitude of the SDSS DR16 quasar catalogue. The figure clearly shows how the \cite{Temple2021} templates provide a better description of the SDSS quasar optical colours than the \textsc{Prospector-}$\beta$ model with the \cite{Nenkova2008,Nenkova2008b} templates. The residual differences between the \cite{Temple2021} templates colours and the SDSS quasar optical colours can be attributed to the fact that we are comparing mock apparent magnitudes (blue contours) against observed ones affected by photometric noise (red contours).

\section{Photometric noise impact on SOM cell assignment algorithm}
\label{appendix:noise_impact_on_som}

Photometric noise may impact the SOM cell assignment as shown in \cite{McCullough2023}. To test how much our conclusions on the fraction of cell change are impacted by photometric noise, we introduced COSMOS2020-like photometric errors in the cell assignment algorithm. We find that adding the median COSMOS2020 photometric errors change the fractions of galaxies each component moves to a different cell by at most $1-2\%$. This implies that the variation in the SOM cell assignment we see is completely driven by the galaxy population colour variation and that a survey with COSMOS-like photometric precision,  for instance Rubin-LSST (see Appendix \ref{appendix:comparison_cosmos_rubin_photometry}), is introducing a noise in the assignment algorithm that is not impacting the conclusions we see in this work.

\section{COSMOS2020 and Rubin-LSST simulated photometric depths}
\label{appendix:comparison_cosmos_rubin_photometry}

The COSMOS2020 \citep{Weaver2022} reference photometric redshift catalogue contains homogeneous and robust multi-wavelength photometry for a sample of 1.7 million sources. In our work, we used the photometric errors of the COSMOS2020 catalogue as representatives of the ones expected for Rubin-LSST data. In order to support this claim, we compared the depths at $3\sigma$ measured in a $3 ''$ diameter aperture shown in Fig. 3 of \cite{Weaver2022} with those generated using the LSST Error model described in Appendix B of \cite{Crenshaw2024}. The latter depths refer to coadded images for stationary sources after 10 years. The Rubin-LSST $g,r,i,z,y$ bands depths are similar to the ones in Fig. 3 of \cite{Weaver2022}, while the Rubin-LSST u-band depth is roughly one magnitude shallower than the COSMOS2020 one.

\end{appendix}


\end{document}